\documentclass[fleqn,usenatbib]{mnras}

\usepackage{newtxtext,newtxmath}
\usepackage{multicol} 

\usepackage[T1]{fontenc}

\DeclareRobustCommand{\VAN}[3]{#2}
\let\VANthebibliography\thebibliography
\def\thebibliography{\DeclareRobustCommand{\VAN}[3]{##3}\VANthebibliography}

\newcommand{\casa}{\textsc{\large casa}}
\newcommand{\getsf}{\textsc{\large getsf}}
\newcommand{\getsources}{\textsc{\large getsources}}
\newcommand{\scousepy}{\textsc{\large scousepy}}
\newcommand{\astrodendro}{\textsc{\large astrodendro}}
\newcommand{\acorns}{\textsc{\large acorns}}
\newcommand{\tablenotemarknew}[1]{$^{#1}$}%

\newcommand{\massrate}{$M_{\odot}$\,yr$^{-1}$}

\newcommand{\hii}{H\textsc{ii}~}
\newcommand{\msun}{$ M_\odot$}

\newcommand{\kms}{km\,s$^{-1}$}
\newcommand{\jybeam}{Jy\,beam$^{-1}$}
\newcommand{\mjybeam}{mJy\,beam$^{-1}$}

\newcommand{\degree}{$^{\circ}$}
\newcommand{\parcsec}{\mbox{$.\!\!\arcsec$}}
\newcommand{\ssstyle}{\scriptscriptstyle}

\newcommand{\htcop}{H$^{13}$CO$^+$}
\newcommand{\hcop}{HCO$^+$}
\newcommand{\hcn}{HCN}
\newcommand{\htcn}{H$^{13}$CN}
\newcommand{\sio}{SiO}
\newcommand{\cch}{CCH}
\newcommand{\chtoh}{CH$_3$OH}
\newcommand{\hctn}{HC$_3$N}
\newcommand{\cs}{CS}
\newcommand{\so}{SO}
\newcommand{\htcs}{H$_2$CS}
\newcommand{\nthp}{N$_2$H$^+$}

\usepackage{graphicx}	
\usepackage{amsmath}	
\usepackage[figuresleft]{rotating}   
\usepackage{adjustbox}
\usepackage{pdflscape}
\usepackage[flushleft]{threeparttable}

\title[Steady Accretion in SDC335]{ATOMS: ALMA Three-millimeter Observations of Massive Star-forming regions -- XV. Steady Accretion from Global Collapse to Core Feeding in Massive Hub-filament System SDC335}

\author[Xu et al.]{
Feng-Wei Xu,$^{1,2}$
Ke Wang,$^{1}$ \thanks{Contact e-mail: \href{mailto:kwang.astro@pku.edu.cn}{kwang.astro@pku.edu.cn}}
Tie Liu,$^{3}$ 
Paul F. Goldsmith,$^{4}$ 
Qizhou Zhang,$^{5}$ 
Mika Juvela,$^{6}$ 
Hong-Li Liu,$^{7}$ 
\newauthor
Sheng-Li Qin,$^{7}$ 
Guang-Xing Li,$^{8}$ 
Anandmayee Tej,$^{9}$ 
Guido Garay,$^{10}$ 
Leonardo Bronfman,$^{10}$ 
Shanghuo Li,$^{11}$ 
\newauthor
Yue-Fang Wu,$^{2,1}$ 
Gilberto C. Gómez,$^{12}$ 
Enrique Vázquez-Semadeni,$^{12}$ 
Ken'ichi Tatematsu,$^{13}$ 
Zhiyuan Ren,$^{14}$ 
\newauthor
Yong Zhang,$^{15}$ 
L. Viktor Toth,$^{16}$ 
Xunchuan Liu,$^{3}$ 
Nannan Yue,$^{1}$ 
Siju Zhang,$^{1}$ 
Tapas Baug,$^{17}$ 
Namitha Issac,$^{18}$ 
\newauthor
Amelia M. Stutz,$^{19}$ 
Meizhu Liu,$^{7}$ 
Gary A. Fuller,$^{20,21}$ 
Mengyao Tang,$^{22}$ 
Chao Zhang,$^{23}$ 
Lokesh Dewangan,$^{24}$ 
\newauthor
Chang Won Lee,$^{25,11}$ 
Jianwen Zhou,$^{14}$ 
Jinjin Xie,$^{3}$ 
Wenyu Jiao,$^{1,2}$ 
Chao Wang,$^{1,2}$ 
Rong Liu,$^{14}$ 
Qiuyi Luo,$^{3}$ 
\newauthor
Archana Soam,$^{18}$ 
and Chakali Eswaraiah$^{26}$ 
}

\date{Update until \today}

\pubyear{2022}

\begin{document}

\label{firstpage}
\pagerange{\pageref{firstpage}--\pageref{lastpage}}
\maketitle

\begin{abstract}
We present ALMA Band-3/7 observations towards ``the Heart'' of a massive hub-filament system (HFS) SDC335, to investigate its fragmentation and accretion. At a resolution of $\sim0.03$\,pc, 3\,mm continuum emission resolves two massive dense cores MM1 and MM2, with $383(^{\ssstyle+234}_{\ssstyle-120})$\,\msun~(10--24\% mass of ``the Heart'') and $74(^{\ssstyle+47}_{\ssstyle-24})$\,\msun, respectively. With a resolution down to $0.01$\,pc, 0.87\,mm continuum emission shows MM1 further fragments into six condensations and multi-transition lines of \htcs~provide temperature estimation. The relation between separation and mass of condensations at a scale of 0.01\,pc favors turbulent Jeans fragmentation where the turbulence seems to be scale-free rather than scale-dependent. We use the \htcop\,$J=1-0$ emission line to resolve the complex gas motion inside ``the Heart'' in position-position-velocity space. We identify four major gas streams connected to large-scale filaments, inheriting the anti-clockwise spiral pattern. Along these streams, gas feed the central massive core MM1. Assuming an inclination angle of $45(\pm15)^{\circ}$ and a \htcop~abundance of $5(\pm3)\times10^{-11}$, the total mass infall rate is estimated to be $2.40(\pm0.78)\times10^{-3}$\,\massrate, numerically consistent with the accretion rates derived from the clump-scale spherical infall model and the core-scale outflows. The consistency suggests a continuous, near steady-state, and efficient accretion from global collapse, therefore ensuring core feeding. Our comprehensive study of SDC335 showcases the detailed gas kinematics in a prototypical massive infalling clump, and calls for further systematic and statistical studies in a large sample.
\end{abstract}

\begin{keywords}
star: formation - ISM: kinematics and dynamics - ISM: clouds - stars: formation - stars: protostars
\end{keywords}

\twocolumn

\section{Introduction}\label{sec:intro}
\subsection{Overview: how do high-mass stars gain their mass?}
High-mass stars ($>8$\,\msun) play a major role in the energy budget of galaxies via their radiation, wind, and supernova events, but the picture of their formation remains unclear. 
Generally speaking, high-mass star formation is a complex process but the key question is how a massive star gains its mass (i.e., mass assembly). For instance, assuming a core-to-star efficiency of 50\%, a massive star of 10\,M$_\odot$ would require a core of at least 20\,M$_\odot$. However, under typical conditions (e.g., sound speed of $c_s=0.2$\,\kms and gas number density $n_{\mathrm{H}}=10^5$\,cm$^{-3}$), the Jeans mass is only $\sim$\,0.2\,M$_\odot$ \citep{2021ApJ...912..156K}. Therefore, it is unclear how a core with more than 100 Jeans masses would survive fragmentation rather than giving rise to hundreds of low-mass cores. More specifically, the mass assembly involves two critical physical process: accretion and fragmentation, whose roles need to be distinguished. 

Two possible models have been proposed to explain the process. 
The first one is the ``turbulent core'' model \citep{2003ApJ...585..850M}, predicting that the high-mass prestellar cores are supported against collapse and fragmentation by a large degree of turbulence and/or strong magnetic fields. In other words, the ``turbulent core'' model suggests a monolithic collapse, as a scaled-up version of low-mass star formation. The second one is the ``competitive accretion'' model \citep{2001MNRAS.323..785B,2004MNRAS.349..735B}, where massive stars start with initial Jeans fragments and then grow their mass via Bondi-Hoyle accretion or regulation of tidal field in a successful competition with others \citep{2018ApJ...861...14C}.

There have been numerous observational studies aimed at testing the aforementioned two theoretical ideas. In the perspective of ``binary opposition'', massive prestellar cores should serve as a key discriminator between the two proposed theoretical models mentioned above. Many observations have searched for massive prestellar cores \citep[e.g.][]{2009ApJ...696..268Z,2011ApJ...733...26Z,2011ApJ...726...97L,2012ApJ...745L..30W,WangK2014,2014ApJ...796L...2C,2017A&A...604A..74S,2017ApJ...834..193K,2018sf2a.conf..311L,2019A&A...626A.132M,2019ApJ...886...36S,2019ApJ...886..102S,2019ApJ...886..130L}. 
However, few convincing candidates have been found until the present, and the high-mass prestellar core is becoming the \textit{Holy Grail} of star formation studies.

In another way, many researches have focused on massive clumps associated with infrared dark clouds (IRDCs), whose temperature can be as low as 15\,K \citep{2021SCPMA..6479511X}.
The reason is that the fragmentation of molecular gas and core properties are both time-dependent, and evolve over time as physical conditions in the cloud vary during star formation. 
So it is challenging to pinpoint the physical conditions that give rise to the fragmentation observed. 
These investigations help reveal the initial conditions of massive star formation. 
For example, \citet{2009ApJ...696..268Z} conducted arcsecond resolution studies of the IRDC G28.34+0.06 with the Submillimeter Array and found that dense cores giving rise to massive stars are much more massive than the thermal Jeans mass of the clump.
This discovery challenges the notion in the competitive accretion model where massive stars should arise from cores of thermal Jean mass. 
The larger core mass in the fragments demands additional support from turbulence and magnetic fields, since the gas temperatures in IRDCs are typically between 10 to 20\,K \citep[e.g.,][]{2006A&A...450..569P,2008ApJ...672L..33W,2012ApJ...745L..30W,2021SCPMA..6479511X}. 
On the other hand, observations also find that the mass of these cores does not contain sufficient material to form a massive star, and the cores typically continue to fragment when observed at higher angular resolution \citep{WangK2011,WangK2014,2015ApJ...804..141Z}. 
Therefore, the idea of monolithic collapse for massive star formation does not match the observations. 
More recent observations of IRDCs with the Atacama Large Millimeter/submillimeter Array (ALMA) routinely reach a mass sensitivity far below the thermal Jeans mass \citep{2015ApJ...804..141Z,2019ApJ...886...36S,2019ApJ...886..102S} and detect low-mass cores in the clumps that are compatible with the thermal Jeans mass. These cores may form low-mass stars in a cluster. However, massive cores in the protocluster remain to be much larger than the thermal Jeans mass.

To summarize, these observations point to a picture of massive star formation in which dense cores continue to gain material from the molecular clump while the embedded protostar undergoes accretion. 
This scenario is somewhat similar to the competitive accretion. 
However, it differs in two important aspects. First, dense cores harboring massive stars are more massive than the thermal Jeans mass, and second, accretion is likely dominated by gas accretion in response to gravity rather than Bondi-Hoyle accretion.

\subsection{Hub-filament systems: laboratories to study mass transfer}
Both observations and simulations have indicated another possible mechanism for massive star formation. Recent observations show that the filaments detected in our Galaxy are well-established as some of the main sites for star formation \citep{2010A&A...518L.100M,2010A&A...518L.102A,2014prpl.conf...27A,2015MNRAS.450.4043W,WangK2016,2016A&A...590A...2S,2022ApJS..259...36G}. In addition, the filaments are observed to funnel material to cores \citep{2010A&A...520A..49S,2012ApJ...745...61L,2013ApJ...766..115K,2014A&A...561A..83P,2018ApJ...852...12Y,2018MNRAS.473.4220L,2019MNRAS.485.4686W,2022ApJ...926..165L}, which can largely ease the burden of having to accumulate an excessive amount of mass during the prestellar core phase. 
As such, additional mass can be accumulated later for forming a massive star during its accretion phase along filaments \citep[e.g.,][]{WangKe2018,2021ApJ...912..156K}, which is also suggested by numerical studies \citep{GV14, 2020ApJ...900...82P, Naranjo+22}.

Hub-filament system (HFS) is defined as a junction of three or more filaments \citep[the junction is called ``hub'' with a higher density;][]{2009ApJ...700.1609M}. 
In recent ten years, HFS have been frequently observed to form high-mass stars \citep{2012A&A...543L...3H,2012ApJ...745...61L,2016ApJ...824...31L,2013A&A...555A.112P,2014A&A...561A..83P,2018ApJ...852...12Y,ATOMS_XI_Zhou}. Moreover, converging flows along the filaments channel gas to the hub where star formation is often most active \citep{2010ApJ...725...17G}. 
Such a scenario is similar to the traditional ``clump-fed'' picture \citep{2009MNRAS.400.1775S} in the sense that the gas mass reservoir is extended over a far larger spatial scale than the core in which the star has formed or is forming.

Theoretically, the nature of the flow feeding the hubs should be related to the origin of the HFS. 
The ``global hierarchical collapse'' \citep[GHC;][]{VS+19} model attributes the formation of the HFS to anisotropic gravitational contraction from the cloud to the filament scale. 
It is well known that during pressureless evolution, a triaxial spheroid contracts first along its shortest dimension \citep{Lin+65}, forming sheets and then filaments. 
Since molecular clouds are known to contain a large number of Jeans masses, the GHC model further assumes that the evolution is dominated by gravity: filaments form (with fluctuations along them) by contraction from the cloud scale, and the gravity of the fluctuations redirects the gas in the filaments towards them, causing the fluctuations to accrete from mass in the filaments, becoming hubs \citep{GV14}. This process may then repeat itself within the hubs when they acquire several Jeans masses, causing their subsequent fragmentation and the formation of streamers that feed young stellar objects and their disks, thus constituting a hierarchy of collapses connected through filaments. In the GHC model, filaments continuously accrete from their environment and direct gas flows toward the clumps at the positions where two or more filaments meet, i.e., the next level in the hierarchy.
Simulations show that this flow redirection happens smoothly, without the presence of a shock, which should lead to specific chemical signatures \citep{Gomez+22}.

Although HFSs have already been observed to efficiently transfer mass to the clump center \citep{2013A&A...555A.112P,2014A&A...561A..83P,2015ApJ...804..141Z,2016ApJ...824...31L,2018ApJ...852...12Y,2018ApJ...855....9L,2020ApJ...903...13D}, the detailed process of mass transfer and fragmentation remain unclear. To be specific, in which way is the mass transferred inwards and how efficient is the transfer process?

\subsection{SDC335: a prototypical hub-filament system}
\begin{figure*}
\centering
\includegraphics[scale=0.5]{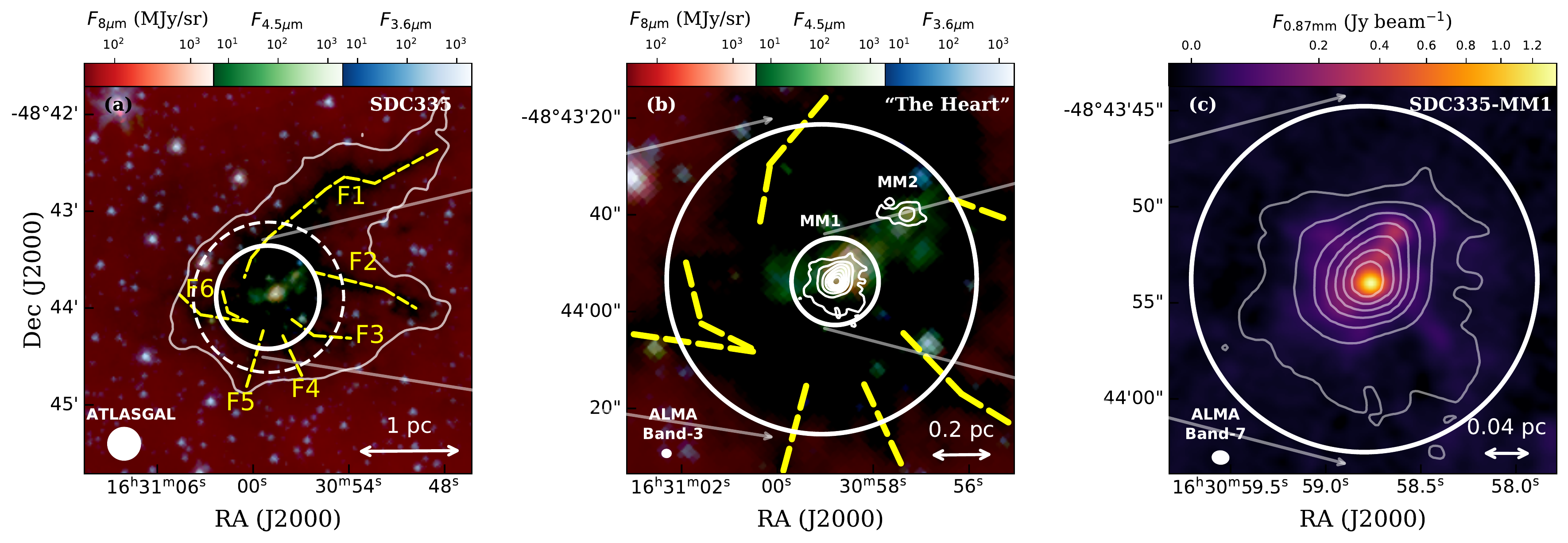}
\caption{Successive zoom-ins from (a) SDC335, to (b) ``the Heart'', and to (c) the massive dense core SDC335-MM1. 
(a) The background color map is the composite (Red/Green/Blue: Spitzer 8/4.5/3.6\,$\mu$m) images of SDC335, with Red/Green/Blue in logarithmic stretch. The ATLASGAL 870\,$\mu$m continuum emission is overlaid as a single contour level of $5\sigma=0.75$\,\jybeam\,with the APEX beam of 21\arcsec. Six converging filaments F1--F6, identified from the Spitzer extinction and dense gas N$_2$H$^+$ \citep{2013A&A...555A.112P}, are marked with yellow dashed lines. Two white circles show the ALMA Band-3 primary beam response of 12m+ACA combined data: the dashed one is 32\arcsec~for 20\% and the solid one is 46.5\arcsec~for 50\%. (b) The zoom-in version of the left panel towards ``the Heart'', with the same background color map. Two white circles show the ALMA primary beam responses respectively at 50\% for Band-3 (outer) and Band-7 (inner). The white contours of 3\,mm continuum emission follows the power-law levels of [1.0, 3.3, 7.6, 14.2, 23.4, 35.1, 49.6, 67.0]\,\mjybeam. (c) The zoom-in version of the middle panel, towards SDC335-MM1. The white contours and the white circle are the same as those in the middle panel. The background color map shows the ALMA 0.87\,mm continuum emission, in a square-root stretch to highlight the weak emission. The beams for ATLASGAL 870\,$\mu$m, ALMA Band-3, or ALMA Band-7 are shown on the left bottom and the scale bars are shown on the right bottom. \label{fig:zoomin}}
\end{figure*}

To address these questions, we use the data from new ALMA observations to investigate the gas motions in a prototypical global-collapse HFS, the IRDC SDC335.579-0.292 \citep[hereafter SDC335;][]{2009A&A...505..405P}. Located at 3.25\,kpc \citep[based on the method from][]{2016ApJ...823...77R} and containing $3.7\times10^3$\,\msun \citep{2021MNRAS.508.2964A}, SDC335 (also IRAS 16272-4837) is a well-studied massive star-forming region \citep{2002ApJ...579..678G,2013A&A...555A.112P,2015A&A...577A..30A,2021A&A...645A.142A,2021ApJ...909..199O,2022ApJ...929...68O}. 

Seen in absorption against the mid-infrared background (Figure\,\ref{fig:zoomin}, left panel), SDC335 covers approximately 2.4\,pc at its widest extent and displays six filamentary arms (yellow dashed lines), which converge towards the infrared-bright source at its centre (the hub). Inside the hub, SDC335 harbours one of the most massive millimeter cores \citep[SDC335-MM1 with $\sim$\,500\,\msun;][]{2013A&A...555A.112P} observed in the Milky Way. At a resolution of $\sim1000$\,au by ALMA, SDC335-MM1 further fragments into at least five sources, while molecular line emission is detected in two of the continuum sources, ALMA1 and ALMA3 \citep{2021ApJ...909..199O}. With a more extended array configuration (a resolution of $\sim200$\,au), \citet{2022ApJ...929...68O} found that a binary system was forming inside ALMA1, with a nearby bow-like structure ($\lesssim1000$\,au) which could add an additional member to the stellar system. However, those studies lack appropriate temperature tracers so the analyses of fragmentation is limited.

Using Mopra and ALMA Band-3 data, \citet{2013A&A...555A.112P} found that the whole SDC335 cloud is in the process of global collapse with a mass infall rate totalling $\dot M_{\mathrm{inf}} \simeq 2.5(\pm 1.0)\times 10^{-3}$\,\massrate. 
Recently, \citet{2021A&A...645A.142A} used the properties of molecular outflows to infer the accretion rates of about $1.4(\pm 0.1) \times 10^{-3}$\,\massrate~for the driving protostellar sources, suggesting a nearly-continuous flow of material from cloud to core scales \citep{2021ApJ...909..199O}. 
Despite these signposts of massive star formation, a weak emission of $\sim0.3$\,mJy at 6\,GHz has been detected towards SDC335 \citep{2015A&A...577A..30A}, suggesting that we are witnessing the early stage of massive cluster formation \citep{2013A&A...555A.112P}.

Our new ALMA observations, both at Band-3 and Band-7, were pointed towards the innermost 1\,pc of SDC335, i.e., ``the SDC335 Heart'' (the filament hub of the cloud, hereafter ``the Heart''). The Band-7 data, with both 0.8\arcsec\,resolution and well-designed multi-transition \htcs\,lines, help study the fragmentation of the most massive dense core MM1. Since no previous studies of SDC335 have resolved the dense gas kinematics at the intermediate scale ($\sim$\,0.1--1\,pc), our Band-3 data with a high angular resolution of $2$\arcsec\,and well-chosen spectral line \htcop\,$J=1-0$, focus on dense gas kinematics, attempting to build a bridge connecting the clump-scale (1\,pc) global collapse and the core-scale (0.1\,pc) gas feeding. 
The paper is organized as follows. First, we introduce the observations in Section\,\ref{sec:obs}. Second, we present the result of the ALMA continuum emission and discussion of fragmentation in Section\,\ref{sec:continuum}.Then, we present the study of gas kinematics of ``the Heart'' in Section\,\ref{sec:feedcore}. Finally, we present our conclusions in Section\,\ref{sec:conclude}.

\section{Observations}\label{sec:obs}
``The Heart'' was observed as part of ``ALMA Three-millimeter Observations of Massive Star-forming regions'' (ATOMS, Project ID: 2019.1.00685.S; PI: Tie Liu). The 12m+ACA combined data have a synthesised beam size of 1\parcsec94\,$\times$\,2\parcsec17 ($\sim$\,0.03\,pc). The maximum recoverable scale (MRS) reaches $\sim$\,87\arcsec~($\sim$\,1.37\,pc). The primary beam of the combined data at responses of 50\% and 20\% are $32\arcsec$ and $46.5\arcsec$, respectively. The continuum sensitivity is 0.2\,\mjybeam~corresponding to $\sim$\,1.6\,\msun~at the distance of SDC335, and a temperature of 23\,K. More details about the ALMA Band-3 observations and data reduction are referred to Appendix\,\ref{app:alma-b3}. 
In addition, ``the Heart'' of SDC335 was targeted in the ALMA 0.87\,mm (Band-7) survey (Project ID: 2017.1.00545.S; PI: Tie Liu) using the 12-m array alone. The reduced continuum data has a beam of 0\parcsec82\,$\times$\,0\parcsec67 ($\sim$\,0.01\,pc), a MRS of $8.45\arcsec$ ($\sim$\,0.13\,pc), and a primary beam of $9\arcsec$ at 50\%. The continuum sensitivity is $\sim$\,1.5\,mJy\,beam$^{-1}$ (i.e. $\sim$\,0.07\,\msun~at the source distance and a temperature of 23\,K). 
More details of the ALMA Band-7 observations and data reduction are referred to Appendix\,\ref{app:alma-b7}.
Besides, the ATCA 6, 8, 23, and 25\,GHz continuum emission was obtained from \citet{2015A&A...577A..30A}. 
The 3.6, 4.5, and 8.0\,$\mu$m images were retrieved from the Spitzer Archive. 
We also used the Planck+ATLASGAL 870\,$\mu$m data\footnote{ATLASGAL Survey Website: \href{https://atlasgal.mpifr-bonn.mpg.de/cgi-bin/ATLASGAL_DATASETS.cgi}{Reduced and Calibrated Maps}} at a resolution of $\sim$\,21\arcsec.

Figure\,\ref{fig:zoomin}(b) shows the ALMA 3\,mm continuum emission towards ``the Heart'' with power-law contour levels\footnote{Hereafter, the power-law levels refer to the levels that start at $5\sigma$ and end at $I_\mathrm{peak}$, increasing in steps following the power-law $f(n)=3\times n^p + 2$ where $n=1,2,3,...N$ and $p$ is determined from $D=3\times N^p+2$ ($D=I_\mathrm{peak}/\sigma$: the dynamical range; $N$: the number of contour levels).} of [1.0, 3.3, 7.6, 14.2, 23.4, 35.1, 49.6, 67.0]\,\mjybeam. The SDC335 clump is well resolved into two cores coincide with Spitzer mid-infrared nebulosity. They correspond to MM1 and MM2 identified at 3.2\,mm by \citep{2013A&A...555A.112P}. Using semi-automatic source extraction method \casa~\textit{imfit}, we identify two bright sources and calculate their fluxes presented in Table\,\ref{tab:fit_3mm_measure}. 
Although the location, morphology, and size of the two dense cores in our data are consistent with those in \citet{2013A&A...555A.112P}, the integrated fluxes of MM1 and MM2 measured in the same frequency are $\sim$\,50\% and $\sim$\,100\% higher, respectively. 
The reason could be that the previous ALMA 12m-alone data at an MRS of $\sim$\,36\arcsec~filtered out larger-scale emission than our new 12m+ACA combined data at an MRS of $\sim$\,60\arcsec. 
Such difference highlights the necessity of including short baseline data for accurate measurement of the flux from extended emission. 
The argument is strengthened based on 12m-alone continuum data at an MRS of $\sim$\,20\arcsec: we derive the fluxes of MM1 and MM2 of $113.7(\pm4.1)$\,mJy and $16.5(\pm1.4)$\,mJy which are $10\%$ and $30\%$ larger than those reported in \citet{2013A&A...555A.112P}. 
We note that because our new ALMA data have higher sensitivity, the two dense cores (but especially MM2) are observed to have extended boundaries and so greater total fluxes. Thus, sizes and fluxes obtained from data with limited sensitivity should be treated with caution.

\section{Dust Continuum Cores}
\label{sec:continuum}

\subsection{Constraining MM1 and MM2 parameters from multi-band observations}
Assuming that the dense cores are in local thermodynamic equilibrium (LTE) and that the dust emission is optically thin, the core masses are then calculated using,
\begin{equation}\label{eq:core_mass}
	M_{\mathrm{core}} = R_\mathrm{gd}\frac{F^\mathrm{int}_\nu D^2}{\kappa_\nu B_\nu (T_\mathrm{dust})},
\end{equation}
where $F^\mathrm{int}_\nu$ is the measured integrated dust emission flux of the core, $R_\mathrm{gd}$ is the gas-to-dust mass ratio (assumed to be 100), $D$ is the distance (3.25\,kpc), $B_\nu (T_\mathrm{dust})$ is the Planck function at a given dust temperature $T_\mathrm{dust}$ \citep{2008A&A...487..993K}, and $\kappa_{\nu}$ is the opacity assumed to be 0.1\,cm$^2$\,g$^{-1}$ at $\lambda\sim$3\,mm \citep{2021MNRAS.508.2964A}.

\citet{2015A&A...577A..30A} used four radio bands (6, 8, 23 and 25\,GHz) observed with the Australia Telescope Compact Array (ATCA) to build the centimeter-wavelength SEDs of MM1 and MM2. To estimate the contribution of the free-free emission at 3\,mm (100\,GHz), we extrapolate the integrated flux and derive the free-free contamination, $F_{\mathrm{ff}}^{\mathrm{\ssstyle MM1}} = 2.4$\,mJy and $F_{\mathrm{ff}}^{\mathrm{\ssstyle MM2}} = 0.66$\,mJy for MM1 and MM2, respectively. We note that the detected radio sources are unresolved or marginally resolved, i.e., the source size is much smaller than that of maximum recoverable scale, so the missing flux should be negligible. In other words, the extrapolated free-free contribution at 3\,mm should not be influenced by the missing flux. If so, the free-free emission is negligible ($<0.5\%$) for both dense cores and is not considered further.

Dense cores are fitted by 2D Gaussian profiles and the results are summarized in Table\,\ref{tab:fit_3mm_measure}. 
The fundamental measurements give the FWHM of the major and minor axis, $\theta_{\rm maj}$ and $\theta_{\rm min}$ from 2D Gaussian fitting ellipses in angular unit. 
Following \citet{2010ApJS..188..123R,2013A&A...549A..45C}, the angular radius can be calculated as the geometric mean of the deconvolved major and minor axes:
\begin{equation}\label{eq:theta_source}
    \theta_{\rm source} = \eta\left[\left(\sigma^2_{\rm maj}-\sigma^2_{\rm bm}\right)\left(\sigma^2_{\rm min}-\sigma^2_{\rm bm}\right)\right]^{1/4},
\end{equation}
where $\sigma_{\rm maj}$ and $\sigma_{\rm min}$ are calculated from $\theta_{\rm maj}/\sqrt{8\ln2}$ and $\theta_{\rm min}/\sqrt{8\ln2}$ respectively. 
The $\sigma_{\rm bm}$ is the averaged dispersion size of the beam (i.e., $\sqrt{\theta_{\rm bmaj}\theta_{\rm bmin}/(8\ln2)}$ where $\theta_{\rm bmj}$ and $\theta_{\rm bmin}$ are the FWHM of the major and minor axis of the beam). $\eta$ is a factor that relates the dispersion size of the emission distribution to the angular radius of the object determined. 
We have elected to use a value of $\eta=2.4$, which is the median value derived for a range of models consisting of a spherical, emissivity distribution \citep{2010ApJS..188..123R}. The results are similar to the deconvolved sizes from \casa~\textit{imfit}, but we choose to the former one to analytically present how deconvolution works here.
Therefore, the source physical size is derived from $R_{\rm source} = \theta_{\rm source} \times D$ and the results are shown in the second column of Table\,\ref{tab:physical_parameter_3mm}.

Following the method given by \citet{2021MNRAS.508.2964A}, the temperatures of protostellar cores MM1 and MM2 are estimated to be $65.3(\pm2.2)$\,K and $58.6(\pm2.0)$\,K, from Herschel 70\,$\mu$m emission. The details can be found in Appendix\,\ref{app:estimate_temperature}. We note that the derived temperature is an average value although the central gas temperature can be $\gtrsim$100\,K from \citet{2022MNRAS.511.3463Q}. A more accurate mass estimation can be made if a temperature profile is given.

Substituting all of these values into Equation \ref{eq:core_mass}, we obtain the gas masses of $383(^{\ssstyle+234}_{\ssstyle-120})$\,\msun~and $74(^{\ssstyle+47}_{\ssstyle-24})$\,\msun~for MM1 and MM2, respectively. 
The mean particle number density of each core is calculated using,
\begin{equation}
	n = \frac{M_{\rm gas}}{(4/3) \pi\mu m_{\ssstyle \rm H}R_{\rm source}^3},
\end{equation}
where $\mu=2.37$ is the mean molecular weight per free particle \citep{2008A&A...487..993K}, and $m_{\rm H}$ is the mass of a hydrogen atom. Following the footnote at Page 12 in \citet{WangK2014}, we simply call free particle volume density as volume density throughout this paper.
The major sources of uncertainty in the mass calculation come from the gas-to-dust mass ratio and the dust opacity. 
We adopt the uncertainties derived by \citet{2017ApJ...841...97S} of 23\% for the gas-to-dust mass ratio and of 28\% for the dust opacity, contributing to $\sim$\,36\% uncertainty of specific dust opacity. 
The uncertainties of flux, temperature, and distance (10\%) were included. Monte-Carlo methods were adopted for uncertainty estimation and the $1\sigma$ confidence intervals are given to the mass estimation. The results are summarized in Table\,\ref{tab:physical_parameter_3mm}.

\begin{table}
\centering
\caption{Physical parameters of dense cores}\label{tab:physical_parameter_3mm}
\renewcommand\arraystretch{1.2}
\begin{tabular}{ccccc} 
\hline
\hline
Dense Core &  $R_{\rm source}$ & $T_{\rm core}$\tablenotemarknew{a} & $M_{\rm core}$\tablenotemarknew{b} & $n$\tablenotemarknew{b} \\
           &        (pc)            &          (K)        &     (\msun)   & (cm$^{-3}$)      \\
\hline
SDC335-MM1        &           0.039       &        65.3(2.2)    &   $383(^{\ssstyle+234}_{\ssstyle-120})$       & $2.7(^{\ssstyle+1.6}_{\ssstyle-0.9})\times 10^7$ \\
SDC335-MM2        &           0.047       &        58.6(2.0)    &   $74(^{\ssstyle+47}_{\ssstyle-24})$          & $3.0(^{\ssstyle+1.8}_{\ssstyle-1.0})\times 10^6$ \\
\hline
\multicolumn{5}{l}{$^{a.}$ The uncertainty is given by error propagation of the 70\,$\mu$m flux.}\\
\multicolumn{5}{l}{$^{b.}$ The confidence interval $\mu(^{+\sigma}_{-\sigma})$ is given.}\\
\end{tabular}
\end{table}

\subsection{Mass concentration of SDC335}\label{subsec:concentrate}
MM1 and MM2, the two most prominent dense cores, are located at ``the Heart'' of SDC335 where the filaments intersect. 
MM1 is one of the most massive, compact protostellar cores ever observed in the Galaxy \citep{2013A&A...555A.112P}. 
The new ALMA observations together with those from the literature, show that MM1 has mass $\sim$\,400\msun (this work) to $\sim$900\,\msun \citep{2021MNRAS.508.2964A}. The factor of two difference in mass mainly arises from temperature uncertainty. Based on the method in Appendix\,\ref{app:protoT}, with the same Herschel 70\,$\mu$m flux, larger radius leads to lower temperature. \citet{2021MNRAS.508.2964A} uses \astrodendro~and derive a radius of 0.156\,pc, which is about four times larger than ours. Normalized to their radius, our temperature and then mass of MM1 is the same as \citet{2021MNRAS.508.2964A}.

Following the definition in \citet{2021MNRAS.508.2964A}, the fraction of the clump mass contained within its most massive core is,
\begin{equation}
	f_\mathrm{MMC} = \frac{M_\mathrm{MMC}}{M_\mathrm{clump}},
\end{equation}
which reveals that SDC335 contains 10\%--24\% of the dense gas in MM1. 
Among a sample of 35 clumps, SDC335 shows the highest $f_\mathrm{MMC}$, indicating the highest efficiency of forming massive dense cores \citep{2021MNRAS.508.2964A}. The high concentration could be resulted from the continuity of mass accretion reported by \citep{2021A&A...645A.142A} and in this work. 
In Section\,\ref{subsec:continuous}, we will address the connection in detail.

\subsection{Fragmentation of MM1}
\label{subsec:fragments}
\begin{figure*}
\centering
\includegraphics[scale=0.32]{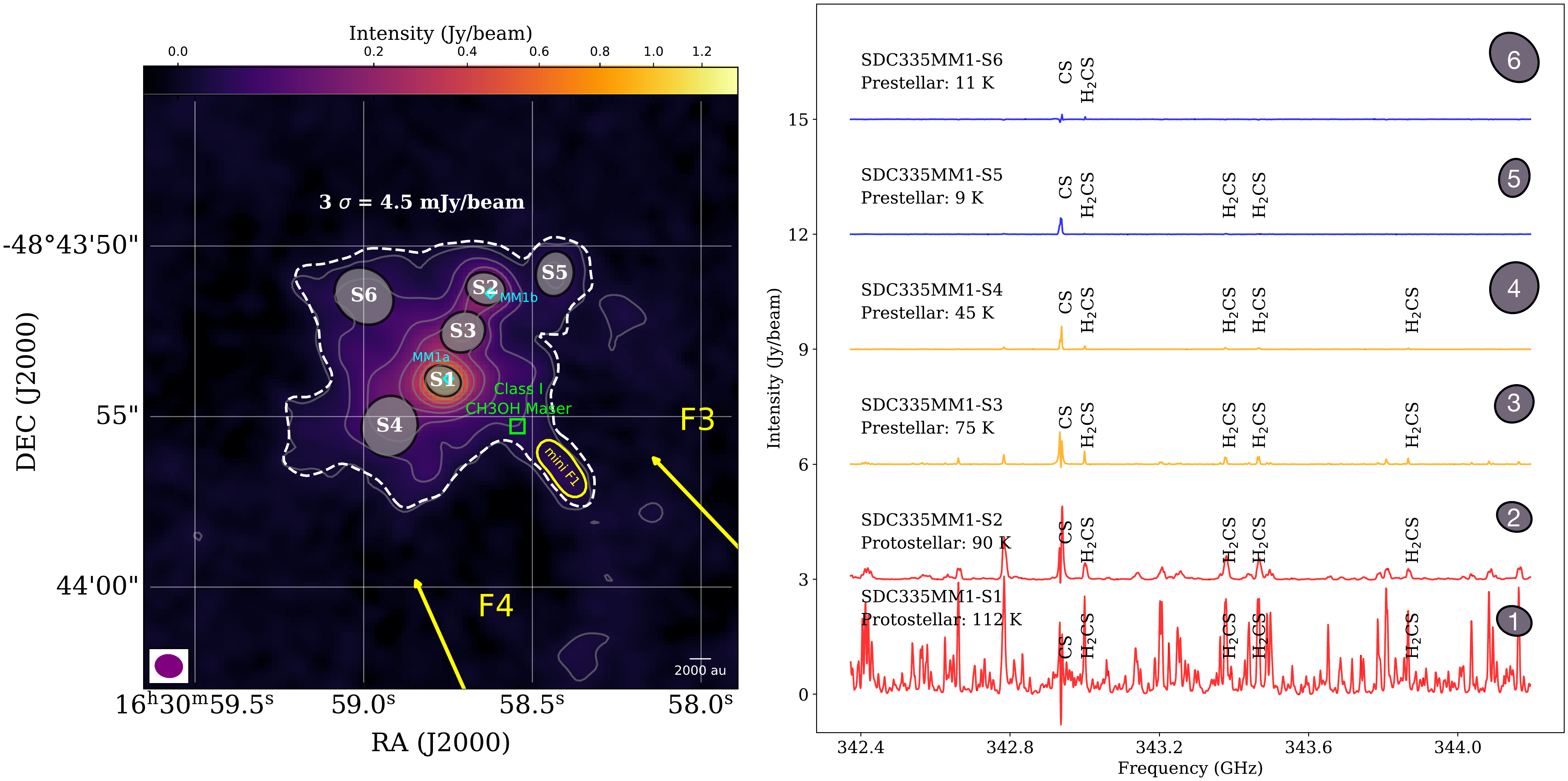}
\caption{\textit{Left}: ALMA Band-7 aggregated 0.87\,mm continuum emission map of SDC335-MM1. The white power-law contour levels are [7.5, 28.1, 71.5, 142.7, 245.9, 384.6, 562.2, 781.6, 1045.5, 1356.5]\,\mjybeam. We highlight the $3\sigma = 4.5$\,\mjybeam~level with white dashed lines. Six condensations S1--S6 are marked with their fitted FWHM ellipses. S1 and S2 are spatially associated with two UC\hii regions MM1a and MM1b shown with blue diamonds, respectively. One filament, called mini F1, found by \getsf~is shown by the yellow elongated shape. The beam size of ALMA Band-7 observations is indicated by the purple ellipse at the lower left and the scale bar is shown at the lower right. \textit{Right}: Core-averaged spectra (offset for clarity) of SPW\,31 from the six condensations marked as S1--S6. 
The colors of the spectra represent their evolutionary phases: those in red are protostellar while those yellow and blue are prestellar. 
The blue spectra do not have a sufficient number of detected \htcs~transitions, so RADEX simulation is used to estimate the temperature. \label{fig:MM1_b7}}
\end{figure*}

In Figure\,\ref{fig:MM1_b7}, ALMA 0.87\,mm continuum emission of SDC335-MM1 shows further fragmentation. Between two frequently used automatic source extraction algorithm \getsf\footnote{\getsf~is publicly available: \href{https://irfu.cea.fr/Pisp/alexander.menshchikov/}{https://irfu.cea.fr/Pisp/alexander.menshchikov/}} \citep{2021A&A...649A..89M} and dendrogram algorithm \astrodendro\footnote{\astrodendro is publicly available: \href{http://www.dendrograms.org/}{http://www.dendrograms.org/}} \citep{2008ApJ...679.1338R}, we choose \getsf~because it: 1) can deal with uneven background and rms noise; 2) can disjoin the blended sources/filaments; 3) can extract extended emission features which \astrodendro~fails to find in our test.

We set the smallest scale to be the size of the circularized beam (0\parcsec82) and the largest scale to be the MRS (8\parcsec45). After running \getsf, we extract six sources, named S1--S6, and one filament named mini F1. 
The mini F1 has the similar extension direction as two large-scale filaments F3 and F4, shown with yellow arrows in Figure\,\ref{fig:zoomin}. Between both, F3 is more spatially correlated with mini F1. Besides, \citet{2021A&A...645A.142A} detected Class-I methanol maser indicated by the green box on the left panel of Figure\,\ref{fig:MM1_spectra}. Given the collisionally excited pumping mechanism responsible for Class-I masers, \citet{2021A&A...645A.142A} argued that the maser is produced by the interaction between outflow and infalling material along the large-scale filaments. Therefore, the potential explanation of the origin of mini F1 could be shock-heated gas flow at the end of the large-scale filament F3. In Section\,\ref{subsubsec:bridge}, we further prove F3 is connected with MM1 by a clear gas stream C2 along which gas is transferred inwards. Meanwhile, the detection of the mini F1 further strengthens the argument of \citet{2021A&A...645A.142A}.

In the rest of the paper, we focus on S1--S6. The fundamental measurements/parameters are included in Table\,\ref{tab:fit_0.87mm_measure}. 
We also extract averaged spectral window 31 for the six condensations, which are shown on the right of Figure\,\ref{fig:MM1_b7}. 
We classify S1 and S2 as protostellar cores since they are spatially coincident with masers, UC\hii regions MM1a and MM1b, and molecular outflows \citep{2015A&A...577A..30A,2021A&A...645A.142A}. 
The central heating sources also excite the abundant line forests as seen from red lines on the right of Figure\,\ref{fig:MM1_b7}. 
The rest have no association with any star forming activities and are then classified as potential prestellar cores, shown with yellow or blue colors. We note that the CS lines seem broad in S5, which could be from contamination of outflows by the protostar in S2.
The difference is that yellow spectra have enough solid detection of \htcs~transition lines while blue ones have only one transition line. We have different strategies to estimate their temperature described in Appendix\,\ref{app:estimate_temperature}.

By substituting the FWHM from Table\,\ref{tab:fit_0.87mm_measure} as well as the ALMA Band-7 beam sizes ($\theta_\mathrm{maj}=0\parcsec82$ and $\theta_\mathrm{min}=0\parcsec67$) into Equations\,\ref{eq:theta_source}, we can calculate the physical radii of the six condensations and the results are listed in the second column of Table\,\ref{tab:physical_parameter_0.87mm}. Hereafter, we use the term ``condensation'' ($\sim$0.01\,pc) following the convention of \citet{WangK2014} since these sources have radii of $\sim$\,0.008--0.013\,pc (at half maximum).

\subsubsection{Jeans fragmentation}
\label{subsubsec:Jeans_theory}
If clump fragmentation is governed by thermal Jeans instabilities, the initially homogeneous gas fragments into smaller objects defined by the Jeans length ($\lambda_\mathrm{J}$) and the Jeans mass ($M_{\mathrm{J}}$),
\begin{equation}\label{eq:Jeans_length}
\lambda_\mathrm{J} = \sigma_\mathrm{th}\left(\frac{\pi}{G\rho}\right)^{1/2} = 0.016\,\mathrm{pc}\left(\frac{T}{60\,\mathrm{K}}\right)^{1/2} \left(\frac{n}{1.0\times10^7\,\mathrm{cm}^{-3}}\right)^{-1/2},
\end{equation}
and
\begin{equation}\label{eq:Jeans_mass}
\begin{split}
M_\mathrm{J} & = \frac{4\pi\rho}{3}\left(\frac{\lambda_\mathrm{J}}{2}\right)^3 = \frac{\pi^{5/2}}{6}\frac{\sigma_\mathrm{th}^3}{\sqrt{G^3\rho}} \\ & = 1.3\,M_\odot\left(\frac{T}{60\,\mathrm{K}}\right)^{3/2} \left(\frac{n}{1.0\times10^7\,\mathrm{cm}^{-3}}\right)^{-1/2},
\end{split}
\end{equation}
where $\rho$ is the mass density and $\sigma_\mathrm{th}$ is the thermal velocity dispersion given by
\begin{equation}\label{eq:sigma_thermal}
\sigma_\mathrm{th} = \left(\frac{k_\mathrm{B}T}{\mu m_\mathrm{H}}\right)^{1/2} = 0.42\,\mathrm{km\,s}^{-1}\,\left(\frac{\mu}{2.37}\right)^{-1/2} \left(\frac{T}{60\,\mathrm{K}}\right)^{1/2}.
\end{equation}
Equation\,\ref{eq:sigma_thermal} is normalized to $\mu=2.37$ and $T=60$\,K. Equations\,\ref{eq:Jeans_length} and \ref{eq:Jeans_mass} are both normalized to $T=60$\,K and $n=1.0\times10^7$\,cm$^{-3}$. 
Substituting the temperature and density of MM1 from Table\,\ref{tab:physical_parameter_3mm}, the thermal Jeans length and thermal Jeans mass are estimated to be $l_{J,\mathrm{th}}=0.01$\,pc and $M_{J,\mathrm{th}}=0.89$\,\msun, respectively.

However, a number of studies have included the contributions of turbulence. For example, \citet{2009ApJ...696..268Z,WangK2011,WangK2014} found that the observed masses of fragments within massive infrared dark cloud clumps are often more than 10\,\msun, an order of magnitude larger than the thermal Jeans mass of the clump. 
This much larger mass was explained by the ``turbulent Jeans fragmentation'' theory, in which microscopic turbulence contributes to the velocity dispersion. 
Similar results were also found in other young pre/protocluster regions 
\citep[e.g.][]{2011A&A...530A.118P,2019ApJ...886..130L,2022MNRAS.510.5009L,ATOMS_XII_Saha}.

To evaluate the turbulence inside massive dense core MM1, we extract seven dense-gas-tracer lines \citep{2020MNRAS.496.2821L} from MM1. We apply different models to fit the spectra. 
We assume: 1) \htcn\,$J=1-0$ has a narrow hyperfine structure (HFS) component as well as a Gaussian outflow component to explain the extended line wing; 2) CS\,$J=2-1$, SO\,$J=3_2-2_1$, and \hctn\,$J=11-10$ have one narrow and one wide Gaussian velocity component to explain the extended line wing; 3) \htcop\,$J=1-0$ and CCH\,$N_{J,F}=1_{3/2,2}-0_{1/2,1}$ have two narrow Gaussia components to explain the double-peak profile; 4) \chtoh\,$J=2_{1,1}-1_{1,0}$ has a single Gaussian component. See Appendix\,\ref{app:fit_spectral_lines} for additional information on \htcn\,$J=1-0$ HFS fitting details. The fitting results are shown in Figure\,\ref{fig:MM1_spectra}.

Among the seven lines, \chtoh, CS, SO, and \hctn~are contaminated by emission from the outflow and show wide line wings and have low velocity resolution ($\geq1$\,\kms). 
Therefore, the fitted linewidths are systematically overestimated. The CCH, \htcop, and \htcn~lines show consistent linewidths. 
We note that although \htcn~is potentially contaminated by the outflow, the high velocity resolution of $\sim$0.2\,\kms~helps disentangle outflow component from the dense and narrow component. 
Figure\,\ref{fig:dense_gas} shows the integrated maps of three dense-gas tracers CCH, \htcop, and \htcn. 
All three species have good spatial correlation with dense cores. 
Therefore, we take the arithmetic mean value of $\Delta V$ among three species as the linewidth of the dense cores. Therefore, MM1 has the line width FWHM $\Delta V_\mathrm{mol} = 2.40$\,\kms, i.e. total velocity dispersion $\sigma_\mathrm{tot,mol} = 1.02$\,\kms.

Substituting $\mu=30$ (the molecular mass for \htcop; CCH and \htcn~have the similar values) in Equation\,\ref{eq:sigma_thermal} gives the thermal velocity dispersion $\sigma_\mathrm{th,mol}=0.134$\,\kms, which contributes little to the total linewidth $\sigma_\mathrm{tot,mol}$. 
The non-thermal velocity dispersion $\sigma_\mathrm{nt,mol}=\sqrt{\sigma^2_\mathrm{tot,mol}-\sigma^2_\mathrm{th,mol}}\simeq1.01$\,\kms.
To take the contribution of turbulence into account, the temperature $T$ in equations\,\ref{eq:Jeans_length} and \ref{eq:Jeans_mass} should be replaced by the effective temperature,
\begin{equation}
	T_\mathrm{eff} = \frac{\mu m_\mathrm{H}}{k_\mathrm{B}}(c_s^2+\sigma^2_\mathrm{nt}).
\end{equation}
For MM1, the effective temperature is $\simeq$\,360\,K, resulting in a turbulent Jeans length $l_{J,\mathrm{turb}}=0.024$\,pc and mass $M_{J,\mathrm{turb}}=11.5$\,\msun.

\begin{table*}
\centering
\renewcommand\arraystretch{1.2}
\caption{Physical parameters of extracted sources from SDC335-MM1}
\label{tab:physical_parameter_0.87mm}
\begin{tabular}{ccccccc}
\hline
Source ID &  $R$\tablenotemarknew{a} & $T_\mathrm{d}$ & $M_\mathrm{d}$ & $n$\tablenotemarknew{d} & Notes \\ 
          & ($\times10^{-2}$\,pc) & (K) & (\msun) & (cm$^{-3}$) & \\
\hline
SDC335MM1-S1 & 1.6 & 112($\pm6$)\tablenotemarknew{b} & $23.2(^{\ssstyle+14.6}_{\ssstyle-7.3})$ & $2.5(^{\ssstyle+1.6}_{\ssstyle-0.8})\times10^7$ & Protostellar: associated with UC\hii region MM1a \\
SDC335MM1-S2 & 1.7 & 90($\pm9$)\tablenotemarknew{b} & $4.3(^{\ssstyle+2.8}_{\ssstyle-1.4})$ & $3.7(^{\ssstyle+2.4}_{\ssstyle-1.2})\times10^6$ & Protostellar: associated with UC\hii region MM1b \\
SDC335MM1-S3 & 2.0 & 75($\pm17$)\tablenotemarknew{b} & $5.9(^{\ssstyle+4.7}_{\ssstyle-2.0})$ & $3.1(^{\ssstyle+2.5}_{\ssstyle-1.0})\times10^6$ & Protostellar? Warm but no star forming activities \\
SDC335MM1-S4 & 2.8 & 45($\pm10$)\tablenotemarknew{b} & $9.3(^{\ssstyle+7.9}_{\ssstyle-3.2})$ & $1.8(^{\ssstyle+1.6}_{\ssstyle-0.6})\times10^6$ & Protostellar? Warm but no star forming activities \\
SDC335MM1-S5 & 1.9 & 9.5($\pm0.9$)\tablenotemarknew{c} & $13.4(^{\ssstyle+10.0}_{\ssstyle-4.4})$ & $7.6(^{\ssstyle+5.8}_{\ssstyle-2.6})\times10^6$ & Prestellar \\
SDC335MM1-S6 & 2.7 & 11.2($\pm1.1$)\tablenotemarknew{c} &$36.8(^{\ssstyle+27.5}_{\ssstyle-12.1})$ & $7.7(^{\ssstyle+5.5}_{\ssstyle-2.6})\times10^6$ & Prestellar \\
\hline 
\multicolumn{6}{l}{$^{a.}$ The beam-deconvolved radius.} \\
\multicolumn{6}{l}{$^{b.}$ The dust temperature is assumed to be equal to the molecular rotational temperature of \htcs.} \\
\multicolumn{6}{l}{$^{c.}$ The dust temperature is estimated from kinetic temperature estimated with RADEX modeling (see Section\,\ref{app:S5-6}).} \\
\multicolumn{6}{l}{$^{d.}$ The average density in spherical assumption.} \\
\end{tabular}
\end{table*}

\subsubsection{Temperature and mass estimation of condensations}
\label{subsubsec:condensation_mass}
We choose the thioformaldehyde (\htcs) lines to compute rotational temperature. The spectral window 31 of ALMA Band-7 observations (cf. Table\,\ref{tab:h2cs}) covers multiple components of the \htcs\,$J=10-9$ transition with velocity resolution of 0.986\,\kms. Besides, the upper energy levels $E_u$ of \htcs~multi-transition lines ranges from 90 to 420\,K in our frequency coverage, making \htcs~a well suited thermometer for SDC335.
We extract the spectral cubes and average spectra for the six sources, as shown in Figure\,\ref{fig:MM1_b7}.
For S1, S2, S3, and S4, at least four emission lines of \htcs\,$J=10-9$ are detected, which can be used for estimating the rotational temperature. We list the transitions (quantum numbers), frequencies, upper energy level as well as line strengths in columns 1--5 of Table\,\ref{tab:h2cs}.

We then use the eXtended \casa~Line Analysis Software Suite \citep[\textsc{XCLASS}\footnote{https://xclass.astro.uni-koeln.de/};][]{2017A&A...598A...7M} for line temperature calculation (more details in Appendix\,\ref{app:S1-4}). The fitted temperature for each source is taken as ``average temperature'' for further analysis. Both S3 and S4 show a significantly higher temperature than that of traditional prestellar cores. They are within the massive dense core SDC335-MM1 and spatially adjacent to the embedded protostellar cores S1 and S2, probably heated by the radiation from protostars (Xu et al. in preparation). Such heating mechanism could be responsible for the formation of massive prestellar cores in protoclusters \citep{2011ApJ...740...74K,2013ApJ...766...97M}. 

For both S5 and S6, only the lowest excitation line is detected. Under the assumption of $T_\mathrm{dust}=T_\mathrm{kin}$, we can estimate the kinematic temperature by NLTE mock grids (more details in Appendix\,\ref{app:S5-6}). Finally, the temperature of the six condensations S1--S6 are listed in Table\,\ref{tab:physical_parameter_0.87mm}.

The mass of condensations can be estimated using Equation\,\ref{eq:core_mass}, substituting the Band-7 frequency $\nu=350$\,GHz and dust opacity $\kappa_\mathrm{0.87mm}$. Based on the coagulation model, \citet{1994A&A...291..943O} systematically computed and tabulated the opacity of dust in dense protostellar cores between 1\,$\mu$m and 1.3~mm, at different gas densities and molecular depletion ratios. 
A value of 1.89\,cm$^2$\,g$^{-1}$ is adopted for $\kappa_\mathrm{0.87mm}$, which is interpolated from the given table in \citet{1994A&A...291..943O}, assuming the MRN \citep*{1977ApJ...217..425M} dust grain model with thin ice mantles and a gas density of $10^6$\,cm$^{-3}$. 
The mass and number density as well as their $1\sigma$ uncertainty are listed in the third and forth column of Table\,\ref{tab:physical_parameter_0.87mm}.

\subsubsection{Thermal versus turbulent Jeans fragmentation}
\label{subsubsec:which_mechanism}

\begin{figure*}
    \centering
    \includegraphics[scale=0.38]{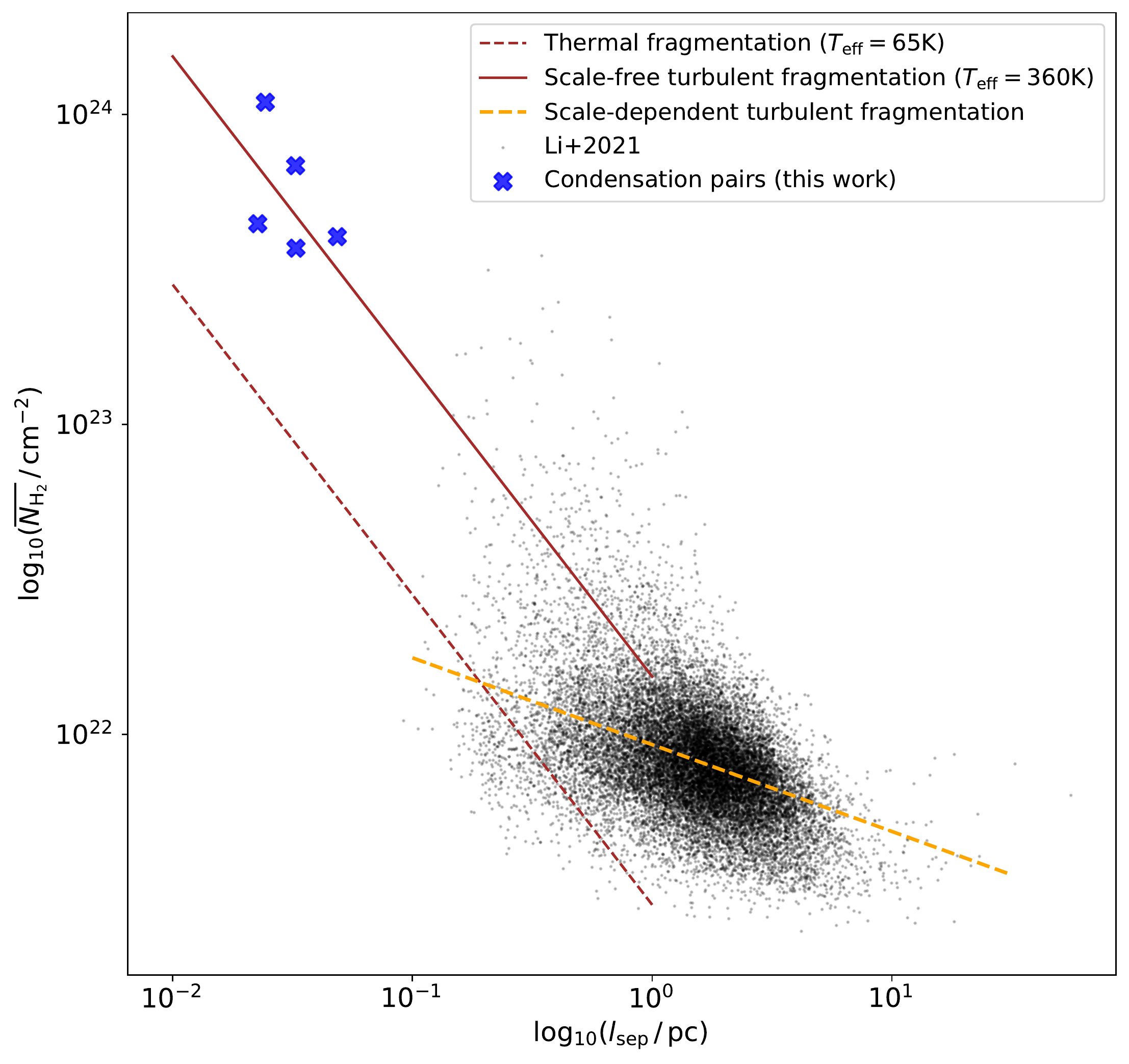}
    \includegraphics[scale=0.38]{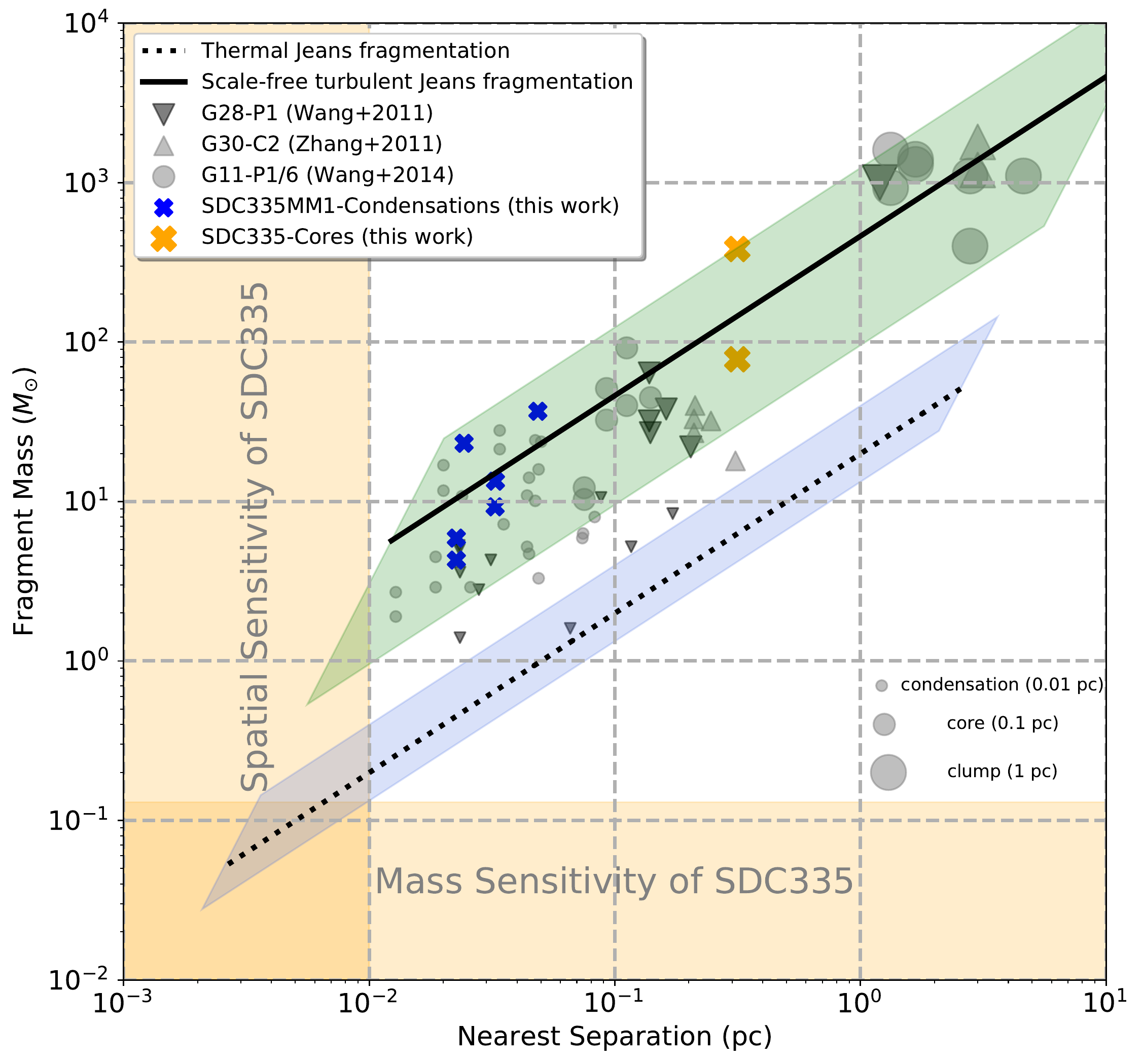}
    \caption{\textit{Left}: the relation between the fragment separation $l_\mathrm{sep}$ and column density $\overline{N_{\mathrm{H}_2}}$, where $l_\mathrm{sep}$ is measured from the length of the edges, and $\overline{N_{\mathrm{H}_2}}$ is the mean column density along the edges. The black data points are from \citet{2021ApJ...916...13L,2021ApJ...918L...4C}. The orange dashed line fits the black points with a scaling relation $\Sigma_\mathrm{gas}\propto l_\mathrm{sep}^{-0.28}$, indicating the scale-free turbulent fragmentation is dominant. The red dotted line is indicated by thermal Jeans fragmentation, with effective temperature of 65\,K. Blue crosses are from this work, which is fitted by the brown solid line. The fitting result shows a quasi-thermal turbulent fragmentation with an effective temperature of 360\,K (see Section\,\ref{subsubsec:Jeans_theory}); \textit{Right}: the relation between fragment mass and nearest separation. The grey downwards triangles are the data points of G28-P1 \citep{WangK2011}. The grey upwards triangles are the data points of G30-C2 \citep{2011ApJ...733...26Z}. The grey circles are the data points of G11-P1 and G11-P6 \citep{WangK2014}. The blue crosses and orange crosses are the data points of condensations and cores in this work. The orange shaded regions show the sensitivity and resolution limit of the SDC335 Band-7 observations. The dotted line shows thermal Jeans fragmentation with $T=15$\,K and $n=[10^2,10^8]$\,cm$^{-3}$, and the blue shaded region corresponds to the same density range but with $T=[10,30]$\,K. The solid line shows a scale-free turbulent Jeans fragmentation with effective temperature $T_\mathrm{eff}$ of 140\,K (total velocity dispersion $\sigma=0.7$\,\kms) and the same density range. The green shaded region corresponds to the same density range but with $T_\mathrm{eff}=[46,413]$\,K (i.e., $\sigma=[0.4,1.2]$\,\kms). The sizes indicate the physical scales of grey data points: the smallest are condensations ($\sim0.01$\,pc), the middle are cores($\sim0.1$\,pc), and the largest are clumps($\sim1$\,pc). This figure shows clearly that the hierarchical fragmentation in SDC335 are dominated by turbulence over thermal pressure.}
    \label{fig:NH2_Lsep}
\end{figure*}

In order to quantify core separations to compare with the Jeans lengths, we used the minimum spanning tree (MST) method customized by \citet{WangK2016} \footnote{https://github.com/pkugyf/MST/} to generate a set of straight lines connecting a set of points (the center of condensations in this case). 
We set the edge weight to be proportional to the physical separation between two points, which minimizes the sum of the lengths. Figure\,\ref{fig:connections} displays the connections from MST algorithm for the ALMA Band-7 observation of SDC335-MM1. 
The separation $l_\mathrm{sep}$ between condensations is $\simeq$\,0.03\,pc on average, ranging from 0.022 to 0.048\,pc.

For the thermal support to be effective, the separation of condensation
is connected to the column density by \citep{2021ApJ...916...13L},
\begin{equation}\label{eq:sigma_l}
    \Sigma_\mathrm{gas} = \frac{c_s^2}{Gl_\mathrm{sep}}.
\end{equation}
We measure the $\Sigma_\mathrm{gas}$, i.e., the column density averaged along every edge between two condensations, which are shown in blue points in the left panel of Figure\,\ref{fig:NH2_Lsep}. 
These blue points can't be fitted by the red dotted lines, so their fragmentation can't be explained by thermal Jeans fragmentation unless the temperature is as large as 360\,K. 
The black data points in the left panel Figure\,\ref{fig:NH2_Lsep} are from a dense core sample of a high-mass star formation region Cygnus-X \citep{2021ApJ...918L...4C}, based on the column density map constructed by \citet{2019ApJS..241....1C} using the \getsources~software \citep{2012A&A...542A..81M}. The typical scale is between 0.1--10\,pc. The fitted orange dotted line gives a scaling relation of $\Sigma_\mathrm{gas}\propto l_\mathrm{sep}^{-0.28}$, indicating a scale-dependent turbulence $\sigma_v=0.87(l_\mathrm{sep}/1.0\mathrm{pc})^{0.36}$ \citep{2021ApJ...916...13L}. However, the fitted slope of blue points ($\simeq-1$) as shown by the red solid line, is much steeper than that of the scale-dependent turbulence ($\simeq-0.28$). Instead, if the turbulence plays a similar role as thermal motions, then the effective temperature inside the MM1 is as high as 360\,K or $c_s\simeq1.12$\,\kms (see Section\,\ref{subsubsec:condensation_mass}). In this case, turbulence is independent of scale, so we call it scale-free turbulent fragmentation, to distinguish the scale-dependent turbulent fragmentation observed at a larger scale.

Another straightforward way to analyze fragmentation is in a mass-separation diagram suggested by \cite{WangK2014}. Here, the condensations (S1--S6, blue crosses) are shown in the right panel of Figure\,\ref{fig:NH2_Lsep}, where fragment mass is plotted against the separation to the nearest fragment (nearest separation, hereafter). To compare with, the grey triangles \citep[G28 condensations;][]{WangK2011} and grey circles \citep[G11 condensations;][]{WangK2014} are plotted. The shaded blue region represents the thermal Jeans fragmentation domain (``thermJ-domain'' hereafter) where the adopted temperature and density (equations\,\ref{eq:Jeans_length}, \ref{eq:Jeans_mass}) are in ranges $T=[10,30]$\,K and $n=[10^2,10^8]$\,cm$^{-3}$. The shaded green region represents the turbulent Jeans fragmentation domain (``turbJ-domain'' hereafter) with the same density range and effective temperature range $T_\mathrm{eff}=[72,646]$\,K (i.e. total velocity dispersion $\sigma=[0.5,1.5]$\,\kms). We find that most condensations are located in ``turbJ-domain'' while quite a few are located in ``thermJ-domain''. 
In the case of SDC335-MM1, the condensations have $\bar M_\mathrm{conden}$\,(15.4\,\msun)\,$\gtrsim M_{J,\mathrm{turb}}$\,(11.5\,\msun)\,$> M_{J,\mathrm{therm}}$\,(0.9\,\msun) and $l_{J,\mathrm{th}}$ (0.01\,pc) < $ l_{J,\mathrm{turb}}$ (0.024\,pc) $\lesssim \bar l_\mathrm{sep}$. The black solid line of panel (b) highlight the $\sigma=1.1$\,\kms, which is the total line width of SDC335-MM1. The good correspondence between the line and the data points further strengthens the ``turbJ-domain'' argument.
Moreover, panels (a) and (b) in Figure\,\ref{fig:zoomin} together show a clear hierarchical fragmentation where SDC335 first fragments into two dense cores MM1 and MM2, while MM1 further fragments into six condensations. As shown in the right panel of Figure\,\ref{fig:NH2_Lsep}, the dense cores (MM1 and MM2, orange crosses) of SDC335 favors ``turbJ-domain'' rather than ``thermJ-domain''. In other words, the hierarchical fragmentation in SDC335 should be regulated by turbulent Jeans fragmentation.

However, it is important to note that fragmentation is a time-varying process: the condition for Jeans fragmentation changes with time. This is because both temperature and density increase as massive clumps evolve toward star formation. For example, the thermal Jeans length can be as large as 0.66\,pc for a cloud with an initial density of $10^3$\,cm$^{-3}$ at $T=10$\,K.
On the other hand, the mass and separation of condensations can be time-dependent. SDC335 is undergoing global collapse at the present time. Following the overall gravitational contraction of the massive dense core MM1, the fragments may move closer to each other, contributing to smaller separations compared to the initial one. Besides, the condensations can accrete mass from the massive dense core, whose mass comes from the natal clump by so-called intermittent gas inflows \citep{2018ARA&A..56...41M}. 
In the case of SDC335, the derived accretion rate $\simeq2.4\times10^{-3}$\,\massrate~contributes to 700\,\msun~per free-fall time-scale \citep[$3\times10^5$\,yrs;][]{2013A&A...555A.112P}, enough to double the mass of MM1. 
Above all, the time-varying process \citep{2018A&A...617A.100B} make it more difficult to tell which controls the fragmentation.

\section{Continuous gas inflow feeds massive dense cores}
\label{sec:feedcore}

\subsection{Dissecting velocity structure inside ``the Heart''}
\label{subsec:dissect}
\citet{2013A&A...555A.112P} used ALMA 3\,mm mosaic observations to study the kinematics inside SDC335, where the spatial distribution of \nthp\,$J=1-0$ emission is similar to the dust extinction mapped in the Spitzer images.
The six filaments F1--F6 identified in extinction map all have dense gas counterparts with similar morphology \citep{2013A&A...555A.112P}.
Such similarity demonstrates how efficient \nthp\,$J=1-0$ is in tracing the network of parsec-long filaments seen in dust extinction \citep[also see example in][]{2021RAA....21...24Y}. 

Although the velocity is coherent within each filament, the kinematics inside ``the Heart'' are much more complicated. 
Specifically, from the ALMA \nthp\,$J=1-0$ observations, \citet{2013A&A...555A.112P} argue that two separate velocity components are present close to MM2, while the broad asymmetric line profiles around MM1 suggest their blending, as observed in other massive cores \citep{2011ApJ...740L...5C}.
Kinematically, the gas traced by \nthp\,$J=1-0$ at the centre of the cloud appears to be composed of a mix of gas originated from two main filaments, F1 and F2 \citep{2013A&A...555A.112P}.

However, limited by the difficulty of hyperfine structure fitting when multiple velocity components are blended, as well as the limited angular resolution ($\sim5\arcsec$), the kinematics are not fully determined in \citet{2013A&A...555A.112P}. 
With the new ALMA Band-3 observations, we use \htcop\,$J=1-0$ to dissect the structure of ``the Heart'' in position-position-velocity (p--p--v) space.

\subsubsection{Justification of the choice of molecular line}
\label{ssubsec:justify}
\htcop\,$J=1-0$ has a high critical density $n_\mathrm{crit}$ of $6.2\times10^4$\,cm$^{-3}$ and effective excitation density of $n_\mathrm{eff}$ of $3.9\times10^4$\,cm$^{-3}$ at 10\,K \citep{2015PASP..127..299S}. 
With the volume density of $5.0(\pm 0.6)\times10^4$\,cm$^{-3}$ \citep{2013A&A...555A.112P}, ``the Heart'' can excite \htcop\,$J=1-0$ strongly enough to be detected (maximum SNR of $\sim$50 and mean SNR of $\sim$5 in our case), without being excessively optically thick. 
The detailed calculation of optical depth $\tau$(\htcop) can be found in Appendix~\ref{app:tau_h13cop}. 
As seen from Figure\,\ref{fig:h13cop_tau}, the optical depth $\tau$(\htcop) is universally thin, even for the densest part nearby the MM1. Hereafter, we approximate \htcop\,$J=1-0$ emission with Gaussian fits.

\citet{2017A&A...604A..74S} found that the spatial distribution of the \htcop\,$J=1-0$ emission is tightly correlated with the column density of the dense gas revealed by the Herschel data.
Very recently, \citet{2022ApJ...926..165L} used ALMA \htcop\,$J=1-0$ to identify 13 narrow gas filaments in a massive infrared dark cloud, NGC 6334S. \citet{ATOMS_XI_Zhou} also used ALMA \htcop\,$J=1-0$ to identify HFSs in proto-clusters. 
These results justify our choice of using \htcop\,$J=1-0$ to trace the kinematics within ``the Heart''.

\subsubsection{Multi-component spectral line decomposition}
\begin{figure}
\centering
\includegraphics[scale=0.43]{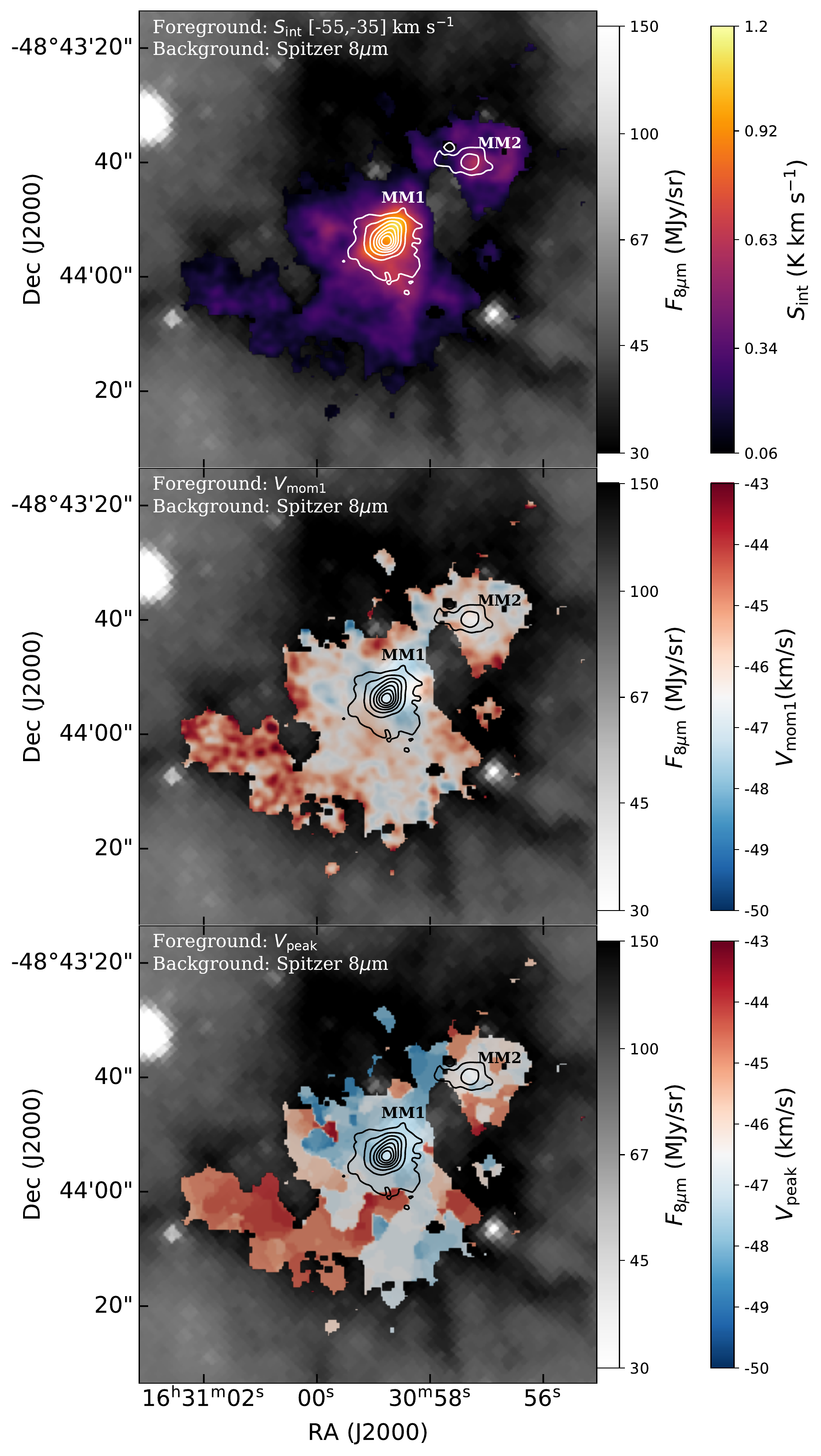}
\caption{The background grayscale maps show the Spitzer 8\,$\mu$m emission of ``the Heart''. The ALMA Band-3 continuum emission is shown with the white/black contours following the power-law levels of [1.0, 3.3, 7.6, 14.2, 23.4, 35.1, 49.6, 67.0]\,\mjybeam. The continuum image without primary beam correction is used, for a uniform noise over the field of view. Overlaid color maps are different from top to bottom and marked at the upper left in each panel. 
Top: the moment 0 map (integrated interval [-55,-35]\,\kms) of \htcop\,$J=1-0$. Middle: the moment 1 map of \htcop\,$J=1-0$. 
Bottom: the velocity at peak of \htcop\,$J=1-0$. The pixels with SNR < 5 are masked. Two colorbars are shown for each panels: the first one (grayscale) is for background Spitzer 8\,$\mu$m in a log stretch; the second one (color-scheme) on a linear scale is for the overlaid moment 0 map or velocity maps. \label{fig:h13cop_emission}}
\end{figure}

In the complex environment of ``the Heart'', multiple velocity components cause multi-Gaussian profiles of \htcop\,$J=1-0$. As seen from Figure~\ref{fig:h13cop_emission}, the top panel shows a good spatial agreement between the Spitzer 8\,$\mu$m extinction and the integrated \htcop\,$J=1-0$ emission. 
More importantly, the moment-1 map ($V_\mathrm{mom1}$ map) in the middle panel shows a marginal velocity difference across ``the Heart''. 
Such a difference of velocity is much clearer in the bottom panel when the velocity at peak intensity ($V_\mathrm{peak}$ map) is shown. 
The advantage of the $V_\mathrm{peak}$ map is to highlight the major component in each pixel which would be averaged out in the $V_\mathrm{mom1}$ map. 
The $V_\mathrm{peak}$ map shows the velocity complexity of ``the Heart'', demanding more comprehensive methods to decompose inherent structures.

To address this challenge, we use Semi-automated multi-COmponent Universal Spectral-line fitting Engine Python Implementation \citep[\scousepy][]{2019MNRAS.485.2457H}. 
Compared to canonical moment analysis, the multi-component decomposition provides an improved description of complex kinematics \citep{2016MNRAS.457.2675H}, particularly important for a clustered environment such as SDC335.
We compile a brief introduction of \scousepy~and the process of data reduction in Appendix\,\ref{app:decompose}.

After running \scousepy~we obtain a reduced spectral cube where the spectrum in each pixel is decomposed into $N$ Gaussian function components (i.e. $3\times N$ free parameters including peak intensity, centroid velocity, and velocity dispersion).
In total, 8873 pixels have solutions and 17098 components are extracted, so there are about two components in each pixel on average. The universal multi-components show the kinematic complexity of ``the Heart''.

\subsubsection{Clustering of the components in p--p--v space}
To study the kinematics from the decomposed \htcop\,$J=1-0$ data, we use Python-based algorithm Agglomerative Clustering for ORganising Nested Structures \citep[\acorns;][]{2019MNRAS.485.2457H} to ``re-assemble'' the decomposed Gaussian data set. 
To analyze clustering in p--p--v space, for two data points to be classified as ``linked'', we require that: 1) the Euclidean distance should be no greater than the beam size (2.4\,arcsec or $\sim$6 pixel size); and 2) the absolute difference in both measured centroid velocity and velocity dispersion should be no greater than the velocity resolution of 0.21\,\kms.
These criteria reflect the observational limits in three dimensions and the clustering results should show coherent structures in p--p--v space beyond such limit.

\citet{2019MNRAS.485.2457H} expands the nomenclature used in dendrograms: a cluster is called a ``tree'' and the whole tree system is called a ``forest'', itself containing numerous trees. 
Each tree may or may not then be further subdivided into branches or leaves. 
Trees with no substructure are also classified as leaves. 
After running \acorns, 12 trees are found and four dominant trees \#0, \#1, \#3, and \#7 contain $\sim$\,75.7\% of voxels (see the left panel in Figure\,\ref{fig:acorns_output}).

\subsection{Major and minor gas streams}
\label{subsec:stream}

\begin{figure*}
\centering
\includegraphics[scale=0.4]{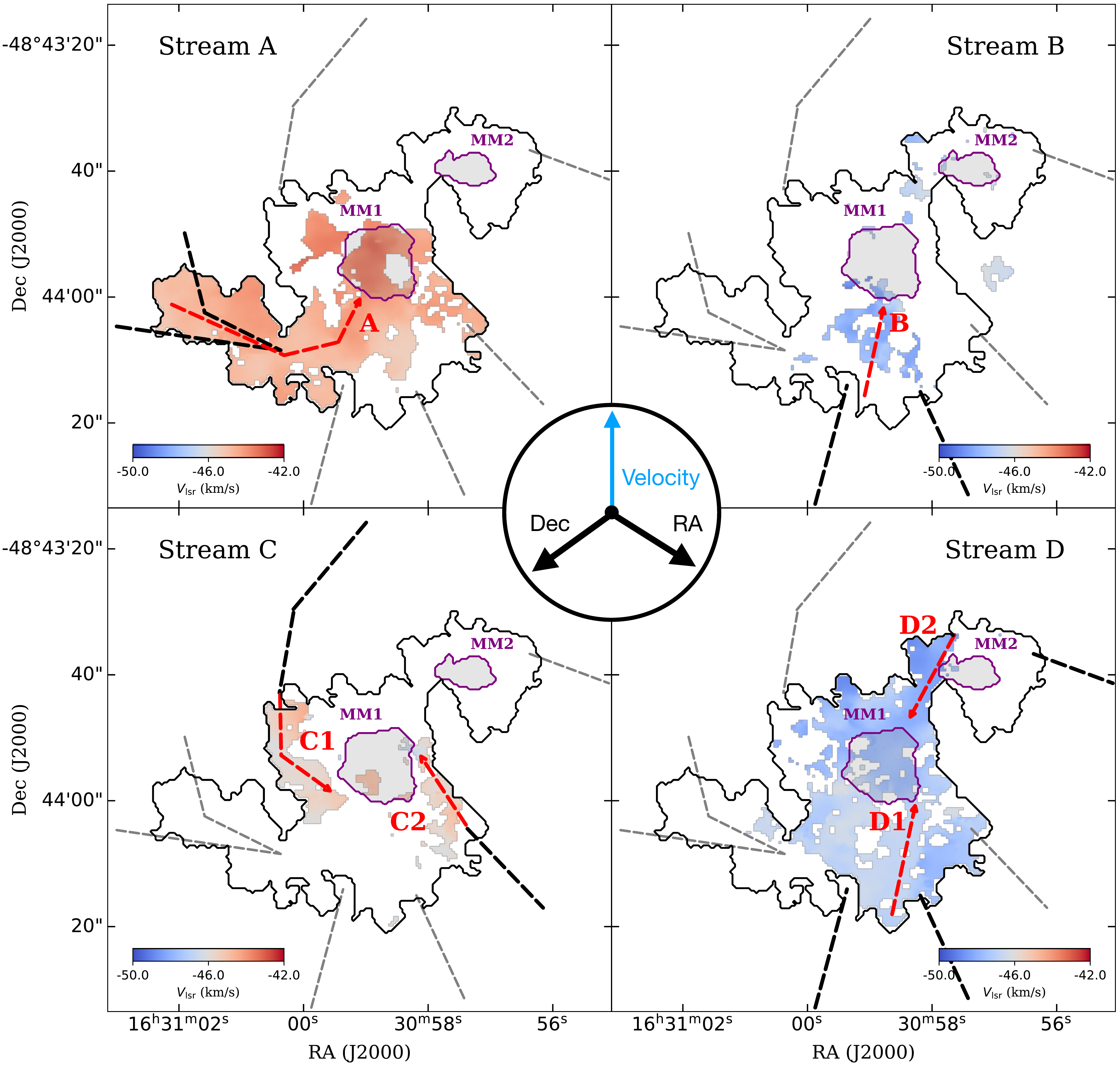}
\caption{The velocity maps of the four major streams are normalized to linear span from -50 to -42\,\kms~in colorscale. The velocity at each pixel is given by the centroid velocity of decomposed Gaussian component (refer to Figure\,\ref{fig:stream_integrate} for integrated flux map). The black solid lines depict 5$\sigma$ level of \htcop\,$J=1-0$. The six groups of dashed gray lines are the filaments identified from the Spitzer 8\,$\mu$m extinction map (the same as in the middle panel of Figure\,\ref{fig:zoomin}). The black bold dashed lines mark the filaments which are assumed to be responsible for the streams in each panel. Streams including two major ones (A and B) and four minor ones (C1, C2, D1, and D2) are marked as red bold dashed vectors. The massive dense cores MM1 and MM2 together with their boundaries ($F_\mathrm{cont,3mm}=1.0$\,\mjybeam) are indicated in purple. The central circle shows how we visualize the 3D cube: blue axis is the collapsing axis, where velocities of pixels are projected on the plane spanned by two black axes.
\label{fig:stream_shift}}
\end{figure*}

\begin{figure*}
\centering
\includegraphics[scale=0.4]{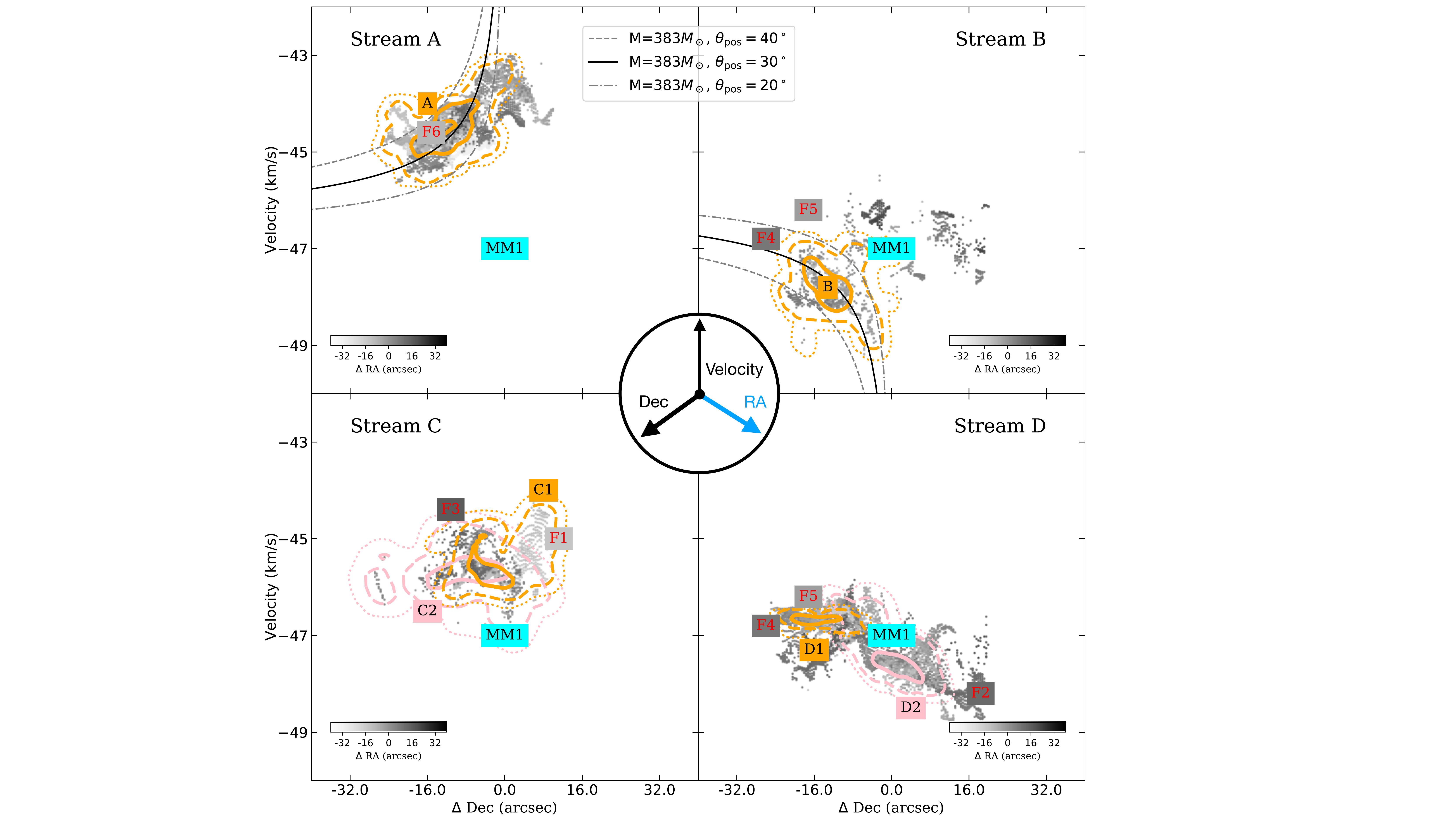}
\caption{The ``$\Delta$ Dec vs. Velocity'' plot of four major streams are shown with scattering points. Each point represents one Gaussian component of the decomposed \htcop\,$J=1-0$ emission, with its color normalized by the $\Delta$ RA. The location of massive dense core MM1 is marked by the cyan color. Two major streams A and B are marked by orange contours. The three contour levels represent the $1\sigma$ (solid line), $2\sigma$ (dashed line), and $3\sigma$ (dotted line) probability distribution of scatter-points. Similarly, minor streams C1 and D1 are marked by the orange contours and minor streams C2 and D2 are marked by the pink colors. In each panel, ``F1--F6'' marks the end of the large-scale filaments which are assumed to be responsible for the streams. The RA, DEC location as well as velocity of filament ends are measured from \nthp~data from \citet{2013A&A...555A.112P}. In the upper two panels, the stream A and B are respectively fitted by free-fall models (a central mass of 383\,\msun) with inclination angles of $\theta=30^{\circ}$ (solid black line; best fitting), $\theta=40^{\circ}$ (dash-dotted black line), and $\theta=20^{\circ}$ (dash-dotted gray line). The central circle marks how we visualize the 3D data cube: blue axis is the collapsing axis, where the data points are projected on the plane spanned by two black axes. \label{fig:PyV}}
\end{figure*}

\subsubsection{Identification and nomenclature}
We denote the four dominant trees \#0, \#1, \#3, and \#7 as the major ``streams'' A, B, C, and D due to their large proportion in voxels (75.7\%). Figure\,\ref{fig:stream_integrate} shows the integrated flux of the four streams which occupy most of \htcop\,$J=1-0$ emission regions outlined by the black solid contour level (5$\sigma$, the same as Figure\,\ref{fig:h13cop_emission}). Figure\,\ref{fig:stream_shift} shows the velocity at each pixel for the four streams. The velocity at each pixel is given by the centroid velocity of the decomposed Gaussian component.
Furthermore, we project these points both on the ``$\Delta$ Dec vs. Velocity'' and ``$\Delta$ RA vs. Velocity'' planes, as shown in Figure\,\ref{fig:PyV} and \ref{fig:PxV}.

Identified from \acorns, the four major streams should be both spatially and velocity coherent, which are shown in Figure\,\ref{fig:stream_shift} and Figure\,\ref{fig:stream_integrate}, respectively.
One can also refer to Figure\,\ref{fig:PyV} and \ref{fig:PxV} for the projected p--p--v scatter-plots of the streams viewed from both Right Accession (RA) and Declination (Dec). 
Stream B has hierarchical structures, in which a minor component consist over 70\% of the voxels. Because only this minor component shows a good velocity coherent structure, we simply call it stream B hereafter.
The stream C has two separate stream components: the eastern one (C1) and the western one (C2), both of which are two leaves in one tree (major stream C). So we call them minor stream C1 and C2 hereafter. 
Similarly, two leaves in the major stream D are named minor stream D1 and D2. The mentioned stream A, B, C1, C2, D1, and D2 are marked as red dashed arrows in Figure\,\ref{fig:stream_shift} and Figure\,\ref{fig:stream_integrate}, which are of our interest in the rest of the paper. In the Figure\,\ref{fig:PxV}, streams B, C1 and D1 are marked as orange contours while the minor streams C2 and D2 are marked as pink contours. The different colors in one panel are only used to distinguish two minor streams easily.

\subsubsection{Bridge the large-scale filaments and the massive core MM1}
\label{subsubsec:bridge}
\begin{figure*}
\centering
\includegraphics[scale=0.4]{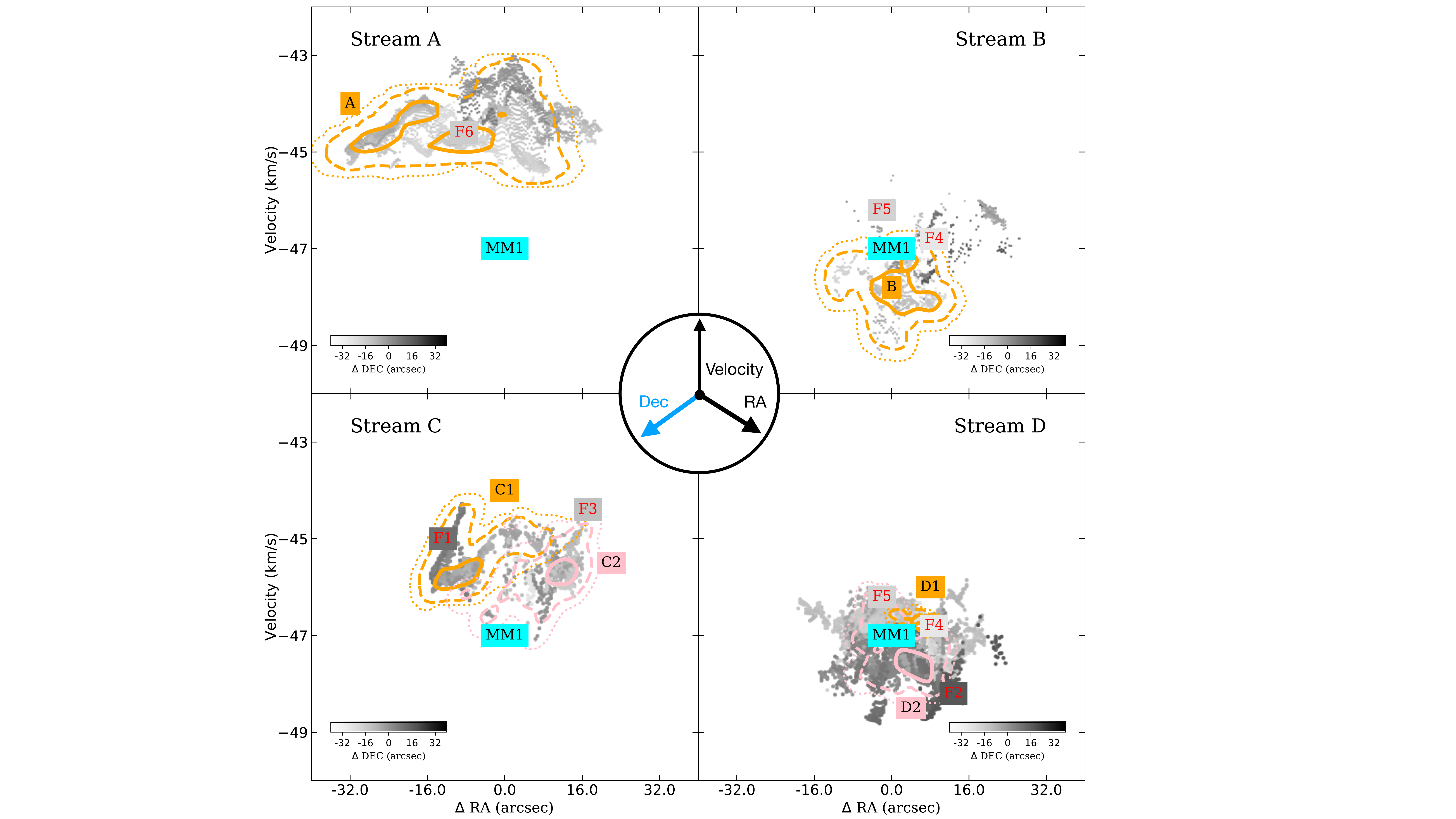}
\caption{The ``$\Delta$ RA vs. Velocity'' plot of four major streams are shown with scattering points. Each point represents one Gaussian component of the decomposed \htcop\,$J=1-0$ emission, with its color normalized by the centroid velocity on a linear span from -50 to -42\,\kms. The location of massive dense core MM1 is marked by the purple color. Two major streams A and B are marked by orange contours. The three contour levels represent the $1\sigma$ (solid line), $2\sigma$ (dashed line), and $3\sigma$ (dotted line) probability distribution of scatter-points. Similarly, minor streams C1 and D1 are marked by the orange contours and minor streams C2 and D2 are marked by the pink colors. In each panel, ``F1--F6'' marks the end of the large-scale filaments which are assumed to be responsible for the streams. The RA, DEC location as well as velocity of filament ends are measured from \nthp~data from \citet{2013A&A...555A.112P}. The central circle marks how we visualize the 3D data cube: blue axis is the collapsing axis, where the data points are projected on the plane spanned by two black axes. \label{fig:PxV}}
\end{figure*}

The streams are spatially correlated or connected with large-scale filaments F1--F6 seen in mid-infrared extinction and \nthp~identified by \citet{2013A&A...555A.112P}. 
F1--F6 themselves already show a spiral pattern, consistent with a rotation anti-clockwise. More interestingly, this global rotation appears to be connected (further to smaller scales) to the streams revealed by red dashed lines in Figure\,\ref{fig:stream_shift} and \ref{fig:stream_integrate}. To better visualize the connection in a p--p--v space, we measure the location of RA, DEC, and velocity of the six large-scale filaments from \nthp~data in \citet{2013A&A...555A.112P} and then mark them in Figure\,\ref{fig:PyV} and \ref{fig:PxV}. The color of markers indicates the value in the collapsed axis (RA or Dec). 
Specifically, part of stream A is spatially correlated with the eastern filament F6; the systematic velocity of the stream A at the outermost end is consistent with that of F6 $V_\mathrm{F6}\simeq-44.6$\,\kms (see in the upper left panel of Figure\,\ref{fig:PxV}). 
The stream B is spatially connected to two southern filaments F4 and F5. However, if free-fall model assumed (see the black curves in the upper right panel of Figure\,\ref{fig:PyV}), then the velocity gradient prefers the connection with F4 rather than F5.
The stream C1 and C2 are spatially connected to the F1 and F3 respectively, which are clearly seen in the lower left panel of both Figure\,\ref{fig:stream_shift} and \ref{fig:PxV}.
The stream D1 can have the same origin as B from F4 and F5, but they actually have a distinct velocity difference. As seen from the orange contours in the upper and lower right panel of Figure\,\ref{fig:PyV} and \ref{fig:PxV}, the stream D1 has a coherent velocity of $\bar V_\mathrm{D}=-46.7$\,\kms~but the stream B has a wide velocity range [-49,-47]\,\kms. 
The minor stream D2 may come from the western filament F2 but the F2 meets the other massive dense core MM2 first.

All the streams are spatially connected to the central massive dense core MM1. Interestingly, the morphology of major stream A and two minor streams C1 and C2 show spiral-like features which were observed in other high-mass star forming regions \citep[e.g.][]{2015ApJ...804...37L,2017MNRAS.467L.120M,2018MNRAS.478.2505I,2019A&A...628A...6S,2019A&A...629A..81T,2020ApJ...905...25G,2021ApJ...915L..10S}. These authors believe that the intrinsic velocity gradients are due to the gas flows along the structures. In the case of ``the Heart'', the large-scale filaments are thought to transfer material inwards \citep{2013A&A...555A.112P} and the identified molecular outflows are evident for ongoing accretion towards MM1 \citep{2021A&A...645A.142A,2021ApJ...909..199O,2022ApJ...929...68O}. 
If the gas streams are physically connected to the central massive dense core MM1, then these streams are likely to bridge the large-scale gas inflow along the filaments and play a role in small-scale gas feeding towards the massive dense core MM1 \citep{2021A&A...645A.142A}.

In Figure\,\ref{fig:PyV}, streams A and B show a clear acceleration towards the central regions. We assume: 1) the most massive core MM1 drives the free-fall acceleration; 2) gas is accelerated along the elongated direction of the stream. Therefore, the gas should follow a free-fall model \citep[cf.][]{2020NatAs...4.1158P,2021ApJ...923L..20C},
\begin{equation}
\frac{v(r)-v_0}{\sin\theta_\mathrm{pos}} = \sqrt{\frac{2GM}{\cos\theta_\mathrm{pos}(r-r_0)}},
\end{equation}
where $\theta_\mathrm{pos}$ is the inclination angle of the gas stream with respect to the plane-of-sky, $v_0$ is the velocity at the starting position of the stream, and $r_0$ is the landing point of the stream (i.e. the position of MM1). We fix the mass of MM1 of 383\,\msun~and adjust the parameters $\theta_\mathrm{pos}$, $v_0$, and $r_0$ to fit the P-V distributions of the infalling stream and the fitting results are shown in Figure\,\ref{fig:PyV}. As a result, $\theta_\mathrm{pos}=-30^{\circ}$ for stream A and $\theta_\mathrm{pos}=+30^{\circ}$ for stream B, both of which are carried with uncertainties of $10^{\circ}$. The ``+/-'' sign means the angle is pointed ``out of/into'' the plane of sky. The angles are particularly useful for further calculations like mass accretion rate in Section\,\ref{subsec:continuous}.

We estimate the total gas mass in the four major streams from \htcop\,$J=1-0$ emission following a standard procedure \citep{2012ApJ...756...60S}. 
First, the column density at Pixel $(i,j)$ for a linear, rigid rotor in the optically thin regime, assuming a filling factor of unity, can be calculated from \citep{2021ApJ...915L..10S},
\begin{equation}
\begin{split}
N_{ij} &= \frac{3k_B}{8\pi^3B_\mathrm{rot}\mu_\mathrm{dm}^2}\frac{T_{\mathrm{ex},ij}+hB_\mathrm{rot}/3k_\mathrm{B}}{(J+1)}\frac{\exp(E_J/k_\mathrm{B})}{[1-\exp(-h\nu/k_\mathrm{B}T_{\mathrm{ex},ij})]}\\
&\times \frac{1}{[J(T_{\mathrm{ex},ij})-J(T_\mathrm{bg})]}\int T_{b,ij}\, dv,
\end{split}
\end{equation}
where $k_\mathrm{B}$ is the Boltzmann constant, $h$ is the Planck constant, $T_{\mathrm{ex},ij}$ is the excitation temperature at Pixel $(i,j)$ calculated from Section\,\ref{app:tau_h13cop}, $\nu$ is the transition frequency (86.754288\,GHz), $\mu_\mathrm{dm}$ is the permanent dipole moment of the molecule (3.89\,Debye), $J$ is the rotational quantum number of the lower state, $E_J=hB_\mathrm{rot}J(J+1)$ is the energy in the level $J$, $B_\mathrm{rot}$ is the rotational constant of the molecule (43.377302\,GHz), $T_{b,ij}$ is the brightness temperature at Pixel $(i,j)$, and $T_\mathrm{bg}=2.73$\,K is the background temperature. 
$J(T)$ is defined as
\begin{equation}
J(T) = \frac{h\nu}{k_\mathrm{B}}\frac{1}{e^{h\nu/k_\mathrm{B}T}-1}.
\end{equation}
The column density is then converted into mass using
\begin{equation}
M_\mathrm{stream} = (X_{\mathrm{H}^{13}\mathrm{CO}^+})^{-1} m_{\mathrm{H}_2} A D^2 \sum^{\mathrm{stream}}_{i,j} N_{ij}(\mathrm{H}^{13}\mathrm{CO}^+),
\end{equation}
where $A$ is the angular area of a pixel (0\parcsec4$\times$0\parcsec4) and $D$ is the distance (3.25\,kpc). $X_{\mathrm{H}^{13}\mathrm{CO}^+}=$[\htcop/H$_2$] is the \htcop~to molecular hydrogen abundance ratio and $m_{\mathrm{H}_2}$ is the mass of a hydrogen molecule. The sum is over the all the pixels within the stream. 
The most uncertain value is the abundance $X_{\mathrm{H}^{13}\mathrm{CO}^+}$. \citet{2013ApJ...777..157H} used MALT90 data of 333 high-mass star-forming regions and derived the abundance of $1.28\times10^{-10}$. \citet{2020ApJ...901...31L} used APEX observation of G34.43+00.24 to estimate clump-averaged \htcop~abundance to be $9\times10^{-12}$. 
\citet{2013A&A...555A.112P} used 1D non-LTE RATRAN radiation transfer code to model the spectral line \htcop\,$J=1-0$ from Mopra observations towards SDC335 and derived the abundance of $5\times10^{-11}$. Very recently, \citet{2022ApJ...926..165L} found a very similar value of $5.4\times10^{-11}$ in a massive infrared dark cloud NGC6334S.
Because our case study target is the same as \citet{2013A&A...555A.112P}, we use the abundance and uncertainty $5^{\ssstyle+3}_{\ssstyle-3}\times10^{-11}$. 
The estimated masses are shown in the second column of Table\,\ref{tab:stream}.

We note that part of streams A and D contain emission from MM1 after running \acorns. To avoid contamination from the dense core, we crudely mask the data within MM1 when discussing the streams. The masked data accounts for 12\% of the voxels by number and 47\% by mass.

\begin{table*}
\centering
\renewcommand\arraystretch{1.2}
\caption{The physical parameters of streams}
\label{tab:stream}
\begin{tabular}{cccccc} 
\hline
\hline
Stream & $M_\mathrm{stream}$ & $\Delta V_\parallel$\tablenotemarknew{a} & Length & $\nabla V_\parallel$\tablenotemarknew{b} & $\dot M_\mathrm{stream}$\tablenotemarknew{d} \\
 & (\msun) & (\kms) & (pc) & (\kms~pc$^{-1}$) & ($M_\odot$\,kyr$^{-1}$) \\
\hline
A & 140 & 2.0 & 0.58 & $5.94(^{\ssstyle+3.48}_{\ssstyle-1.85})$ & $0.85(^{\ssstyle+0.42}_{\ssstyle-0.42})(^{\ssstyle+0.50}_{\ssstyle-0.26})$ \\
\hline
B & 40 & 2.0 & 0.24 & $14.66(^{\ssstyle+8.6}_{\ssstyle-4.6})$ & $0.6(^{\ssstyle+0.3}_{\ssstyle-0.3})(^{\ssstyle+0.35}_{\ssstyle-0.19})$ \\
\hline
C & 55 & - & - & - & - \\
C1 & 32 & 1.8 & 0.24 & $7.62(^{\ssstyle+5.58}_{\ssstyle-3.22})$ & $0.25(^{\ssstyle+0.12}_{\ssstyle-0.12})(^{\ssstyle+0.18}_{\ssstyle-0.1})$ \\
C2 & 23 & 2.0 & 0.22 & $9.07(^{\ssstyle+6.64}_{\ssstyle-3.83})$ & $0.21(^{\ssstyle+0.10}_{\ssstyle-0.10})(^{\ssstyle+0.16}_{\ssstyle-0.09})$ \\
\hline
D & 285 & - & - & - \\
D1 & 34 & 0.5 & 0.28 & $1.76(^{\ssstyle+1.29}_{\ssstyle-0.74})$ & $0.06(^{\ssstyle+0.03}_{\ssstyle-0.03})(^{\ssstyle+0.04}_{\ssstyle-0.02})$ \\
D2 & 100 & 1.0 & 0.24 & $4.23(^{\ssstyle+3.10}_{\ssstyle-1.79})$ & $0.43(^{\ssstyle+0.21}_{\ssstyle-0.21})(^{\ssstyle+0.32}_{\ssstyle-0.18})$ \\
\hline
Total\tablenotemarknew{c} & 369 & - & - & - & $2.40(\pm0.78)$ \\
\hline
\multicolumn{5}{l}{$^{a.}$ Velocity difference along the stream.} \\
\multicolumn{5}{l}{$^{b.}$ Velocity gradient along the stream.} \\
\multicolumn{5}{l}{$^{c.}$ Total streams thought to connect to MM1.} \\
\multicolumn{5}{l}{$^{d.}$ The uncertainty of individual gas stream consists of two: the former} \\
\multicolumn{5}{l}{$^{}$ from mass and the latter from inclination angle. The $1\sigma$ uncertainty} \\
\multicolumn{5}{l}{$^{}$ of total mass infall rate is calculated from Monte Carlo runs.} \\
\end{tabular}
\end{table*}

\subsection{Continuous gas flow towards MM1}
\label{subsec:continuous}
\subsubsection{Measurements of the streams show ``continuity''}
We measure the velocity difference $\Delta V$ and gradient $\nabla V$ from Figure\,\ref{fig:stream_shift} as well as the Figure\,\ref{fig:PyV} and \ref{fig:PxV}.
We measure streams A and B directly since they show coherent p--p--v structure. For streams C and D, we split the major streams into minor ones (C into C1/C2 and D into D1/D2) to measure $\Delta V$ and $\nabla V$.

If the velocity gradients are mainly caused by inflows, we can estimate the accretion rate ($\dot M_\parallel$) along the gas streams following \citet{2013ApJ...766..115K},
\begin{equation}
\dot M_\parallel = \frac{\nabla V_{\parallel} M_\mathrm{stream}} {\tan\theta},
\end{equation}
where $\nabla V_\parallel$ is the velocity gradient, $M_\mathrm{stream}$ is the mass of the stream, and $\theta$ is the plane-of-sky inclination angle of gas streams. We adopt $\theta=30(\pm10)^\circ$ for streams A and B, and $\theta=45(\pm15)^\circ$ for others. Considering the dominant uncertainty from the abundance $X_{\mathrm{H}^{13}\mathrm{CO}^+}$, we include 50\% uncertainty in $M_\mathrm{stream}$.
The mass inflow rate as well as its uncertainty for each stream is shown in the last column of Table\,\ref{tab:stream}. In the last row of Table\,\ref{tab:stream}, we aggregate all the streams that are thought to transfer mass into MM1 and estimate the mean value as well as $1\sigma$ uncertainty determined by Monte Carlo runs. 
As a result, the mass inflow rate along the streams is $\dot M_\mathrm{stream} = 2.40(\pm0.78)\times10^{-3}$\,\massrate.

With this value, an intriguing comparison can be made with the derived mass infall rate, $\dot M_\mathrm{infall}$, for the whole SDC335 cloud. \citet{2013A&A...555A.112P} assumed that SDC335 was undergoing a global collapse and that the central region (i.e. ``the Heart'') contributed to the majority of the infalling mass. 
They estimated $\dot M_\mathrm{infall}\simeq2.5(\pm1.0)\times10^{-3}$\,\massrate. 
The striking consistency of this value with what we derive in this work tightens the determination of the mass infall rate within ``the Heart'', since these are two independent methods. The consistency also suggests that \htcop\,$J=1-0$ is good at tracing infalling gas in high-mass star formation regions. \citet{2021A&A...645A.142A} considered the total mass accretion rate towards protostars in MM1, $\dot M_\mathrm{tot,acc}$ to be $1.4(\pm 0.1)\times10^{-3}$\,\massrate~from outflow analyses. These authors argue that SDC335 has a continuous infall of material from cloud/clump scale (a few pc), funnelled onto the accretion disk scale (<\,0.01\,pc), and driving energetic outflows. The mass inflow rate derived from our new ALMA observation, in an intermediate scale, further proves the continuity. 

\subsubsection{Some implications of ``continuity''}
First, continuity means that the mass inflow rate at different radii is constant. This serves as a basic assumption for the turbulence-regulated gravitational collapse model in \citet{2018MNRAS.477.4951L} where the ``continuity'' provided an extra equation to finally derive the density profile in the form of $\rho\propto r^{-2}$. From the perspective of a single observed case, we justify one of the assumptions of the theoretical model.

Second, continuity along with the high mass fraction of the most massive core $f_\mathrm{MMC}$ in SDC335 shows us an early picture of high-mass star formation. 
As suggested in \citet{2021MNRAS.508.2964A}, a massive HFS (i.e. SDC335) accretes mass from its surroundings, and meanwhile feeds its most massive core (i.e. MM1) more efficiently at an early stage. The continuum emission shows that the most massive core MM1 contains 10--24\% of the total mass of the clump, giving a high mass concentration in the dense structures (see Section\,\ref{subsec:concentrate}). With the present detailed kinematic studies, we suggest that the high mass concentration can be due to the continuous gas inflow from clump scale to core scale. 
Interestingly, the ``bathtub'' model \citep[see simplified version in][for the galactic bathtub]{2017MmSAI..88..533B,2010ApJ...718.1001B} provides a possible physical explanation for the causality. 
For a star-forming region that is powered by a constant inflow of gas $\dot M_\mathrm{acc}$, we assume a critical gas density threshold $n_\mathrm{dense}=10^4$\,cm$^{-3}$ \citep{2010ApJ...724..687L}, above which stars form on a local free fall timescale $\tau_\mathrm{ff}\simeq4\times10^5$\,yrs \citep{2012ApJ...745...69K}. 
We then have
\begin{equation}
    \frac{\mathrm{d}M_\mathrm{dense}}{\mathrm{d}t} = \dot M_\mathrm{acc} - \frac{M_\mathrm{dense}}{\tau_\mathrm{ff}},
\end{equation}
where $M_\mathrm{dense}$ is the mass contained in the dense cores and $M_\mathrm{dense}/\tau_\mathrm{ff}$ shows how fast is the dense gas depleted if no more gas were accreted. 
Including a non-zero accretion term, the solution becomes
\begin{equation}
    M_\mathrm{dense} = \dot M_\mathrm{acc} \tau_\mathrm{ff}\left[1-\exp\left(-\frac{t}{\tau_\mathrm{ff}}\right)\right].
\end{equation}
We can identify two phases: 1) for $t \ll \tau_\mathrm{ff}$, the dense gas increases linearly with time as $M_\mathrm{dense} \sim \dot M_\mathrm{acc} \times t$; 2) for $t\gtrsim\tau_\mathrm{ff}$, the dense gas mass approaches a constant value as $M_\mathrm{dense} \rightarrow \dot M_\mathrm{acc}\times\tau_\mathrm{ff}$\footnote{The bathtub has a maximum volume, and so does the dense core. This is why we call it the ``bathtub'' model.}. 
For continuous accretion in SDC335, a large amount of material is directly fed into the massive dense core MM1, leading to a high mass concentration. However, once the continuity is broken (by stellar feedback, for example), then the accretion rate drops, leading to a lower mass concentration in dense structures.
This is consistent with \citet{2021MNRAS.508.2964A}, where it is found that infrared dark clumps usually have a higher mass concentration while the opposite is true for infrared bright clumps. However, to specify the connection between both types of clumps, similar detailed studies of ``continuous accretion'' should be carried out encompassing a larger sample.

Last, as suggested by \citet{2021A&A...645A.142A}, a near continuous flow of material from clump to core scale would have implications for high-mass star forming models. Under the competitive accretion \citep{2001MNRAS.323..785B} and GHC \citep{VS+19} models, massive cores are fed by the parental clump, so collapse of the clump should regulate core growth. In the case of SDC335, the natal mass reservoir grows itself and at the same time feeds the cores by multiple streams as seen in \htcop\,$J=1-0$. 
Located at the at the center of SDC335, MM1 enjoys more infalling material than MM2 and then becomes more massive, consistent with a competitive accretion.
If so, the growth of massive dense core in SDC335 prefers ``competitive accretion'' and GHC models rather than ``turbulent core'' model since the latter predicts an isolated massive core.
Even though it is only a single case, this comprehensive study details how mass is transferred inwards and highlights the advantage of the interferometers such as ALMA. 
Moreover, the study of such a prototype of global collapse as SDC335 should encourage a systematic study of kinematics of massive clumps, especially at early evolutionary stages.


\subsection{What does a global ``blue profile'' mean?}
A general prediction of collapsing models is ``blue profile'', a line asymmetry with the peak skewed to the blue side for an optically thick line, while an optically thin line must peak at the velocity of the absorption (usually a dip) of the optically thick line to rule out the possibility of two velocity components \citep{2003ApJ...592L..79W}. 
Furthermore, one can simulate such blue profile in a simple model where the optically thick lines show red-shifted self-absorption and therefore blue-shifted double-peaked line profile \citep[e.g.][]{1993ApJ...404..232Z,1996ApJ...465L.133M}. Detailed tests of consistency between observations of infall asymmetry and models of collapse require maps in both optically thick and thin spectral lines. The maps can reveal the center, shape, and extent of the zone of infall asymmetry and allow comparison to models \citep{2000prpl.conf..217M}. Mapping observations are also needed to discriminate infall motion from rotation and bipolar outflows \citep{1999ApJ...526..788L,2007ApJ...669L..37W}.

Mapping observations of optically thick lines sometimes reveal a universal or global ``blue profile'' \citep[e.g.][]{2007ApJ...669L..37W,2010A&A...520A..49S,2013A&A...555A.112P,2015MNRAS.450.1926H,2016MNRAS.456.2681Q}, which is interpreted as global gravitational collapse by simple radiation transfer model. 
In the case of SDC335, the Mopra observation of \hcop\,$J=1-0$ emission line shows a global ``blue profile'', while the \htcop\,$J=1-0$ emission line shows marginally a single Gaussian profile. \citet{2013A&A...555A.112P} took it as an evidence to exclude the double velocity component and demonstrated that the ``blue profile'' was due to the self-absorption when the gas fall inwards. 
However, in our new ALMA observations with high angular resolution, the dense gas tracer \htcop\,$J=1-0$ shows unquestionably multiple velocity components throughout the entire map of ``the Heart'' of SDC335, which is identified to be a highly complex region with several ppv-coherent streams rather than a quasi-spherical morphology seen in the large scale (see Section\,\ref{subsec:dissect}). Although these streams are transferring gas inwards which corresponds to what large-scale ``blue profile'' said, yet they detail how and especially in what morphology the gas falls inwards.

Although SDC335 show a good correspondence between large and small scale, we can still not tell whether a clump is collapsing simply by low-resolution single-dish data. For example, recent ALMA observations towards G286 show the double peaked profiles observed from the single dish can be caused by the relative motions of two sub-clumps and outflows \citep{2021MNRAS.508.4639Z}. And in another example of G12.42, the ``blue profile'' is not due to an infall motion but two velocity components \citep{ATOMS_XII_Saha}.
Numerical simulations also show the blue asymmetry of optically thick lines is not significantly correlated with actual line-of-sight motions in the cloud. 
The spectra can be more complex, which makes it difficult to unambiguously interpret any observed spectral asymmetries in terms of a collapsing motion \citep{ngVLA_Juvela}. 
Therefore, we suggest great caution when interpreting global collapse from global ``blue profiles'' towards distant massive clumps in single dish observations, even with mapping observations.
This is simply because that the coarse resolution cannot resolve the complex gas motions within those distant massive clumps.

\section{Conclusion}
\label{sec:conclude}
We have presented new ALMA observations of infrared dark cloud SDC335 both in Band-3 and Band-7 from the ATOMS survey. We have focused on the densest part of SDC335, ``the Heart''. 
At a resolution of $\sim$\,0.03\,pc, we use 3\,mm dust continuum emission to study the physical properties of massive dense cores ($\lesssim0.1$\,pc) MM1 and MM2. Furthermore, we use 0.87\,mm continuum emission as well as multi-transition of \htcs~to study the fragmentation of MM1 down to a scale of 0.01\,pc. 
We use \htcop\,$J=1-0$ at a velocity resolution of 0.2\,\kms~to study the gas kinematics inside ``the Heart'', serving as the bridge between the clump-scale global collapse and the core-scale gas feeding the protostellar core. 
Our main results are as follows:

\begin{enumerate}
	\item Two dense cores are detected in 3\,mm dust continuum emission, with physical size smaller than 0.1\,pc. Both are massive: MM1 with $383(^{\ssstyle+234}_{\ssstyle-120})$\,\msun~and MM2 with $74(^{\ssstyle+47}_{\ssstyle-24})$\,\msun. The mass concentration of SDC335 within its most massive core MM1 is as high as 10\%-24\%.
	
	\item From the 0.87\,mm continuum data, the massive core MM1 is further fragmented into six condensations S1--S6. S1 and S2 are protostellar, and are spatially coincident with UC\hii regions identified in ATCA centimeter continuum. he relation between separation and mass of condensations favor turbulent Jeans fragmentation where the turbulence seems to be scale-free rather than scale-dependent as it is on large scales. 
	
	\item For the first time, we use the ALMA \htcop\,$J=1-0$ line to map the complex gas motions inside ``the Heart''. The emission of \htcop\,$J=1-0$ is almost always optically thin, even in the densest portions of MM1. In this case, the observed multi-peaked profiles are due to multiple velocity components. With the decomposition of spectral lines and clustering algorithm, we identify the four major gas streams A, B, C, and D. Due to the nature of adopted clustering algorithm, each stream is coherent in p--p--v space. Streams A and C both show spiral-like morphology. Stream C is composed of two minor streams C1 and C2, while the stream D is also composed of two minor ones: D1 and D2. Those major and minor streams are spatially correlated or connected with the large-scale filaments which is identified from the Spitzer mid-infrared map.
	
	\item Streams A, B, C1, C2, D1, and D2 are connected to the most massive core MM1. If the velocity gradients trace the mass inflow along the streams, then the total mass inflow rate is $2.40(\pm0.78)\times10^{-3}$\,\massrate, which is consistent with the value derived from an independent method of spherical global collapse by \citet{2013A&A...555A.112P}. The consistency not only suggests that \htcop\,$J=1-0$~line is a good tracer of inflowing gas, but also implies a nearly continuous flow from large-scale cloud collapse (SDC335, $\sim$\,1\,pc) to smaller-scale core feeding (MM1, $\lesssim$\,0.1\,pc). We suggest the high mass concentration in MM1 is due to the high efficiency of mass inflow.
	
	\item In the perspective of high angular resolution, ALMA sees multiple curving gas streams moving inwards and feeding the core, corresponding to what large-scale blue profiles told. Although SDC335 keeps the coherence from large to small scale, we still suggest great caution and high-resolution observations when interpreting so-called global ``blue profiles''.
\end{enumerate}

Our comprehensive study of SDC335 showcases the detailed gas kinematics in a prototypical massive infalling clump, and calls for a further systematic and statistical studies.

\section*{Acknowledgements}
We thank an anonymous referee for the thorough review and critical comments that helped us to improve the scientific content and the presentation of this paper.
FWX and KW acknowledge support from the China Manned Space Project (CMS-CSST-2021-A09, CMS- CSST-2021-B06), the National Key Research and Development Program of China (2017YFA0402702, 2019YFA0405100), the National Science Foundation of China (11973013, 11721303, 12033005), and the High-Performance Computing Platform of Peking University. 
TL acknowledges the support by NSFC through grants No.12073061 and No.12122307, the international partnership program of Chinese academy of sciences through grant No.114231KYSB20200009, and Shanghai Pujiang Program 20PJ1415500. 
This research was carried out in part at the Jet Propulsion Laboratory, which is operated by the California Institute of Technology under a contract with the National Aeronautics and Space Administration (80NM0018D0004). 
HLL is supported by National Natural Science Foundation of China (NSFC) through the grant No.12103045. 
GG and LB gratefully acknowledge support by the ANID BASAL projects ACE210002 and FB210003. 
GCG acknowledges support by UNAM-PAPIIT IN103822 grant. 
KT was supported by JSPS KAKENHI (Grant Number 20H05645). 
AS gratefully acknowledges support by the Fondecyt Regular (project code 1220610), and ANID BASAL projects ACE210002 and FB210003. 
GAF acknowledges support from the Collaborative Research Centre 956, funded by the Deutsche Forschungsgemeinschaft (DFG) project ID 184018867. 
CWL is supported by the Basic Science Research Program through the National Research Foundation of Korea (NRF) funded by the Ministry of Education, Science and Technology (NRF-2019R1A2C1010851), and by the Korea Astronomy and Space Science Institute grant funded by the Korea government (MSIT) (Project No. 2022-1-840-05).
CE acknowledges the financial support from grant RJF/2020/000071 as a part of Ramanujan Fellowship awarded by Science and Engineering Research Board (SERB), Department of Science and Technology (DST), Government of India.

This research makes use of \textsc{\large astropy}, a community-developed core \textsc{\large python} package for Astronomy \citep{2018AJ....156..123A}. This research uses \getsf, a multi-scale, multi-wavelength extraction algorithm of sources and filaments \citep{2021A&A...649A..89M}. This research uses \scousepy, a multi-component spectral line decomposition \citep{2020ascl.soft03004H}. This research uses \acorns, an unsupervised clustering algorithm \citep{2020ascl.soft03003H}. This research makes use of \textsc{\large montage}, funded by the National Science Foundation under Grant Number ACI-1440620, and previously funded by the National Aeronautics and Space Administration's Earth Science Technology Office, Computation Technologies Project, under Cooperative Agreement Number NCC5-626 between NASA and the California Institute of Technology.

This paper makes use of the following ALMA data: ADS/JAO.ALMA 2019.1.00685.S and 2017.1.00545.S. ALMA is a partnership of ESO (representing its member states), NSF (USA), and NINS (Japan), together with NRC (Canada), MOST and ASIAA (Taiwan), and KASI (Republic of Korea), in cooperation with the Republic of Chile. The Joint ALMA Observatory is operated by ESO, AUI/NRAO, and NAOJ.

\section*{Data Availability}
The ALMA archival data in this paper can be downloaded from ADS/JAO.ALMA 2019.1.00685.S and 2017.1.00545.S at \href{https://almascience.eso.org/aq/}{ALMA science archive}. The Spitzer GLIMPSE legacy survey data can be downloaded at \href{https://irsa.ipac.caltech.edu/irsaviewer/}{NASA/IPAC Infrared Science Archive}.



\bibliographystyle{mnras}
\bibliography{bridge} 

\begin{thebibliography}{}
\makeatletter
\relax
\def\mn@urlcharsother{\let\do\@makeother \do\$\do\&\do\#\do\^\do\_\do\%\do\~}
\def\mn@doi{\begingroup\mn@urlcharsother \@ifnextchar [ {\mn@doi@}
  {\mn@doi@[]}}
\def\mn@doi@[#1]#2{\def\@tempa{#1}\ifx\@tempa\@empty \href
  {http://dx.doi.org/#2} {doi:#2}\else \href {http://dx.doi.org/#2} {#1}\fi
  \endgroup}
\def\mn@eprint#1#2{\mn@eprint@#1:#2::\@nil}
\def\mn@eprint@arXiv#1{\href {http://arxiv.org/abs/#1} {{\tt arXiv:#1}}}
\def\mn@eprint@dblp#1{\href {http://dblp.uni-trier.de/rec/bibtex/#1.xml}
  {dblp:#1}}
\def\mn@eprint@#1:#2:#3:#4\@nil{\def\@tempa {#1}\def\@tempb {#2}\def\@tempc
  {#3}\ifx \@tempc \@empty \let \@tempc \@tempb \let \@tempb \@tempa \fi \ifx
  \@tempb \@empty \def\@tempb {arXiv}\fi \@ifundefined
  {mn@eprint@\@tempb}{\@tempb:\@tempc}{\expandafter \expandafter \csname
  mn@eprint@\@tempb\endcsname \expandafter{\@tempc}}}

\bibitem[\protect\citeauthoryear{{Anderson} et~al.,}{{Anderson}
  et~al.}{2021}]{2021MNRAS.508.2964A}
{Anderson} M.,  et~al., 2021, \mn@doi [\mnras] {10.1093/mnras/stab2674}, \href
  {https://ui.adsabs.harvard.edu/abs/2021MNRAS.508.2964A} {508, 2964}

\bibitem[\protect\citeauthoryear{{Andr{\'e}} et~al.,}{{Andr{\'e}}
  et~al.}{2010}]{2010A&A...518L.102A}
{Andr{\'e}} P.,  et~al., 2010, \mn@doi [\aap] {10.1051/0004-6361/201014666},
  \href {https://ui.adsabs.harvard.edu/abs/2010A&A...518L.102A} {518, L102}

\bibitem[\protect\citeauthoryear{{Andr{\'e}}, {Di Francesco}, {Ward-Thompson},
  {Inutsuka}, {Pudritz}  \& {Pineda}}{{Andr{\'e}}
  et~al.}{2014}]{2014prpl.conf...27A}
{Andr{\'e}} P.,  {Di Francesco} J.,  {Ward-Thompson} D.,  {Inutsuka} S.~I.,
  {Pudritz} R.~E.,   {Pineda} J.~E.,  2014, in {Beuther} H.,  {Klessen} R.~S.,
  {Dullemond} C.~P.,   {Henning} T.,  eds, Protostars and Planets VI. p.~27
  (\mn@eprint {arXiv} {1312.6232}),
  \mn@doi{10.2458/azu\_uapress\_9780816531240-ch002}

\bibitem[\protect\citeauthoryear{{Astropy Collaboration} et~al.,}{{Astropy
  Collaboration} et~al.}{2018}]{2018AJ....156..123A}
{Astropy Collaboration} et~al., 2018, \mn@doi [\aj] {10.3847/1538-3881/aabc4f},
  \href {https://ui.adsabs.harvard.edu/abs/2018AJ....156..123A} {156, 123}

\bibitem[\protect\citeauthoryear{{Avison}, {Peretto}, {Fuller},
  {Duarte-Cabral}, {Traficante}  \& {Pineda}}{{Avison}
  et~al.}{2015}]{2015A&A...577A..30A}
{Avison} A.,  {Peretto} N.,  {Fuller} G.~A.,  {Duarte-Cabral} A.,  {Traficante}
  A.,   {Pineda} J.~E.,  2015, \mn@doi [\aap] {10.1051/0004-6361/201425041},
  \href {https://ui.adsabs.harvard.edu/abs/2015A&A...577A..30A} {577, A30}

\bibitem[\protect\citeauthoryear{{Avison} et~al.,}{{Avison}
  et~al.}{2021}]{2021A&A...645A.142A}
{Avison} A.,  et~al., 2021, \mn@doi [\aap] {10.1051/0004-6361/201936043}, \href
  {https://ui.adsabs.harvard.edu/abs/2021A&A...645A.142A} {645, A142}

\bibitem[\protect\citeauthoryear{{Barrow}, {Bhavsar}  \& {Sonoda}}{{Barrow}
  et~al.}{1985}]{1985MNRAS.216...17B}
{Barrow} J.~D.,  {Bhavsar} S.~P.,   {Sonoda} D.~H.,  1985, \mn@doi [\mnras]
  {10.1093/mnras/216.1.17}, \href
  {https://ui.adsabs.harvard.edu/abs/1985MNRAS.216...17B} {216, 17}

\bibitem[\protect\citeauthoryear{{Baug} et~al.,}{{Baug}
  et~al.}{2020}]{2020ApJ...890...44B}
{Baug} T.,  et~al., 2020, \mn@doi [\apj] {10.3847/1538-4357/ab66b6}, \href
  {https://ui.adsabs.harvard.edu/abs/2020ApJ...890...44B} {890, 44}

\bibitem[\protect\citeauthoryear{{Beuther} et~al.,}{{Beuther}
  et~al.}{2018}]{2018A&A...617A.100B}
{Beuther} H.,  et~al., 2018, \mn@doi [\aap] {10.1051/0004-6361/201833021},
  \href {https://ui.adsabs.harvard.edu/abs/2018A&A...617A.100B} {617, A100}

\bibitem[\protect\citeauthoryear{{Bonnell}, {Bate}, {Clarke}  \&
  {Pringle}}{{Bonnell} et~al.}{2001}]{2001MNRAS.323..785B}
{Bonnell} I.~A.,  {Bate} M.~R.,  {Clarke} C.~J.,   {Pringle} J.~E.,  2001,
  \mn@doi [\mnras] {10.1046/j.1365-8711.2001.04270.x}, \href
  {https://ui.adsabs.harvard.edu/abs/2001MNRAS.323..785B} {323, 785}

\bibitem[\protect\citeauthoryear{{Bonnell}, {Vine}  \& {Bate}}{{Bonnell}
  et~al.}{2004}]{2004MNRAS.349..735B}
{Bonnell} I.~A.,  {Vine} S.~G.,   {Bate} M.~R.,  2004, \mn@doi [\mnras]
  {10.1111/j.1365-2966.2004.07543.x}, \href
  {https://ui.adsabs.harvard.edu/abs/2004MNRAS.349..735B} {349, 735}

\bibitem[\protect\citeauthoryear{{Bouch{\'e}} et~al.,}{{Bouch{\'e}}
  et~al.}{2010}]{2010ApJ...718.1001B}
{Bouch{\'e}} N.,  et~al., 2010, \mn@doi [\apj] {10.1088/0004-637X/718/2/1001},
  \href {https://ui.adsabs.harvard.edu/abs/2010ApJ...718.1001B} {718, 1001}

\bibitem[\protect\citeauthoryear{{Burkert}}{{Burkert}}{2017}]{2017MmSAI..88..533B}
{Burkert} A.,  2017, \memsai, \href
  {https://ui.adsabs.harvard.edu/abs/2017MmSAI..88..533B} {88, 533}

\bibitem[\protect\citeauthoryear{{Cao}, {Qiu}, {Zhang}, {Wang}, {Hu}  \&
  {Liu}}{{Cao} et~al.}{2019}]{2019ApJS..241....1C}
{Cao} Y.,  {Qiu} K.,  {Zhang} Q.,  {Wang} Y.,  {Hu} B.,   {Liu} J.,  2019,
  \mn@doi [\apjs] {10.3847/1538-4365/ab0025}, \href
  {https://ui.adsabs.harvard.edu/abs/2019ApJS..241....1C} {241, 1}

\bibitem[\protect\citeauthoryear{{Cao}, {Qiu}, {Zhang}, {Wang}  \&
  {Xiao}}{{Cao} et~al.}{2021}]{2021ApJ...918L...4C}
{Cao} Y.,  {Qiu} K.,  {Zhang} Q.,  {Wang} Y.,   {Xiao} Y.,  2021, \mn@doi
  [\apjl] {10.3847/2041-8213/ac1947}, \href
  {https://ui.adsabs.harvard.edu/abs/2021ApJ...918L...4C} {918, L4}

\bibitem[\protect\citeauthoryear{{Chen} et~al.,}{{Chen}
  et~al.}{2021}]{2021ApJ...923L..20C}
{Chen} X.,  et~al., 2021, \mn@doi [\apjl] {10.3847/2041-8213/ac3ec8}, \href
  {https://ui.adsabs.harvard.edu/abs/2021ApJ...923L..20C} {923, L20}

\bibitem[\protect\citeauthoryear{{Chira}, {Smith}, {Klessen}, {Stutz}  \&
  {Shetty}}{{Chira} et~al.}{2014}]{2014MNRAS.444..874C}
{Chira} R.-A.,  {Smith} R.~J.,  {Klessen} R.~S.,  {Stutz} A.~M.,   {Shetty} R.,
   2014, \mn@doi [\mnras] {10.1093/mnras/stu1497}, \href
  {https://ui.adsabs.harvard.edu/abs/2014MNRAS.444..874C} {444, 874}

\bibitem[\protect\citeauthoryear{{Contreras} et~al.,}{{Contreras}
  et~al.}{2013}]{2013A&A...549A..45C}
{Contreras} Y.,  et~al., 2013, \mn@doi [\aap] {10.1051/0004-6361/201220155},
  \href {https://ui.adsabs.harvard.edu/abs/2013A&A...549A..45C} {549, A45}

\bibitem[\protect\citeauthoryear{{Contreras} et~al.,}{{Contreras}
  et~al.}{2018}]{2018ApJ...861...14C}
{Contreras} Y.,  et~al., 2018, \mn@doi [\apj] {10.3847/1538-4357/aac2ec}, \href
  {https://ui.adsabs.harvard.edu/abs/2018ApJ...861...14C} {861, 14}

\bibitem[\protect\citeauthoryear{{Csengeri}, {Bontemps}, {Schneider}, {Motte},
  {Gueth}  \& {Hora}}{{Csengeri} et~al.}{2011}]{2011ApJ...740L...5C}
{Csengeri} T.,  {Bontemps} S.,  {Schneider} N.,  {Motte} F.,  {Gueth} F.,
  {Hora} J.~L.,  2011, \mn@doi [\apjl] {10.1088/2041-8205/740/1/L5}, \href
  {https://ui.adsabs.harvard.edu/abs/2011ApJ...740L...5C} {740, L5}

\bibitem[\protect\citeauthoryear{{Cyganowski} et~al.,}{{Cyganowski}
  et~al.}{2014}]{2014ApJ...796L...2C}
{Cyganowski} C.~J.,  et~al., 2014, \mn@doi [\apjl]
  {10.1088/2041-8205/796/1/L2}, \href
  {https://ui.adsabs.harvard.edu/abs/2014ApJ...796L...2C} {796, L2}

\bibitem[\protect\citeauthoryear{{Dewangan}, {Ojha}, {Sharma}, {Palacio},
  {Bhadari}  \& {Das}}{{Dewangan} et~al.}{2020}]{2020ApJ...903...13D}
{Dewangan} L.~K.,  {Ojha} D.~K.,  {Sharma} S.,  {Palacio} S.~d.,  {Bhadari}
  N.~K.,   {Das} A.,  2020, \mn@doi [\apj] {10.3847/1538-4357/abb827}, \href
  {https://ui.adsabs.harvard.edu/abs/2020ApJ...903...13D} {903, 13}

\bibitem[\protect\citeauthoryear{{Elia} et~al.,}{{Elia}
  et~al.}{2017}]{2017MNRAS.471..100E}
{Elia} D.,  et~al., 2017, \mn@doi [\mnras] {10.1093/mnras/stx1357}, \href
  {https://ui.adsabs.harvard.edu/abs/2017MNRAS.471..100E} {471, 100}

\bibitem[\protect\citeauthoryear{{Galv{\'a}n-Madrid}, {Zhang}, {Keto}, {Ho},
  {Zapata}, {Rodr{\'\i}guez}, {Pineda}  \&
  {V{\'a}zquez-Semadeni}}{{Galv{\'a}n-Madrid}
  et~al.}{2010}]{2010ApJ...725...17G}
{Galv{\'a}n-Madrid} R.,  {Zhang} Q.,  {Keto} E.,  {Ho} P. T.~P.,  {Zapata}
  L.~A.,  {Rodr{\'\i}guez} L.~F.,  {Pineda} J.~E.,   {V{\'a}zquez-Semadeni} E.,
   2010, \mn@doi [\apj] {10.1088/0004-637X/725/1/17}, \href
  {https://ui.adsabs.harvard.edu/abs/2010ApJ...725...17G} {725, 17}

\bibitem[\protect\citeauthoryear{{Garay}, {Brooks}, {Mardones}, {Norris}  \&
  {Burton}}{{Garay} et~al.}{2002}]{2002ApJ...579..678G}
{Garay} G.,  {Brooks} K.~J.,  {Mardones} D.,  {Norris} R.~P.,   {Burton} M.~G.,
   2002, \mn@doi [\apj] {10.1086/342986}, \href
  {https://ui.adsabs.harvard.edu/abs/2002ApJ...579..678G} {579, 678}

\bibitem[\protect\citeauthoryear{{Ge} \& {Wang}}{{Ge} \&
  {Wang}}{2022}]{2022ApJS..259...36G}
{Ge} Y.,  {Wang} K.,  2022, \mn@doi [\apjs] {10.3847/1538-4365/ac4a76}, \href
  {https://ui.adsabs.harvard.edu/abs/2022ApJS..259...36G} {259, 36}

\bibitem[\protect\citeauthoryear{{Goddi}, {Ginsburg}, {Maud}, {Zhang}  \&
  {Zapata}}{{Goddi} et~al.}{2020}]{2020ApJ...905...25G}
{Goddi} C.,  {Ginsburg} A.,  {Maud} L.~T.,  {Zhang} Q.,   {Zapata} L.~A.,
  2020, \mn@doi [\apj] {10.3847/1538-4357/abc88e}, \href
  {https://ui.adsabs.harvard.edu/abs/2020ApJ...905...25G} {905, 25}

\bibitem[\protect\citeauthoryear{{G{\'o}mez} \&
  {V{\'a}zquez-Semadeni}}{{G{\'o}mez} \& {V{\'a}zquez-Semadeni}}{2014}]{GV14}
{G{\'o}mez} G.~C.,  {V{\'a}zquez-Semadeni} E.,  2014, \mn@doi [\apj]
  {10.1088/0004-637X/791/2/124}, \href
  {https://ui.adsabs.harvard.edu/abs/2014ApJ...791..124G} {791, 124}

\bibitem[\protect\citeauthoryear{{G{\'o}mez}, {Walsh}  \& {Palau}}{{G{\'o}mez}
  et~al.}{2022}]{Gomez+22}
{G{\'o}mez} G.~C.,  {Walsh} C.,   {Palau} A.,  2022, \mn@doi [\mnras]
  {10.1093/mnras/stac912}, \href
  {https://ui.adsabs.harvard.edu/abs/2022MNRAS.513.1244G} {513, 1244}

\bibitem[\protect\citeauthoryear{{He} et~al.,}{{He}
  et~al.}{2015}]{2015MNRAS.450.1926H}
{He} Y.-X.,  et~al., 2015, \mn@doi [\mnras] {10.1093/mnras/stv732}, \href
  {https://ui.adsabs.harvard.edu/abs/2015MNRAS.450.1926H} {450, 1926}

\bibitem[\protect\citeauthoryear{{Hennemann} et~al.,}{{Hennemann}
  et~al.}{2012}]{2012A&A...543L...3H}
{Hennemann} M.,  et~al., 2012, \mn@doi [\aap] {10.1051/0004-6361/201219429},
  \href {https://ui.adsabs.harvard.edu/abs/2012A&A...543L...3H} {543, L3}

\bibitem[\protect\citeauthoryear{{Henshaw} et~al.,}{{Henshaw}
  et~al.}{2016}]{2016MNRAS.457.2675H}
{Henshaw} J.~D.,  et~al., 2016, \mn@doi [\mnras] {10.1093/mnras/stw121}, \href
  {https://ui.adsabs.harvard.edu/abs/2016MNRAS.457.2675H} {457, 2675}

\bibitem[\protect\citeauthoryear{{Henshaw} et~al.,}{{Henshaw}
  et~al.}{2019}]{2019MNRAS.485.2457H}
{Henshaw} J.~D.,  et~al., 2019, \mn@doi [\mnras] {10.1093/mnras/stz471}, \href
  {https://ui.adsabs.harvard.edu/abs/2019MNRAS.485.2457H} {485, 2457}

\bibitem[\protect\citeauthoryear{{Henshaw}, {Sokolov}  \& {Ginsburg}}{{Henshaw}
  et~al.}{2020a}]{2020ascl.soft03003H}
{Henshaw} J.,  {Sokolov} V.,   {Ginsburg} A.,  2020a, {acorns: Agglomerative
  Clustering for ORganising Nested Structures} (\mn@eprint {ascl} {2003.003})

\bibitem[\protect\citeauthoryear{{Henshaw}, {Ginsburg}  \& {Riener}}{{Henshaw}
  et~al.}{2020b}]{2020ascl.soft03004H}
{Henshaw} J.,  {Ginsburg} A.,   {Riener} M.,  2020b, {scousepy: Semi-automated
  multi-COmponent Universal Spectral-line fitting Engine} (\mn@eprint {ascl}
  {2003.004})

\bibitem[\protect\citeauthoryear{{Hoq} et~al.,}{{Hoq}
  et~al.}{2013}]{2013ApJ...777..157H}
{Hoq} S.,  et~al., 2013, \mn@doi [\apj] {10.1088/0004-637X/777/2/157}, \href
  {https://ui.adsabs.harvard.edu/abs/2013ApJ...777..157H} {777, 157}

\bibitem[\protect\citeauthoryear{{Izquierdo}, {Galv{\'a}n-Madrid}, {Maud},
  {Hoare}, {Johnston}, {Keto}, {Zhang}  \& {de Wit}}{{Izquierdo}
  et~al.}{2018}]{2018MNRAS.478.2505I}
{Izquierdo} A.~F.,  {Galv{\'a}n-Madrid} R.,  {Maud} L.~T.,  {Hoare} M.~G.,
  {Johnston} K.~G.,  {Keto} E.~R.,  {Zhang} Q.,   {de Wit} W.-J.,  2018,
  \mn@doi [\mnras] {10.1093/mnras/sty1096}, \href
  {https://ui.adsabs.harvard.edu/abs/2018MNRAS.478.2505I} {478, 2505}

\bibitem[\protect\citeauthoryear{{Juvela}, {Mannfors}, {Liu}  \&
  {Toth}}{{Juvela} et~al.}{2022}]{ngVLA_Juvela}
{Juvela} M.,  {Mannfors} E.,  {Liu} T.,   {Toth} L.~V.,  2022, arXiv e-prints,
  \href {https://ui.adsabs.harvard.edu/abs/2022arXiv220801894J} {p.
  arXiv:2208.01894}

\bibitem[\protect\citeauthoryear{{Kauffmann}, {Bertoldi}, {Bourke}, {Evans}  \&
  {Lee}}{{Kauffmann} et~al.}{2008}]{2008A&A...487..993K}
{Kauffmann} J.,  {Bertoldi} F.,  {Bourke} T.~L.,  {Evans} N.~J. I.,   {Lee}
  C.~W.,  2008, \mn@doi [\aap] {10.1051/0004-6361:200809481}, \href
  {https://ui.adsabs.harvard.edu/abs/2008A&A...487..993K} {487, 993}

\bibitem[\protect\citeauthoryear{{Kirk}, {Myers}, {Bourke}, {Gutermuth},
  {Hedden}  \& {Wilson}}{{Kirk} et~al.}{2013}]{2013ApJ...766..115K}
{Kirk} H.,  {Myers} P.~C.,  {Bourke} T.~L.,  {Gutermuth} R.~A.,  {Hedden} A.,
  {Wilson} G.~W.,  2013, \mn@doi [\apj] {10.1088/0004-637X/766/2/115}, \href
  {https://ui.adsabs.harvard.edu/abs/2013ApJ...766..115K} {766, 115}

\bibitem[\protect\citeauthoryear{{Kong}, {Tan}, {Caselli}, {Fontani}, {Liu}  \&
  {Butler}}{{Kong} et~al.}{2017}]{2017ApJ...834..193K}
{Kong} S.,  {Tan} J.~C.,  {Caselli} P.,  {Fontani} F.,  {Liu} M.,   {Butler}
  M.~J.,  2017, \mn@doi [\apj] {10.3847/1538-4357/834/2/193}, \href
  {https://ui.adsabs.harvard.edu/abs/2017ApJ...834..193K} {834, 193}

\bibitem[\protect\citeauthoryear{{Kong}, {Arce}, {Shirley}  \&
  {Glasgow}}{{Kong} et~al.}{2021}]{2021ApJ...912..156K}
{Kong} S.,  {Arce} H.~G.,  {Shirley} Y.,   {Glasgow} C.,  2021, \mn@doi [\apj]
  {10.3847/1538-4357/abefe7}, \href
  {https://ui.adsabs.harvard.edu/abs/2021ApJ...912..156K} {912, 156}

\bibitem[\protect\citeauthoryear{{Krumholz}, {Klein}  \& {McKee}}{{Krumholz}
  et~al.}{2011}]{2011ApJ...740...74K}
{Krumholz} M.~R.,  {Klein} R.~I.,   {McKee} C.~F.,  2011, \mn@doi [\apj]
  {10.1088/0004-637X/740/2/74}, \href
  {https://ui.adsabs.harvard.edu/abs/2011ApJ...740...74K} {740, 74}

\bibitem[\protect\citeauthoryear{{Krumholz}, {Dekel}  \& {McKee}}{{Krumholz}
  et~al.}{2012}]{2012ApJ...745...69K}
{Krumholz} M.~R.,  {Dekel} A.,   {McKee} C.~F.,  2012, \mn@doi [\apj]
  {10.1088/0004-637X/745/1/69}, \href
  {https://ui.adsabs.harvard.edu/abs/2012ApJ...745...69K} {745, 69}

\bibitem[\protect\citeauthoryear{{Lada}, {Lombardi}  \& {Alves}}{{Lada}
  et~al.}{2010}]{2010ApJ...724..687L}
{Lada} C.~J.,  {Lombardi} M.,   {Alves} J.~F.,  2010, \mn@doi [\apj]
  {10.1088/0004-637X/724/1/687}, \href
  {https://ui.adsabs.harvard.edu/abs/2010ApJ...724..687L} {724, 687}

\bibitem[\protect\citeauthoryear{{Lee}, {Myers}  \& {Tafalla}}{{Lee}
  et~al.}{1999}]{1999ApJ...526..788L}
{Lee} C.~W.,  {Myers} P.~C.,   {Tafalla} M.,  1999, \mn@doi [\apj]
  {10.1086/308027}, \href
  {https://ui.adsabs.harvard.edu/abs/1999ApJ...526..788L} {526, 788}

\bibitem[\protect\citeauthoryear{{Levshakov}, {Henkel}, {Reimers}, {Wang},
  {Mao}, {Wang}  \& {Xu}}{{Levshakov} et~al.}{2013}]{2013A&A...553A..58L}
{Levshakov} S.~A.,  {Henkel} C.,  {Reimers} D.,  {Wang} M.,  {Mao} R.,  {Wang}
  H.,   {Xu} Y.,  2013, \mn@doi [\aap] {10.1051/0004-6361/201220354}, \href
  {https://ui.adsabs.harvard.edu/abs/2013A&A...553A..58L} {553, A58}

\bibitem[\protect\citeauthoryear{{Li}}{{Li}}{2018}]{2018MNRAS.477.4951L}
{Li} G.-X.,  2018, \mn@doi [\mnras] {10.1093/mnras/sty657}, \href
  {https://ui.adsabs.harvard.edu/abs/2018MNRAS.477.4951L} {477, 4951}

\bibitem[\protect\citeauthoryear{{Li}, {Klein}  \& {McKee}}{{Li}
  et~al.}{2018}]{2018MNRAS.473.4220L}
{Li} P.~S.,  {Klein} R.~I.,   {McKee} C.~F.,  2018, \mn@doi [\mnras]
  {10.1093/mnras/stx2611}, \href
  {https://ui.adsabs.harvard.edu/abs/2018MNRAS.473.4220L} {473, 4220}

\bibitem[\protect\citeauthoryear{{Li}, {Zhang}, {Pillai}, {Stephens}, {Wang}
  \& {Li}}{{Li} et~al.}{2019}]{2019ApJ...886..130L}
{Li} S.,  {Zhang} Q.,  {Pillai} T.,  {Stephens} I.~W.,  {Wang} J.,   {Li} F.,
  2019, \mn@doi [\apj] {10.3847/1538-4357/ab464e}, \href
  {https://ui.adsabs.harvard.edu/abs/2019ApJ...886..130L} {886, 130}

\bibitem[\protect\citeauthoryear{{Li}, {Cao}  \& {Qiu}}{{Li}
  et~al.}{2021}]{2021ApJ...916...13L}
{Li} G.-X.,  {Cao} Y.,   {Qiu} K.,  2021, \mn@doi [\apj]
  {10.3847/1538-4357/ac01d4}, \href
  {https://ui.adsabs.harvard.edu/abs/2021ApJ...916...13L} {916, 13}

\bibitem[\protect\citeauthoryear{{Li} et~al.,}{{Li}
  et~al.}{2022}]{2022ApJ...926..165L}
{Li} S.,  et~al., 2022, \mn@doi [\apj] {10.3847/1538-4357/ac3df8}, \href
  {https://ui.adsabs.harvard.edu/abs/2022ApJ...926..165L} {926, 165}

\bibitem[\protect\citeauthoryear{{Lin}, {Mestel}  \& {Shu}}{{Lin}
  et~al.}{1965}]{Lin+65}
{Lin} C.~C.,  {Mestel} L.,   {Shu} F.~H.,  1965, \mn@doi [\apj]
  {10.1086/148428}, \href
  {https://ui.adsabs.harvard.edu/abs/1965ApJ...142.1431L} {142, 1431}

\bibitem[\protect\citeauthoryear{{Liu}, {Quintana-Lacaci}, {Wang}, {Ho}, {Li},
  {Zhang}  \& {Zhang}}{{Liu} et~al.}{2012}]{2012ApJ...745...61L}
{Liu} H.~B.,  {Quintana-Lacaci} G.,  {Wang} K.,  {Ho} P. T.~P.,  {Li} Z.-Y.,
  {Zhang} Q.,   {Zhang} Z.-Y.,  2012, \mn@doi [\apj]
  {10.1088/0004-637X/745/1/61}, \href
  {https://ui.adsabs.harvard.edu/abs/2012ApJ...745...61L} {745, 61}

\bibitem[\protect\citeauthoryear{{Liu}, {Galv{\'a}n-Madrid},
  {Jim{\'e}nez-Serra}, {Rom{\'a}n-Z{\'u}{\~n}iga}, {Zhang}, {Li}  \&
  {Chen}}{{Liu} et~al.}{2015}]{2015ApJ...804...37L}
{Liu} H.~B.,  {Galv{\'a}n-Madrid} R.,  {Jim{\'e}nez-Serra} I.,
  {Rom{\'a}n-Z{\'u}{\~n}iga} C.,  {Zhang} Q.,  {Li} Z.,   {Chen} H.-R.,  2015,
  \mn@doi [\apj] {10.1088/0004-637X/804/1/37}, \href
  {https://ui.adsabs.harvard.edu/abs/2015ApJ...804...37L} {804, 37}

\bibitem[\protect\citeauthoryear{{Liu} et~al.,}{{Liu}
  et~al.}{2016}]{2016ApJ...824...31L}
{Liu} T.,  et~al., 2016, \mn@doi [\apj] {10.3847/0004-637X/824/1/31}, \href
  {https://ui.adsabs.harvard.edu/abs/2016ApJ...824...31L} {824, 31}

\bibitem[\protect\citeauthoryear{{Liu} et~al.,}{{Liu}
  et~al.}{2020a}]{2020MNRAS.496.2790L}
{Liu} T.,  et~al., 2020a, \mn@doi [\mnras] {10.1093/mnras/staa1577}, \href
  {https://ui.adsabs.harvard.edu/abs/2020MNRAS.496.2790L} {496, 2790}

\bibitem[\protect\citeauthoryear{{Liu} et~al.,}{{Liu}
  et~al.}{2020b}]{2020MNRAS.496.2821L}
{Liu} T.,  et~al., 2020b, \mn@doi [\mnras] {10.1093/mnras/staa1501}, \href
  {https://ui.adsabs.harvard.edu/abs/2020MNRAS.496.2821L} {496, 2821}

\bibitem[\protect\citeauthoryear{{Liu}, {Sanhueza}, {Liu}, {Zavagno}, {Tang},
  {Wu}  \& {Zhang}}{{Liu} et~al.}{2020c}]{2020ApJ...901...31L}
{Liu} H.-L.,  {Sanhueza} P.,  {Liu} T.,  {Zavagno} A.,  {Tang} X.-D.,  {Wu} Y.,
    {Zhang} S.,  2020c, \mn@doi [\apj] {10.3847/1538-4357/abadfe}, \href
  {https://ui.adsabs.harvard.edu/abs/2020ApJ...901...31L} {901, 31}

\bibitem[\protect\citeauthoryear{{Liu} et~al.,}{{Liu}
  et~al.}{2021}]{2021ApJ...912..148L}
{Liu} X.~C.,  et~al., 2021, \mn@doi [\apj] {10.3847/1538-4357/abee73}, \href
  {https://ui.adsabs.harvard.edu/abs/2021ApJ...912..148L} {912, 148}

\bibitem[\protect\citeauthoryear{{Liu} et~al.,}{{Liu}
  et~al.}{2022}]{2022MNRAS.510.5009L}
{Liu} H.-L.,  et~al., 2022, \mn@doi [\mnras] {10.1093/mnras/stab2757}, \href
  {https://ui.adsabs.harvard.edu/abs/2022MNRAS.510.5009L} {510, 5009}

\bibitem[\protect\citeauthoryear{{Longmore}, {Pillai}, {Keto}, {Zhang}  \&
  {Qiu}}{{Longmore} et~al.}{2011}]{2011ApJ...726...97L}
{Longmore} S.~N.,  {Pillai} T.,  {Keto} E.,  {Zhang} Q.,   {Qiu} K.,  2011,
  \mn@doi [\apj] {10.1088/0004-637X/726/2/97}, \href
  {https://ui.adsabs.harvard.edu/abs/2011ApJ...726...97L} {726, 97}

\bibitem[\protect\citeauthoryear{{Louvet}}{{Louvet}}{2018}]{2018sf2a.conf..311L}
{Louvet} F.,  2018, in {Di Matteo} P.,  {Billebaud} F.,  {Herpin} F.,
  {Lagarde} N.,  {Marquette} J.~B.,  {Robin} A.,   {Venot} O.,  eds, SF2A-2018:
  Proceedings of the Annual meeting of the French Society of Astronomy and
  Astrophysics. p.~Di

\bibitem[\protect\citeauthoryear{{Lu} et~al.,}{{Lu}
  et~al.}{2018}]{2018ApJ...855....9L}
{Lu} X.,  et~al., 2018, \mn@doi [\apj] {10.3847/1538-4357/aaad11}, \href
  {https://ui.adsabs.harvard.edu/abs/2018ApJ...855....9L} {855, 9}

\bibitem[\protect\citeauthoryear{{Mathis}, {Rumpl}  \& {Nordsieck}}{{Mathis}
  et~al.}{1977}]{1977ApJ...217..425M}
{Mathis} J.~S.,  {Rumpl} W.,   {Nordsieck} K.~H.,  1977, \mn@doi [\apj]
  {10.1086/155591}, \href
  {https://ui.adsabs.harvard.edu/abs/1977ApJ...217..425M} {217, 425}

\bibitem[\protect\citeauthoryear{{Maud}, {Hoare}, {Galv{\'a}n-Madrid}, {Zhang},
  {de Wit}, {Keto}, {Johnston}  \& {Pineda}}{{Maud}
  et~al.}{2017}]{2017MNRAS.467L.120M}
{Maud} L.~T.,  {Hoare} M.~G.,  {Galv{\'a}n-Madrid} R.,  {Zhang} Q.,  {de Wit}
  W.~J.,  {Keto} E.,  {Johnston} K.~G.,   {Pineda} J.~E.,  2017, \mn@doi
  [\mnras] {10.1093/mnrasl/slx010}, \href
  {https://ui.adsabs.harvard.edu/abs/2017MNRAS.467L.120M} {467, L120}

\bibitem[\protect\citeauthoryear{{McKee} \& {Tan}}{{McKee} \&
  {Tan}}{2003}]{2003ApJ...585..850M}
{McKee} C.~F.,  {Tan} J.~C.,  2003, \mn@doi [\apj] {10.1086/346149}, \href
  {https://ui.adsabs.harvard.edu/abs/2003ApJ...585..850M} {585, 850}

\bibitem[\protect\citeauthoryear{{McMullin}, {Waters}, {Schiebel}, {Young}  \&
  {Golap}}{{McMullin} et~al.}{2007}]{2007ASPC..376..127M}
{McMullin} J.~P.,  {Waters} B.,  {Schiebel} D.,  {Young} W.,   {Golap} K.,
  2007, in {Shaw} R.~A.,  {Hill} F.,   {Bell} D.~J.,  eds,  Astronomical
  Society of the Pacific Conference Series Vol. 376, Astronomical Data Analysis
  Software and Systems XVI. p.~127

\bibitem[\protect\citeauthoryear{{Men'shchikov}}{{Men'shchikov}}{2021}]{2021A&A...649A..89M}
{Men'shchikov} A.,  2021, \mn@doi [\aap] {10.1051/0004-6361/202039913}, \href
  {https://ui.adsabs.harvard.edu/abs/2021A&A...649A..89M} {649, A89}

\bibitem[\protect\citeauthoryear{{Men'shchikov}, {Andr{\'e}}, {Didelon},
  {Motte}, {Hennemann}  \& {Schneider}}{{Men'shchikov}
  et~al.}{2012}]{2012A&A...542A..81M}
{Men'shchikov} A.,  {Andr{\'e}} P.,  {Didelon} P.,  {Motte} F.,  {Hennemann}
  M.,   {Schneider} N.,  2012, \mn@doi [\aap] {10.1051/0004-6361/201218797},
  \href {https://ui.adsabs.harvard.edu/abs/2012A&A...542A..81M} {542, A81}

\bibitem[\protect\citeauthoryear{{Molet} et~al.,}{{Molet}
  et~al.}{2019}]{2019A&A...626A.132M}
{Molet} J.,  et~al., 2019, \mn@doi [\aap] {10.1051/0004-6361/201935497}, \href
  {https://ui.adsabs.harvard.edu/abs/2019A&A...626A.132M} {626, A132}

\bibitem[\protect\citeauthoryear{{Molinari} et~al.,}{{Molinari}
  et~al.}{2010}]{2010A&A...518L.100M}
{Molinari} S.,  et~al., 2010, \mn@doi [\aap] {10.1051/0004-6361/201014659},
  \href {https://ui.adsabs.harvard.edu/abs/2010A&A...518L.100M} {518, L100}

\bibitem[\protect\citeauthoryear{{Molinari} et~al.,}{{Molinari}
  et~al.}{2016}]{2016A&A...591A.149M}
{Molinari} S.,  et~al., 2016, \mn@doi [\aap] {10.1051/0004-6361/201526380},
  \href {https://ui.adsabs.harvard.edu/abs/2016A&A...591A.149M} {591, A149}

\bibitem[\protect\citeauthoryear{{M{\"o}ller}, {Bernst}, {Panoglou}, {Muders},
  {Ossenkopf}, {R{\"o}llig}  \& {Schilke}}{{M{\"o}ller}
  et~al.}{2013}]{2013A&A...549A..21M}
{M{\"o}ller} T.,  {Bernst} I.,  {Panoglou} D.,  {Muders} D.,  {Ossenkopf} V.,
  {R{\"o}llig} M.,   {Schilke} P.,  2013, \mn@doi [\aap]
  {10.1051/0004-6361/201220063}, \href
  {https://ui.adsabs.harvard.edu/abs/2013A&A...549A..21M} {549, A21}

\bibitem[\protect\citeauthoryear{{M{\"o}ller}, {Endres}  \&
  {Schilke}}{{M{\"o}ller} et~al.}{2017}]{2017A&A...598A...7M}
{M{\"o}ller} T.,  {Endres} C.,   {Schilke} P.,  2017, \mn@doi [\aap]
  {10.1051/0004-6361/201527203}, \href
  {https://ui.adsabs.harvard.edu/abs/2017A&A...598A...7M} {598, A7}

\bibitem[\protect\citeauthoryear{{Motte}, {Bontemps}  \& {Louvet}}{{Motte}
  et~al.}{2018}]{2018ARA&A..56...41M}
{Motte} F.,  {Bontemps} S.,   {Louvet} F.,  2018, \mn@doi [\araa]
  {10.1146/annurev-astro-091916-055235}, \href
  {https://ui.adsabs.harvard.edu/abs/2018ARA&A..56...41M} {56, 41}

\bibitem[\protect\citeauthoryear{{Myers}}{{Myers}}{2009}]{2009ApJ...700.1609M}
{Myers} P.~C.,  2009, \mn@doi [\apj] {10.1088/0004-637X/700/2/1609}, \href
  {https://ui.adsabs.harvard.edu/abs/2009ApJ...700.1609M} {700, 1609}

\bibitem[\protect\citeauthoryear{{Myers}, {Mardones}, {Tafalla}, {Williams}  \&
  {Wilner}}{{Myers} et~al.}{1996}]{1996ApJ...465L.133M}
{Myers} P.~C.,  {Mardones} D.,  {Tafalla} M.,  {Williams} J.~P.,   {Wilner}
  D.~J.,  1996, \mn@doi [\apjl] {10.1086/310146}, \href
  {https://ui.adsabs.harvard.edu/abs/1996ApJ...465L.133M} {465, L133}

\bibitem[\protect\citeauthoryear{{Myers}, {Evans}  \& {Ohashi}}{{Myers}
  et~al.}{2000}]{2000prpl.conf..217M}
{Myers} P.~C.,  {Evans} N.~J. I.,   {Ohashi} N.,  2000, in {Mannings} V.,
  {Boss} A.~P.,   {Russell} S.~S.,  eds, Protostars and Planets IV. p.~217

\bibitem[\protect\citeauthoryear{{Myers}, {McKee}, {Cunningham}, {Klein}  \&
  {Krumholz}}{{Myers} et~al.}{2013}]{2013ApJ...766...97M}
{Myers} A.~T.,  {McKee} C.~F.,  {Cunningham} A.~J.,  {Klein} R.~I.,
  {Krumholz} M.~R.,  2013, \mn@doi [\apj] {10.1088/0004-637X/766/2/97}, \href
  {https://ui.adsabs.harvard.edu/abs/2013ApJ...766...97M} {766, 97}

\bibitem[\protect\citeauthoryear{{Naranjo-Romero}, {V{\'a}zquez-Semadeni}  \&
  {Loughnane}}{{Naranjo-Romero} et~al.}{2022}]{Naranjo+22}
{Naranjo-Romero} R.,  {V{\'a}zquez-Semadeni} E.,   {Loughnane} R.~M.,  2022,
  \mn@doi [\mnras] {10.1093/mnras/stac804}, \href
  {https://ui.adsabs.harvard.edu/abs/2022MNRAS.512.4715N} {512, 4715}

\bibitem[\protect\citeauthoryear{{Olguin} et~al.,}{{Olguin}
  et~al.}{2021}]{2021ApJ...909..199O}
{Olguin} F.~A.,  et~al., 2021, \mn@doi [\apj] {10.3847/1538-4357/abde3f}, \href
  {https://ui.adsabs.harvard.edu/abs/2021ApJ...909..199O} {909, 199}

\bibitem[\protect\citeauthoryear{{Olguin}, {Sanhueza}, {Ginsburg}, {Chen},
  {Zhang}, {Li}, {Lu}  \& {Sakai}}{{Olguin} et~al.}{2022}]{2022ApJ...929...68O}
{Olguin} F.~A.,  {Sanhueza} P.,  {Ginsburg} A.,  {Chen} H.-R.~V.,  {Zhang} Q.,
  {Li} S.,  {Lu} X.,   {Sakai} T.,  2022, \mn@doi [\apj]
  {10.3847/1538-4357/ac5bd8}, \href
  {https://ui.adsabs.harvard.edu/abs/2022ApJ...929...68O} {929, 68}

\bibitem[\protect\citeauthoryear{{Ossenkopf} \& {Henning}}{{Ossenkopf} \&
  {Henning}}{1994}]{1994A&A...291..943O}
{Ossenkopf} V.,  {Henning} T.,  1994, \aap, \href
  {https://ui.adsabs.harvard.edu/abs/1994A&A...291..943O} {291, 943}

\bibitem[\protect\citeauthoryear{{Padoan}, {Pan}, {Juvela}, {Haugb{\o}lle}  \&
  {Nordlund}}{{Padoan} et~al.}{2020}]{2020ApJ...900...82P}
{Padoan} P.,  {Pan} L.,  {Juvela} M.,  {Haugb{\o}lle} T.,   {Nordlund}
  {\r{A}}.,  2020, \mn@doi [\apj] {10.3847/1538-4357/abaa47}, \href
  {https://ui.adsabs.harvard.edu/abs/2020ApJ...900...82P} {900, 82}

\bibitem[\protect\citeauthoryear{{Peretto} \& {Fuller}}{{Peretto} \&
  {Fuller}}{2009}]{2009A&A...505..405P}
{Peretto} N.,  {Fuller} G.~A.,  2009, \mn@doi [\aap]
  {10.1051/0004-6361/200912127}, \href
  {https://ui.adsabs.harvard.edu/abs/2009A&A...505..405P} {505, 405}

\bibitem[\protect\citeauthoryear{{Peretto} et~al.,}{{Peretto}
  et~al.}{2013}]{2013A&A...555A.112P}
{Peretto} N.,  et~al., 2013, \mn@doi [\aap] {10.1051/0004-6361/201321318},
  \href {https://ui.adsabs.harvard.edu/abs/2013A&A...555A.112P} {555, A112}

\bibitem[\protect\citeauthoryear{{Peretto} et~al.,}{{Peretto}
  et~al.}{2014}]{2014A&A...561A..83P}
{Peretto} N.,  et~al., 2014, \mn@doi [\aap] {10.1051/0004-6361/201322172},
  \href {https://ui.adsabs.harvard.edu/abs/2014A&A...561A..83P} {561, A83}

\bibitem[\protect\citeauthoryear{{Pillai}, {Wyrowski}, {Carey}  \&
  {Menten}}{{Pillai} et~al.}{2006}]{2006A&A...450..569P}
{Pillai} T.,  {Wyrowski} F.,  {Carey} S.~J.,   {Menten} K.~M.,  2006, \mn@doi
  [\aap] {10.1051/0004-6361:20054128}, \href
  {https://ui.adsabs.harvard.edu/abs/2006A&A...450..569P} {450, 569}

\bibitem[\protect\citeauthoryear{{Pillai}, {Kauffmann}, {Wyrowski}, {Hatchell},
  {Gibb}  \& {Thompson}}{{Pillai} et~al.}{2011}]{2011A&A...530A.118P}
{Pillai} T.,  {Kauffmann} J.,  {Wyrowski} F.,  {Hatchell} J.,  {Gibb} A.~G.,
  {Thompson} M.~A.,  2011, \mn@doi [\aap] {10.1051/0004-6361/201015899}, \href
  {https://ui.adsabs.harvard.edu/abs/2011A&A...530A.118P} {530, A118}

\bibitem[\protect\citeauthoryear{{Pineda}, {Segura-Cox}, {Caselli},
  {Cunningham}, {Zhao}, {Schmiedeke}, {Maureira}  \& {Neri}}{{Pineda}
  et~al.}{2020}]{2020NatAs...4.1158P}
{Pineda} J.~E.,  {Segura-Cox} D.,  {Caselli} P.,  {Cunningham} N.,  {Zhao} B.,
  {Schmiedeke} A.,  {Maureira} M.~J.,   {Neri} R.,  2020, \mn@doi [Nature
  Astronomy] {10.1038/s41550-020-1150-z}, \href
  {https://ui.adsabs.harvard.edu/abs/2020NatAs...4.1158P} {4, 1158}

\bibitem[\protect\citeauthoryear{{Prim}}{{Prim}}{1957}]{1957BSTJ...36.1389P}
{Prim} R.~C.,  1957, \mn@doi [Bell System Technical Journal]
  {10.1002/j.1538-7305.1957.tb01515.x}, \href
  {https://ui.adsabs.harvard.edu/abs/1957BSTJ...36.1389P} {36, 1389}

\bibitem[\protect\citeauthoryear{{Qin}, {Schilke}, {Wu}, {Liu}, {Wu},
  {S{\'a}nchez-Monge}  \& {Liu}}{{Qin} et~al.}{2016}]{2016MNRAS.456.2681Q}
{Qin} S.-L.,  {Schilke} P.,  {Wu} J.,  {Liu} T.,  {Wu} Y.,  {S{\'a}nchez-Monge}
  {\'A}.,   {Liu} Y.,  2016, \mn@doi [\mnras] {10.1093/mnras/stv2801}, \href
  {https://ui.adsabs.harvard.edu/abs/2016MNRAS.456.2681Q} {456, 2681}

\bibitem[\protect\citeauthoryear{{Qin} et~al.,}{{Qin}
  et~al.}{2022}]{2022MNRAS.511.3463Q}
{Qin} S.-L.,  et~al., 2022, \mn@doi [\mnras] {10.1093/mnras/stac219}, \href
  {https://ui.adsabs.harvard.edu/abs/2022MNRAS.511.3463Q} {511, 3463}

\bibitem[\protect\citeauthoryear{{Reid}, {Dame}, {Menten}  \&
  {Brunthaler}}{{Reid} et~al.}{2016}]{2016ApJ...823...77R}
{Reid} M.~J.,  {Dame} T.~M.,  {Menten} K.~M.,   {Brunthaler} A.,  2016, \mn@doi
  [\apj] {10.3847/0004-637X/823/2/77}, \href
  {https://ui.adsabs.harvard.edu/abs/2016ApJ...823...77R} {823, 77}

\bibitem[\protect\citeauthoryear{{Rosolowsky}, {Pineda}, {Kauffmann}  \&
  {Goodman}}{{Rosolowsky} et~al.}{2008}]{2008ApJ...679.1338R}
{Rosolowsky} E.~W.,  {Pineda} J.~E.,  {Kauffmann} J.,   {Goodman} A.~A.,  2008,
  \mn@doi [\apj] {10.1086/587685}, \href
  {https://ui.adsabs.harvard.edu/abs/2008ApJ...679.1338R} {679, 1338}

\bibitem[\protect\citeauthoryear{{Rosolowsky} et~al.,}{{Rosolowsky}
  et~al.}{2010}]{2010ApJS..188..123R}
{Rosolowsky} E.,  et~al., 2010, \mn@doi [\apjs] {10.1088/0067-0049/188/1/123},
  \href {https://ui.adsabs.harvard.edu/abs/2010ApJS..188..123R} {188, 123}

\bibitem[\protect\citeauthoryear{{Saha} et~al.,}{{Saha}
  et~al.}{2022}]{ATOMS_XII_Saha}
{Saha} A.,  et~al., 2022, \mn@doi [\mnras] {10.1093/mnras/stac2353}, \href
  {https://ui.adsabs.harvard.edu/abs/2022MNRAS.516.1983S} {516, 1983}

\bibitem[\protect\citeauthoryear{{Sanhueza}, {Jackson}, {Foster}, {Garay},
  {Silva}  \& {Finn}}{{Sanhueza} et~al.}{2012}]{2012ApJ...756...60S}
{Sanhueza} P.,  {Jackson} J.~M.,  {Foster} J.~B.,  {Garay} G.,  {Silva} A.,
  {Finn} S.~C.,  2012, \mn@doi [\apj] {10.1088/0004-637X/756/1/60}, \href
  {https://ui.adsabs.harvard.edu/abs/2012ApJ...756...60S} {756, 60}

\bibitem[\protect\citeauthoryear{{Sanhueza}, {Jackson}, {Zhang}, {Guzm{\'a}n},
  {Lu}, {Stephens}, {Wang}  \& {Tatematsu}}{{Sanhueza}
  et~al.}{2017}]{2017ApJ...841...97S}
{Sanhueza} P.,  {Jackson} J.~M.,  {Zhang} Q.,  {Guzm{\'a}n} A.~E.,  {Lu} X.,
  {Stephens} I.~W.,  {Wang} K.,   {Tatematsu} K.,  2017, \mn@doi [\apj]
  {10.3847/1538-4357/aa6ff8}, \href
  {https://ui.adsabs.harvard.edu/abs/2017ApJ...841...97S} {841, 97}

\bibitem[\protect\citeauthoryear{{Sanhueza} et~al.,}{{Sanhueza}
  et~al.}{2019}]{2019ApJ...886..102S}
{Sanhueza} P.,  et~al., 2019, \mn@doi [\apj] {10.3847/1538-4357/ab45e9}, \href
  {https://ui.adsabs.harvard.edu/abs/2019ApJ...886..102S} {886, 102}

\bibitem[\protect\citeauthoryear{{Sanhueza} et~al.,}{{Sanhueza}
  et~al.}{2021}]{2021ApJ...915L..10S}
{Sanhueza} P.,  et~al., 2021, \mn@doi [\apjl] {10.3847/2041-8213/ac081c}, \href
  {https://ui.adsabs.harvard.edu/abs/2021ApJ...915L..10S} {915, L10}

\bibitem[\protect\citeauthoryear{{Schneider}, {Csengeri}, {Bontemps}, {Motte},
  {Simon}, {Hennebelle}, {Federrath}  \& {Klessen}}{{Schneider}
  et~al.}{2010}]{2010A&A...520A..49S}
{Schneider} N.,  {Csengeri} T.,  {Bontemps} S.,  {Motte} F.,  {Simon} R.,
  {Hennebelle} P.,  {Federrath} C.,   {Klessen} R.,  2010, \mn@doi [\aap]
  {10.1051/0004-6361/201014481}, \href
  {https://ui.adsabs.harvard.edu/abs/2010A&A...520A..49S} {520, A49}

\bibitem[\protect\citeauthoryear{{Schw{\"o}rer} et~al.,}{{Schw{\"o}rer}
  et~al.}{2019}]{2019A&A...628A...6S}
{Schw{\"o}rer} A.,  et~al., 2019, \mn@doi [\aap] {10.1051/0004-6361/201935200},
  \href {https://ui.adsabs.harvard.edu/abs/2019A&A...628A...6S} {628, A6}

\bibitem[\protect\citeauthoryear{{Shimajiri} et~al.,}{{Shimajiri}
  et~al.}{2017}]{2017A&A...604A..74S}
{Shimajiri} Y.,  et~al., 2017, \mn@doi [\aap] {10.1051/0004-6361/201730633},
  \href {https://ui.adsabs.harvard.edu/abs/2017A&A...604A..74S} {604, A74}

\bibitem[\protect\citeauthoryear{{Shirley}}{{Shirley}}{2015}]{2015PASP..127..299S}
{Shirley} Y.~L.,  2015, \mn@doi [\pasp] {10.1086/680342}, \href
  {https://ui.adsabs.harvard.edu/abs/2015PASP..127..299S} {127, 299}

\bibitem[\protect\citeauthoryear{{Smith}, {Longmore}  \& {Bonnell}}{{Smith}
  et~al.}{2009}]{2009MNRAS.400.1775S}
{Smith} R.~J.,  {Longmore} S.,   {Bonnell} I.,  2009, \mn@doi [\mnras]
  {10.1111/j.1365-2966.2009.15621.x}, \href
  {https://ui.adsabs.harvard.edu/abs/2009MNRAS.400.1775S} {400, 1775}

\bibitem[\protect\citeauthoryear{{Stutz} \& {Gould}}{{Stutz} \&
  {Gould}}{2016}]{2016A&A...590A...2S}
{Stutz} A.~M.,  {Gould} A.,  2016, \mn@doi [\aap]
  {10.1051/0004-6361/201527979}, \href
  {https://ui.adsabs.harvard.edu/abs/2016A&A...590A...2S} {590, A2}

\bibitem[\protect\citeauthoryear{{Svoboda} et~al.,}{{Svoboda}
  et~al.}{2019}]{2019ApJ...886...36S}
{Svoboda} B.~E.,  et~al., 2019, \mn@doi [\apj] {10.3847/1538-4357/ab40ca},
  \href {https://ui.adsabs.harvard.edu/abs/2019ApJ...886...36S} {886, 36}

\bibitem[\protect\citeauthoryear{{Terebey}, {Chandler}  \& {Andre}}{{Terebey}
  et~al.}{1993}]{1993ApJ...414..759T}
{Terebey} S.,  {Chandler} C.~J.,   {Andre} P.,  1993, \mn@doi [\apj]
  {10.1086/173121}, \href
  {https://ui.adsabs.harvard.edu/abs/1993ApJ...414..759T} {414, 759}

\bibitem[\protect\citeauthoryear{{Trevi{\~n}o-Morales}
  et~al.,}{{Trevi{\~n}o-Morales} et~al.}{2019}]{2019A&A...629A..81T}
{Trevi{\~n}o-Morales} S.~P.,  et~al., 2019, \mn@doi [\aap]
  {10.1051/0004-6361/201935260}, \href
  {https://ui.adsabs.harvard.edu/abs/2019A&A...629A..81T} {629, A81}

\bibitem[\protect\citeauthoryear{{V{\'a}zquez-Semadeni}, {Palau},
  {Ballesteros-Paredes}, {G{\'o}mez}  \&
  {Zamora-Avil{\'e}s}}{{V{\'a}zquez-Semadeni} et~al.}{2019}]{VS+19}
{V{\'a}zquez-Semadeni} E.,  {Palau} A.,  {Ballesteros-Paredes} J.,  {G{\'o}mez}
  G.~C.,   {Zamora-Avil{\'e}s} M.,  2019, \mn@doi [\mnras]
  {10.1093/mnras/stz2736}, \href
  {https://ui.adsabs.harvard.edu/abs/2019MNRAS.490.3061V} {490, 3061}

\bibitem[\protect\citeauthoryear{{Wang}}{{Wang}}{2018}]{WangKe2018}
{Wang} K.,  2018, \mn@doi [Research Notes of the American Astronomical Society]
  {10.3847/2515-5172/aacb29}, \href
  {https://ui.adsabs.harvard.edu/abs/2018RNAAS...2...52W} {2, 52}

\bibitem[\protect\citeauthoryear{{Wang}, {Zhang}, {Pillai}, {Wyrowski}  \&
  {Wu}}{{Wang} et~al.}{2008}]{2008ApJ...672L..33W}
{Wang} Y.,  {Zhang} Q.,  {Pillai} T.,  {Wyrowski} F.,   {Wu} Y.,  2008, \mn@doi
  [\apjl] {10.1086/524949}, \href
  {https://ui.adsabs.harvard.edu/abs/2008ApJ...672L..33W} {672, L33}

\bibitem[\protect\citeauthoryear{{Wang}, {Zhang}, {Wu}  \& {Zhang}}{{Wang}
  et~al.}{2011}]{WangK2011}
{Wang} K.,  {Zhang} Q.,  {Wu} Y.,   {Zhang} H.,  2011, \mn@doi [\apj]
  {10.1088/0004-637X/735/1/64}, \href
  {https://ui.adsabs.harvard.edu/abs/2011ApJ...735...64W} {735, 64}

\bibitem[\protect\citeauthoryear{{Wang}, {Zhang}, {Wu}, {Li}  \&
  {Zhang}}{{Wang} et~al.}{2012}]{2012ApJ...745L..30W}
{Wang} K.,  {Zhang} Q.,  {Wu} Y.,  {Li} H.-b.,   {Zhang} H.,  2012, \mn@doi
  [\apjl] {10.1088/2041-8205/745/2/L30}, \href
  {https://ui.adsabs.harvard.edu/abs/2012ApJ...745L..30W} {745, L30}

\bibitem[\protect\citeauthoryear{{Wang} et~al.,}{{Wang}
  et~al.}{2014}]{WangK2014}
{Wang} K.,  et~al., 2014, \mn@doi [\mnras] {10.1093/mnras/stu127}, \href
  {https://ui.adsabs.harvard.edu/abs/2014MNRAS.439.3275W} {439, 3275}

\bibitem[\protect\citeauthoryear{{Wang}, {Testi}, {Ginsburg}, {Walmsley},
  {Molinari}  \& {Schisano}}{{Wang} et~al.}{2015}]{2015MNRAS.450.4043W}
{Wang} K.,  {Testi} L.,  {Ginsburg} A.,  {Walmsley} C.~M.,  {Molinari} S.,
  {Schisano} E.,  2015, \mn@doi [\mnras] {10.1093/mnras/stv735}, \href
  {https://ui.adsabs.harvard.edu/abs/2015MNRAS.450.4043W} {450, 4043}

\bibitem[\protect\citeauthoryear{{Wang}, {Testi}, {Burkert}, {Walmsley},
  {Beuther}  \& {Henning}}{{Wang} et~al.}{2016}]{WangK2016}
{Wang} K.,  {Testi} L.,  {Burkert} A.,  {Walmsley} C.~M.,  {Beuther} H.,
  {Henning} T.,  2016, \mn@doi [\apjs] {10.3847/0067-0049/226/1/9}, \href
  {https://ui.adsabs.harvard.edu/abs/2016ApJS..226....9W} {226, 9}

\bibitem[\protect\citeauthoryear{{Wareing}, {Falle}  \& {Pittard}}{{Wareing}
  et~al.}{2019}]{2019MNRAS.485.4686W}
{Wareing} C.~J.,  {Falle} S.~A.~E.~G.,   {Pittard} J.~M.,  2019, \mn@doi
  [\mnras] {10.1093/mnras/stz768}, \href
  {https://ui.adsabs.harvard.edu/abs/2019MNRAS.485.4686W} {485, 4686}

\bibitem[\protect\citeauthoryear{{Wu} \& {Evans}}{{Wu} \&
  {Evans}}{2003}]{2003ApJ...592L..79W}
{Wu} J.,  {Evans} Neal~J. I.,  2003, \mn@doi [\apjl] {10.1086/377679}, \href
  {https://ui.adsabs.harvard.edu/abs/2003ApJ...592L..79W} {592, L79}

\bibitem[\protect\citeauthoryear{{Wu}, {Henkel}, {Xue}, {Guan}  \&
  {Miller}}{{Wu} et~al.}{2007}]{2007ApJ...669L..37W}
{Wu} Y.,  {Henkel} C.,  {Xue} R.,  {Guan} X.,   {Miller} M.,  2007, \mn@doi
  [\apjl] {10.1086/522958}, \href
  {https://ui.adsabs.harvard.edu/abs/2007ApJ...669L..37W} {669, L37}

\bibitem[\protect\citeauthoryear{{Wu}, {Tan}, {Christie}, {Nakamura}, {Van Loo}
   \& {Collins}}{{Wu} et~al.}{2017}]{2017ApJ...841...88W}
{Wu} B.,  {Tan} J.~C.,  {Christie} D.,  {Nakamura} F.,  {Van Loo} S.,
  {Collins} D.,  2017, \mn@doi [\apj] {10.3847/1538-4357/aa6ffa}, \href
  {https://ui.adsabs.harvard.edu/abs/2017ApJ...841...88W} {841, 88}

\bibitem[\protect\citeauthoryear{{Xie} et~al.,}{{Xie}
  et~al.}{2021}]{2021SCPMA..6479511X}
{Xie} J.,  et~al., 2021, \mn@doi [Science China Physics, Mechanics, and
  Astronomy] {10.1007/s11433-021-1695-0}, \href
  {https://ui.adsabs.harvard.edu/abs/2021SCPMA..6479511X} {64, 279511}

\bibitem[\protect\citeauthoryear{{Yuan} et~al.,}{{Yuan}
  et~al.}{2018}]{2018ApJ...852...12Y}
{Yuan} J.,  et~al., 2018, \mn@doi [\apj] {10.3847/1538-4357/aa9d40}, \href
  {https://ui.adsabs.harvard.edu/abs/2018ApJ...852...12Y} {852, 12}

\bibitem[\protect\citeauthoryear{{Yue}, {Li}, {Zhang}, {Zhu}, {Henshaw},
  {Mardones}  \& {Ren}}{{Yue} et~al.}{2021}]{2021RAA....21...24Y}
{Yue} N.-N.,  {Li} D.,  {Zhang} Q.-Z.,  {Zhu} L.,  {Henshaw} J.,  {Mardones}
  D.,   {Ren} Z.-Y.,  2021, \mn@doi [Research in Astronomy and Astrophysics]
  {10.1088/1674-4527/21/1/24}, \href
  {https://ui.adsabs.harvard.edu/abs/2021RAA....21...24Y} {21, 024}

\bibitem[\protect\citeauthoryear{{Zhang} \& {Wang}}{{Zhang} \&
  {Wang}}{2011}]{2011ApJ...733...26Z}
{Zhang} Q.,  {Wang} K.,  2011, \mn@doi [\apj] {10.1088/0004-637X/733/1/26},
  \href {https://ui.adsabs.harvard.edu/abs/2011ApJ...733...26Z} {733, 26}

\bibitem[\protect\citeauthoryear{{Zhang}, {Wang}, {Pillai}  \&
  {Rathborne}}{{Zhang} et~al.}{2009}]{2009ApJ...696..268Z}
{Zhang} Q.,  {Wang} Y.,  {Pillai} T.,   {Rathborne} J.,  2009, \mn@doi [\apj]
  {10.1088/0004-637X/696/1/268}, \href
  {https://ui.adsabs.harvard.edu/abs/2009ApJ...696..268Z} {696, 268}

\bibitem[\protect\citeauthoryear{{Zhang}, {Wang}, {Lu}  \&
  {Jim{\'e}nez-Serra}}{{Zhang} et~al.}{2015}]{2015ApJ...804..141Z}
{Zhang} Q.,  {Wang} K.,  {Lu} X.,   {Jim{\'e}nez-Serra} I.,  2015, \mn@doi
  [\apj] {10.1088/0004-637X/804/2/141}, \href
  {https://ui.adsabs.harvard.edu/abs/2015ApJ...804..141Z} {804, 141}

\bibitem[\protect\citeauthoryear{{Zhou}, {Evans}, {Koempe}  \&
  {Walmsley}}{{Zhou} et~al.}{1993}]{1993ApJ...404..232Z}
{Zhou} S.,  {Evans} Neal~J. I.,  {Koempe} C.,   {Walmsley} C.~M.,  1993,
  \mn@doi [\apj] {10.1086/172271}, \href
  {https://ui.adsabs.harvard.edu/abs/1993ApJ...404..232Z} {404, 232}

\bibitem[\protect\citeauthoryear{{Zhou} et~al.,}{{Zhou}
  et~al.}{2021}]{2021MNRAS.508.4639Z}
{Zhou} J.-W.,  et~al., 2021, \mn@doi [\mnras] {10.1093/mnras/stab2801}, \href
  {https://ui.adsabs.harvard.edu/abs/2021MNRAS.508.4639Z} {508, 4639}

\bibitem[\protect\citeauthoryear{{Zhou} et~al.,}{{Zhou}
  et~al.}{2022}]{ATOMS_XI_Zhou}
{Zhou} J.-W.,  et~al., 2022, arXiv e-prints, \href
  {https://ui.adsabs.harvard.edu/abs/2022arXiv220608505Z} {p. arXiv:2206.08505}

\bibitem[\protect\citeauthoryear{{van der Tak}, {Black}, {Sch{\"o}ier},
  {Jansen}  \& {van Dishoeck}}{{van der Tak}
  et~al.}{2007}]{2007A&A...468..627V}
{van der Tak} F.~F.~S.,  {Black} J.~H.,  {Sch{\"o}ier} F.~L.,  {Jansen} D.~J.,
   {van Dishoeck} E.~F.,  2007, \mn@doi [\aap] {10.1051/0004-6361:20066820},
  \href {https://ui.adsabs.harvard.edu/abs/2007A&A...468..627V} {468, 627}

\makeatother
\end{thebibliography}


\clearpage
\onecolumn

\noindent
\textit{Affiliations:}
\\ \hspace*{\fill} \\
\noindent
$^{1}$Kavli Institute for Astronomy and Astrophysics, Peking University, Beijing, 100871, People's Republic of China \\
$^{2}$Department of Astronomy, School of Physics, Peking University, Beijing, 100871, People's Republic of China \\
$^{3}$Shanghai Astronomical Observatory, Chinese Academy of Sciences, 80 Nandan Road, Shanghai 200030, People's Republic of China \\
$^{4}$Jet Propulsion Laboratory, California Institute of Technology, 4800 Oak Grove Drive, Pasadena, CA 91109, USA \\
$^{5}$Center for Astrophysics $|$ Harvard \& Smithsonian, 60 Garden Street, Cambridge, MA 02138, USA \\
$^{6}$Department of Physics, University of Helsinki, PO Box 64, FI-00014 Helsinki, Finland \\
$^{7}$Department of Astronomy, Yunnan University, Kunming 650091, People's Republic of China \\
$^{8}$South-Western Institute for Astronomy Research, Yunnan University, Kunming, People's Republic of China \\
$^{9}$Indian Institute of Space Science and Technology, Thiruvananthapuram 695 547, Kerala, India \\
$^{10}$Departamento de Astronom\'{\i}a, Universidad de Chile, Las Condes, 7591245 Santiago, Chile \\
$^{11}$Max Planck Institute for Astronomy, K\"onigstuhl 17, 69117 Heidelberg, Germany \\
$^{12}$Instituto de Radioastronomía y Astrofísica, Universidad Nacional Autónoma de M\'exico, Apdo. Postal 3-72, Morelia, Michoac\'an, 58089, M\'exico \\
$^{13}$Nobeyama Radio Observatory, National Astronomical Observatory of Japan, National Institutes of Natural Sciences, Nobeyama, Minamimaki, Minamisaku, Nagano 384-1305, Japan \\
$^{14}$National Astronomical Observatories, Chinese Academy of Sciences, Beijing 100101, People’s Republic of China \\
$^{15}$School of Physics and Astronomy, Sun Yat-sen University, 2 Daxue Road, Zhuhai, Guangdong, 519082, People’s Republic of China \\
$^{16}$Department of Astronomy, E\"{o}tv\"{o}s Lor\'{a}nd University, P\'{a}zm\'{a}ny P\'{e}ter s\'{e}t\'{a}ny 1/A, H-1117, Budapest, Hungary \\
$^{17}$S. N. Bose National Centre for Basic Sciences, Block-JD, Sector-III, Salt Lake City, Kolkata 700106, India \\
$^{18}$Indian Institute of Astrophysics, II Block, Koramangala, Bengaluru 560034, India \\
$^{19}$Departamento de Astronom\'ia, Universidad de Concepci\'on, Casilla 160-C, Concepci\'on, Chile \\
$^{20}$Jodrell Bank Centre for Astrophysics, Department of Physics and Astronomy, School of Natural Sciences, The University of Manchester, Oxford Road, Manchester M13 9PL, United Kingdom \\
$^{21}$I. Physikalisches Institut, University of Cologne, Z\"ulpicher Str. 77, 50937 K\"oln, Germany\\
$^{22}$Institute of Astrophysics, School of Physics and Electronic Science, Chuxiong Normal University, Chuxiong 675000, People’s Republic of China \\
$^{23}$Institute of Astronomy and Astrophysics, Anqing Normal University, Anqing 246133, People’s Republic of China \\
$^{24}$Physical Research Laboratory, Navrangpura, Ahmedabad-380 009, India \\
$^{25}$University of Science and Technology, Korea (UST), 217 Gajeongro, Yuseong-gu, Daejeon 34113, Republic of Korea \\
$^{26}$Indian Institute of Science Education and Research (IISER) Tirupati, Rami Reddy Nagar, Karakambadi Road, Mangalam (P.O.), Tirupati 517 507, India

\twocolumn
\appendix
\section{Observations and data reduction}
\subsection{ALMA Band-3}\label{app:alma-b3}
The 3\,mm observations of SDC335 (ALMA source name: I16272-4837) is included in the ALMA Band-3 survey entitled ``ALMA Three-millimeter Observations of Massive Star-forming regions'' (ATOMS; Project ID: 2019.1.00685.S; PI: Tie Liu). 
The Atacama Compact 7-m Array (ACA; Morita Array) observation of SDC335 was conducted on the 12th of November 2019 and the 12-m array (in C43-3 configurations) observation was conducted on the 3rd of November 2019. 
The on-source time is $\sim$\,5 minutes for ACA and $\sim$\,3 minutes for 12-m array, respectively. 
The angular resolution (AR) and maximum recoverable scale (MRS) for ACA are $\sim$\,12\parcsec6 and $\sim$\,87\parcsec4, respectively. 
The AR and MRS for the 12-m array are $\sim$\,1\parcsec7 and $\sim$\,20\parcsec1, respectively. The UV data calibration and imaging were processed using the Common Astronomy Software Applications \citep[\casa][]{2007ASPC..376..127M}. The ACA 7-m data and ALMA 12-m array data were calibrated separately. We then imaged and cleaned jointly the ACA and 12-m array data using the natural weighting (to optimize the signal-to-noise ratio) and taking \textit{pblimit} = 0.2, in the \casa~TCLEAN task, for both continuum images and line cubes. 
Continuum images were created from line-free frequency ranges of SPWs\,7–-8 centred at $\sim$99.4\,GHz while the spectral line cube of each SPW was produced with its native spectral resolution. The 12m+ACA combined continuum image of this source has a beam size of 1\parcsec94\,$\times$\,2\parcsec17 (position angle = 87\degree), and a sensitivity of 0.2\,mJy, which corresponds to a mass sensitivity of $\sim1.6$\,M$_{\odot}$ at the source distance and a dust temperature of $\sim23$\,K.

Spectral windows (SPWs) 1--6 in the lower sideband have a bandwidth of 57.92 MHz, which is slightly narrower than that of 12m-only observations. It's due to the narrower bandwidth of ACA observations. The six SPWs cover dense gas tracers such as the $J=1-0$ transition of \hcop, \htcop, \hcn, and \htcn, shock tracer \sio\,$J=2-1$, and photo-dissociation region (PDR) tracer \cch\,$J=1-0$. The other two wide SPWs 7–-8 in the upper sideband cover a frequency range of 97.530--101.341\,GHz, each with a bandwidth of 1867.70\,MHz, and are used for continuum emission and line surveys. 
We summarize the necessary information of the ALMA Band-3 observations in Table\,\ref{tab:B3+7}.
Column 1 lists the spectral window (SPW). Columns 2--4 list the total bandwidth, spectral resolution, and beam sizes of each SPW. The fifth column lists the averaged RMS noise of the SPW. The species names, transitions, rest frequencies, and upper energies are listed in Columns 6--9. 
The critical density ($n_\mathrm{crit}$), as the density where the rate on spontaneous transitions from the upper level to the lower level is equal to the collisional depopulation rate in a multilevel system, of these transitions at 100\,K are shown in Column 10. In Column 11, we also listed the effective excitation density ($n_\mathrm{eff}$) at 100\,K, the density which results in a molecular line with an integrated intensity of 1\,K\,\kms~\citep{2015PASP..127..299S}.
The notes of species and their transitions are retrieved from Table 2 in \citet{2020MNRAS.496.2790L}.

\subsection{ALMA Band-7}\label{app:alma-b7}
The 0.87\,mm observation of SDC335 (ALMA source name: I16272) is included in the ALMA Band-7 survey named ``How to form high-mass stars in proto-clusters?'' (Project ID: 2017.1.00545.S; PI: Tie Liu). 
The observations were carried out on the 20th of May 2018 in ALMA Cycle 5, using 43 12-m antennas in C43-1 configuration. For all observations, baselines range from 15 to 313.7 meters with 1128 baselines in total.
The on-source time is $\sim$\,3.7\,minutes.
J1650-5044 and J1924-2914 were used as atmosphere calibrators, J1924-2914 was used as a bandpass calibrator. J1924-2914 and J1650-5044 were used as flux and phase calibrators, respectively. 
The pipeline provided by the ALMA team was used to do data calibration in CASA on version 5.1.15. The imaging was conducted by TCLEAN task in CASA 5.3. We ran 3 rounds of phase self-calibration and one round of amplitude self-calibration to improve the images of I16272. In the imaging processes, the deconvolution was set as ``hogbom'', and the weighting parameter is set as ``briggs'' with a robust of 0.5 to balance the sensitivity and angular resolution. In the last round of self-calibration, the primary beam calibration was also conducted with pblimit=0.2.

The four spectral windows SPW\,31, SPW\,29, SPW\,25, and SPW\,27, are respectively centered at 343.2, 345.1, 354.4, and 356.7\,GHz. 
The total of 104 line-free channels are extracted for the aggregated continuum image to achieve a sensitivity of $\sim$\,1.5\,\mjybeam.
The main strong lines include \hcn\,$J=4-3$, \hcop\,$J=4-3$, CO\,$J=3-2$, and CS\,$J=7-6$. 
\hcn\,$J=4-3$ and \hcop\,$J=4-3$ are used as reliable infall tracers \citep{2014MNRAS.444..874C} and CO\,$J=3-2$ can be used for outflows \citep{2020ApJ...890...44B}. Shock tracers include SO\,$^3\Sigma$~$J=8_8-7_7$ and \chtoh\,$J=13_{1,12}-13_{0,13}$. 
High density tracers like \htcn\,$J=4-3$ and CS\,$J=7-6$ can determine core velocity and trace high-density outflows as well. Some sulfur-bearing molecules like \htcs~and SO$_2$ can serve as tracers of rotational envelopes. 
Chemically, SPWs 29 and 31 both cover wide bands for hot-/warm-core molecular lines, including CH$_3$OH, CH$_3$OCH$_3$, and \htcs, have a sufficient number of transitions which can be used for rotation-temperature estimation. Besides, the nitrogen-/oxygen-bearing molecules detected in hot cores are particularly useful to study chemical reaction and evolution in high-mass star formation \citep[][; Liu et al. in preparation]{2022MNRAS.511.3463Q}. The necessary information of the ALMA Band-7 observations are also summarized in Table\,\ref{tab:B3+7} as mentioned in Band-3 observations.

\onecolumn
\begin{landscape}
\begin{table}
\caption{The spectral window (SPW) and the main targeted lines in ALMA Band-3/7 observations \label{tab:B3+7}}
\setlength{\tabcolsep}{4pt}
\begin{tabular}{cccccccccccc}
\hline
\hline
SPW & Bandwidth  & $\Delta v$ & Beam Size & RMS & Species  & Transition  & Rest frequency & $E_u/k$ & $n_\mathrm{crit}$ (100\,K) \tablenotemarknew{a} & $n_\mathrm{eff}$ (100\,K) \tablenotemarknew{a} & Note \\
    & (MHz) & (\kms) & & (\mjybeam) & & & (GHz) & (K) & (cm$^{-3}$) & (cm$^{-3}$) & \\
\hline
\multicolumn{12}{c}{ALMA Band-3 (12m+ACA Combined)} \\
\hline
SPW1 & 57.92 & 0.212 & 2.52\arcsec\,$\times$\,2.28\arcsec & 8.8 & \htcn & $J=1-0$ & 86.340167 & 4.14 & $9.7\times10^4$ & $6.5\times10^4$ & High-density tracer \\
SPW2 & 57.92 & 0.212 & 2.51\arcsec\,$\times$\,2.27\arcsec & 7.2 & \htcn & $J=1-0$ & 86.754288 & 4.16 & $2.0\times10^4$ & $1.1\times10^4$ & High-density and ionization tracer \\
SPW3 & 57.92 & 0.212 & 2.48\arcsec\,$\times$\,2.25\arcsec & 9.2 & \cch & $J=1-0$ & 87.316898 & 4.19 & $1.9\times10^5$ &  & Tracer of photodissociation regions \\
SPW4 & 57.92 & 0.212 & 2.50\arcsec\,$\times$\,2.26\arcsec & 7.2 & \sio & $J=1-0$ & 86.846960 & 6.25 & $1.7\times10^5$ &  & Shock/outflow tracer \\
SPW5 & 57.96 & 0.106 & 2.45\arcsec\,$\times$\,2.22\arcsec & 11.0 & \hcn & $J=1-0$ & 88.631847 & 4.25 & $1.1\times10^5$ & $1.7\times10^3$ & High-density/infall/outflow tracer \\
SPW6 & 57.96 & 0.106 & 2.45\arcsec\,$\times$\,2.19\arcsec & 10.6 & \hcop & $J=1-0$ & 89.188526 & 4.28 & $2.3\times10^4$ & $2.6\times10^2$ & High-density/infall/outflow tracer \\
SPW7 & 1869.70 & 1.695 & 2.20\arcsec\,$\times$\,2.00\arcsec & 3.8 & \chtoh\tablenotemarknew{b} & $J=2_{1,1}-1_{1,0}$ & 97.582798 & 21.56 & $4.8\times10^4$ & $2.6\times10^2$ & High-density/hot cores/shock tracer \\
 & & & & & \cs & $J=2-1$ & 97.980953 & 7.05 & $5.5\times10^4$ & $4.7\times10^3$ & High-density/infall/outflow tracer \\
 & & & & & \so & $J=3_2-2_1$ & 99.299870 & 9.23 & $3.0\times10^5$ & $4.7\times10^3$ & High-density/shock/outflow tracer \\
 & & & & & H$_{\alpha}$ & H$_{40\alpha}$ & 99.022952 & 9.23 & $3.0\times10^5$ & $4.7\times10^3$ & Ionized gas/\hii~region tracer \\
SPW8 & 1869.70 & 1.695 & 2.17\arcsec\,$\times$\,1.95\arcsec & 4.0 & \htcn & $J=11-10$ & 100.07639 & 28.82 & $9.2\times10^4$ & $1.1\times10^4$ & High-density/hot-cores tracer \\
\hline
\multicolumn{12}{c}{ALMA Band-7 (12m Only)} \\
\hline
SPW 25 &  468.75 & 0.239 & 0.81\arcsec\,$\times$\,0.67\arcsec & 8.9 & HCN & $J=4-3$ & 354.5054779 & 42.53 & $9.1\times10^6$ & $3.7\times10^4$ & High-density/infall/outflow tracer \\
SPW 27 &  468.75 & 0.237 & 0.78\arcsec\,$\times$\,0.65\arcsec & 10.4 & HCO$^+$ & $J=4-3$ & 356.7342420 & 42.80 & $2.0\times10^6$ & $4.0\times10^3$ & High-density/infall/outflow tracer \\
SPW 29 & 1875.00 & 0.980 & 0.83\arcsec\,$\times$\,0.68\arcsec & 4.4 & CO & $J=3-2$ & 345.7959900 & 33.19 &                 &                 & Outflow tracer                     \\
       &         &       & & & SO~$^3\Sigma$ & $J=8_8-7_7$ & 344.3106120 & 49.32 &             &                 & High-density/shock/outflow tracer  \\
       &         &       & & & H$^{13}$CN    & $J=4-3$ & 345.3397599 & 41.43 & $7.7\times10^6$ & $1.1\times10^6$ & High-density tracer                \\
SPW 31 & 1875.00 & 0.986 & 0.83\arcsec\,$\times$\,0.68\arcsec & 4.7 & CS & $J=7-6$ & 342.8828503 & 34.32 & $2.6\times10^6$ & $1.3\times10^5$ & High-density/infall/outflow tracer \\
       &         &       & & & CH$_3$OH\tablenotemarknew{b}      & $J=13_{1,12}-13_{0,13}$ & 342.7297810 & 227.47 & &               & High-density/hot cores/shock tracer \\
       &         &       & & & \htcs\tablenotemarknew{b} & $J=10_{0,10}-9_{0,9}$ & 342.9464239 & 90.59 &  &   & High-density/warm cores/disk tracer \\
\hline
\multicolumn{12}{l}{$^{a.}$ The critical density $n_\mathrm{crit}$ and the effective excitation density $n_\mathrm{eff}$ at 100\,K \citep{2015PASP..127..299S}.} \\
\multicolumn{12}{l}{$^{b.}$ Hot/warm core molecules have enough transitions for rotational temperature estimation.} \\
\end{tabular}
\end{table}
\end{landscape}

\twocolumn

\section{Source Extraction Algorithm}
\subsection{CASA imfit}\label{app:casaimfit}
ALMA 3~mm continuum emission of SDC335 shows two prominent dense cores MM1 and MM2. In semi-automatic algorithm \casa~\textit{imfit}, we draw circles to cover most emission of dense cores and then run the 2D Gaussian fitting. By doing so, we make fundamental measurements of both cores and the results are summarized in Table\,\ref{tab:fit_3mm_measure}.

\begin{table*}
\centering
\caption{Fundamental 2D fitting parameters of dense cores from ALMA 3\,mm continuum emission \label{tab:fit_3mm_measure}}
\setlength{\tabcolsep}{5pt} 
\begin{tabular}{cccccccc} 
\hline
\hline
Dense Core &     RA    &     Dec   &  Peak Intensity   & Flux Density & $\theta_\mathrm{maj}\times\theta_\mathrm{min}$\tablenotemarknew{a.} & $\theta_\mathrm{deconv}$\tablenotemarknew{b.} &  Position Angle \\
	       &    (J2000)  &    (J2000)  &    (\mjybeam)     &   (mJy)      &  (\arcsec$\times$\arcsec)   & (\arcsec) &  (\degree) \\
\hline
MM1        & 16:30:58.76 & -48:43:53.5 &      59.8(1.5)    &  146.4(5.4)  &    3.63(10) $\times$ 2.84(7) & 2.47 &   146.4(4.2) \\
MM2        & 16:30:57.29 & -48:43:39.8 &       8.2(6)      &   25.4(2.3)  &    4.13(31) $\times$ 3.15(21) & 2.98 &  89.1(9.7)  \\
\hline
\multicolumn{7}{l}{$^{a.}$ Apparent major and minor axis FWHM from 2D Gaussian fitting, convolved with beam.}\\
\multicolumn{7}{l}{$^{b.}$ FWHM deconvolved with beam.}\\
\end{tabular}
\end{table*}

\subsection{GETSF}\label{app:getsf}
\getsf~is a multi-scale multi-wavelength extraction of sources and filaments algorithm, using separation of the structural components \citep{2021A&A...649A..89M}. It is able to separate three types of structures: sources, filaments, and backgrounds. Refer to \citet{2021A&A...649A..89M} for more details.

In the case of SDC335-MM1, we first circularize the beam into the original major beam size $\theta_\mathrm{maj}$=0\parcsec82. 
We adopt the smallest scale to be the size of the circularized beam, and the largest scale to be the MRS (8\parcsec45). \getsf~extracts six sources and one filament with high reliability. 
We name the six sources as condensations S1--S6 and the filament as mini F1. In the Section\,\ref{subsec:fragments}, we care more about the spherical morphology. 
Therefore, we only list the fundamental measurements in Table\,\ref{tab:fit_0.87mm_measure}.

\begin{table*}
\caption{Sources extracted by \getsf~from ALMA 0.87\,mm continuum emission}
\label{tab:fit_0.87mm_measure}
\centering
\setlength{\tabcolsep}{5pt}
\begin{tabular}{ccccccccc} 
\hline
\hline
Source ID & RA  & DEC & Goodness\tablenotemarknew{a} & Peak intensity & Total flux & $\theta_\mathrm{maj}\times\theta_\mathrm{min}$\tablenotemarknew{b} & $\theta_\mathrm{deconv}$\tablenotemarknew{c} & Position Angle\tablenotemarknew{d} \\
 & (J2000) & (J2000) & & \jybeam & (Jy) & (\arcsec$\times$\arcsec) & (\arcsec) & (\degree) \\
\hline
S1 & 16:30:58.765 & -48:43:53.95 & 1.610E+04 & 1.360(2) & 3.384(10) & 1.081\,$\times$\,0.087 & 0.989 & 67.9 \\
S2 & 16:30:58.639 & -48:43:51.25 & 8.188E+02 & 0.271(3) & 0.494(3)  & 1.159\,$\times$\,0.951 & 1.070 & 73.5 \\
S3 & 16:30:58.706 & -48:43:52.52 & 6.934E+02 & 0.280(3) & 0.555(3)  & 1.333\,$\times$\,1.162 & 1.268 & 118.2 \\
S4 & 16:30:58.923 & -48:43:55.29 & 1.111E+02 & 0.122(4) & 0.484(4)  & 1.798\,$\times$\,1.638 & 1.749 & 165.0 \\
S5 & 16:30:58.434 & -48:43:50.81 & 8.739E+00 & 0.029(4) & 0.065(3)  & 1.319\,$\times$\,1.104 & 1.230 & 168.8 \\
S6 & 16:30:58.999 & -48:43:51.47 & 3.468E+01 & 0.048(2) & 0.025(3)  & 1.832\,$\times$\,1.554 & 1.720 & 50.7 \\
\hline
\multicolumn{7}{l}{$^{a.}$ The reliability of a source \citep[see Equation 42 in][]{2021A&A...649A..89M}.}\\
\multicolumn{7}{l}{$^{b.}$ FWHM of Gaussian fitting, convolved with beam.}\\
\multicolumn{7}{l}{$^{c.}$ FWHM, deconvolved with beam.}\\
\multicolumn{7}{l}{$^{d.}$ Position angle fitted from the intensity moments, excluding noise and background fluctuations.}\\
\end{tabular}
\end{table*}

\section{Temperature estimation}
\label{app:estimate_temperature}

\subsection{The temperature of protostellar cores MM1 and MM2}
\label{app:protoT}
We cross-match the Hi-GAL 70\,$\mu$m Compact Source Catalogue \citep{2016A&A...591A.149M} and then convert the flux densities to bolometric (internal) luminosity using the following relation \citep{2017MNRAS.471..100E},
\begin{equation}
	L_{\mathrm{int}} = 25.6\left(\frac{S_{70\mu\mathrm{m}}}{10\, \mathrm{Jy}}\right) \left(\frac{D}{1\, \mathrm{kpc}}\right)^2 \, L_{\odot},
\end{equation}
where $S_{70\mu\mathrm{m}}$ is the integrated 70\,$\mu$m flux density of the source and $d$ is the distance to the clump. Assuming that the dust emission from a protostellar core is optically thin and is predominantly in the far-infrared, we calculate the mean mass-weighted temperature $\overline{T_\mathrm{d}}$ within core equivalent radius $r = R_{\mathrm{source}}$ \citep{1993ApJ...414..759T},
\begin{equation}
	\overline{T_\mathrm{d}} = \frac{3}{2} T_0 \left(\frac{L_\mathrm{int}}{L_0}\right)^{1/6} \left(\frac{r}{r_0}\right)^{-1/3},
\end{equation}
where $L_\mathrm{int}$ is the source's internal luminosity, $r$ is the core's effective radius, and the reference values are $T_0 = 25$\,K, $L_0 = 520$~$L_{\odot}$, and $r_0=0.032$\,pc. 
The scaling relation above assumes dust opacity index $\beta=2$ and density profile follows $\rho(r)\propto r^{-2}$. 
Assuming dust and gas are well mixed and in thermal equilibrium, we then assign the core temperature $T_{\mathrm{core}} = \overline{T_\mathrm{d}}$. 
The temperatures of MM1 and MM2 are listed in the third column of Table\,\ref{tab:physical_parameter_3mm}.

\subsection{The temperature estimation of S1--S4 by XCLASS}
\label{app:S1-4}

\begin{table*}
\centering
\caption{The parameters of \htcs~multiple transition at ALMA Band-7}
\label{tab:h2cs}
\begin{tabular}{cccccc} 
\hline
Resolved QNs\tablenotemarknew{a.} & Unresolved QNs\tablenotemarknew{a.}& Frequency & $E_u/k$ & $S_\mathrm{ij}\mu^2$ & Notes \\
 & & (GHz) & (K) & (Debye$^2$) & \\
\hline
$J=10_{0,10}-9_{0,9}$ & & 342.94646 & 90.59 & 27.19 & \\
$J=10_{2,9}-9_{2,8}$ & & 343.32189 & 143.30 & 26.10 & Blended with H$_2^{13}$CO \\
$J=10_{2,8}-9_{2,7}$ & & 343.81294 & 143.37 & 26.11 & \\
$J=10_{3,8}-9_{3,7}$ & & 343.40946 & 209.06 & 74.24 & Blended with $10_{3,7}-9_{3,6}$ \\
$J=10_{3,7}-9_{3,6}$ & & 343.41365 & 209.06 & 74.24 & Blended with $10_{3,8}-9_{3,7}$ \\
$J=10_{4,7}-9_{4,6}$ & $J=10_{4,7}-9_{4,6}$ & 343.30891 & 300.95 & 22.84 & Unresolved with $J=10_{4,6}-9_{4,5}$ \\
 & $J=10_{4,6}-9_{4,5}$ & 343.30892 & 300.95 & 22.84 & Unresolved with $J=10_{4,7}-9_{4,6}$ \\
$J=10_{5,6}-9_{5,5}$ & $J=10_{5,6}-9_{5,5}$ & 343.20188 & 418.85 & 61.19 & Unresolved with $10_{5,5}-9_{5,4}$ \\
 & $J=10_{5,5}-9_{5,4}$ & 343.20188 & 418.85 & 61.19 & Unresolved with $10_{5,6}-9_{5,5}$ \\
\hline
\multicolumn{6}{l}{$^{a.}$ Two energy transitions are (un)resolved with laboratory precision. If unresolved, two transitions share the same ``Resolved QN''.}\\
\end{tabular}
\end{table*}

\begin{figure}
\centering
\includegraphics[scale=0.36]{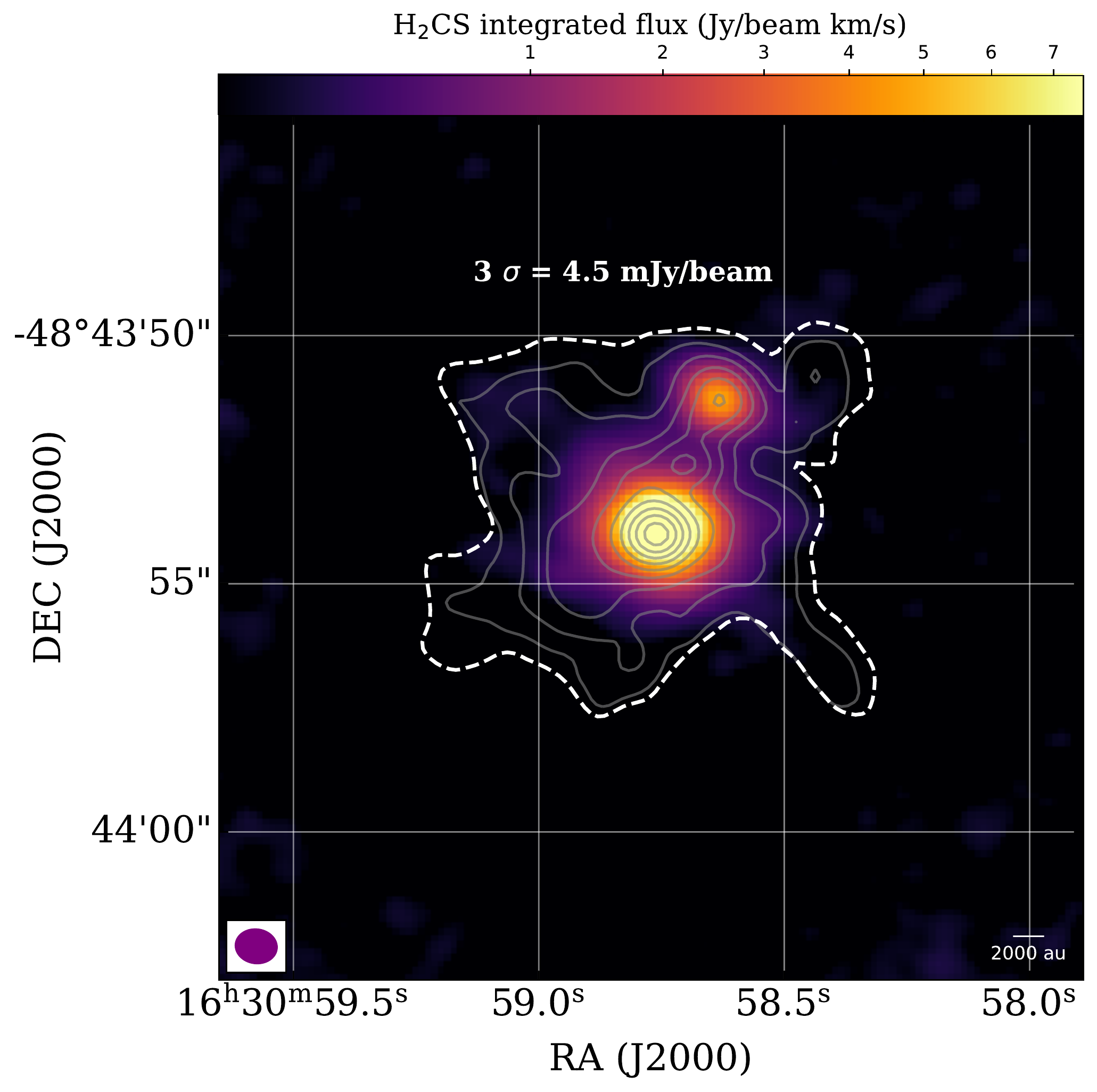}
\caption{The integrated flux of \htcs~main line ($J=10_{0,10}-9_{0,9}$) at rest frequency of $\nu=342.94646$\,GHz, overlaid with contours of ALMA Band-7 continuum emission. The white power-law contour levels are [7.5, 28.1, 71.5, 142.7, 245.9, 384.6, 562.2, 781.6, 1045.5, 1356.5]\,\mjybeam. We highlight the $3\sigma = 4.5$\,\mjybeam~level with white dashed line. \label{fig:h2cs_m0}}
\end{figure}

We first check the correlation between ALMA Band-7 continuum emission and \htcs~emission. In Figure\,\ref{fig:h2cs_m0}, the background color map shows the integrated flux of the $J=10_{0,10}-9_{0,9}$ transition of \htcs~line (the main line) at rest frequency of $\nu=342.94646$\,GHz, overlaid with contours of ALMA Band-7 continuum emission. The integrated flux of the \htcs~main line is well correlated in the dust emission region, ensuring the safety of dust temperature estimation from \htcs~rotational temperature.

One frequently adopted technique to derive temperature from molecular line emission is eXtended \casa~Line Analysis Software Suit \citep[\textsc{XCLASS}\footnote{https://xclass.astro.uni-koeln.de/};][]{2017A&A...598A...7M}. Assuming that the molecular gas in local thermodynamical equilibrium (LTE) condition, \textsc{XCLASS} solves a radiative transfer equation and produces synthetic spectra for specific molecular transitions by taking the source size, beam filling factor, line profile, line blending, excitation, and opacity into account. In the \textsc{XCLASS} modelling, the input parameters are the source size, beam size, line velocity width, velocity offset, rotation temperature, and column density \citep{2017A&A...598A...7M}. The source size is the angular radius deconvolved from the beam. We then set rotation temperatures, column densities, systematic velocity, and velocity widths as free parameters to simulate the observed spectra. To obtain optimized rotation temperature and column density parameters, we employ Modeling and Analysis Generic Interface for eXternal numerical codes \citep[\textsc{MAGIX}][]{2013A&A...549A..21M} for further calculation. The uncertainties of temperature and column density are calculated from Markov Chain Monte Carlo (MCMC) method.

\subsection{RADEX mock grid of S5 and S6}
\label{app:S5-6}

\begin{figure*}
\includegraphics[width=1.0\linewidth]{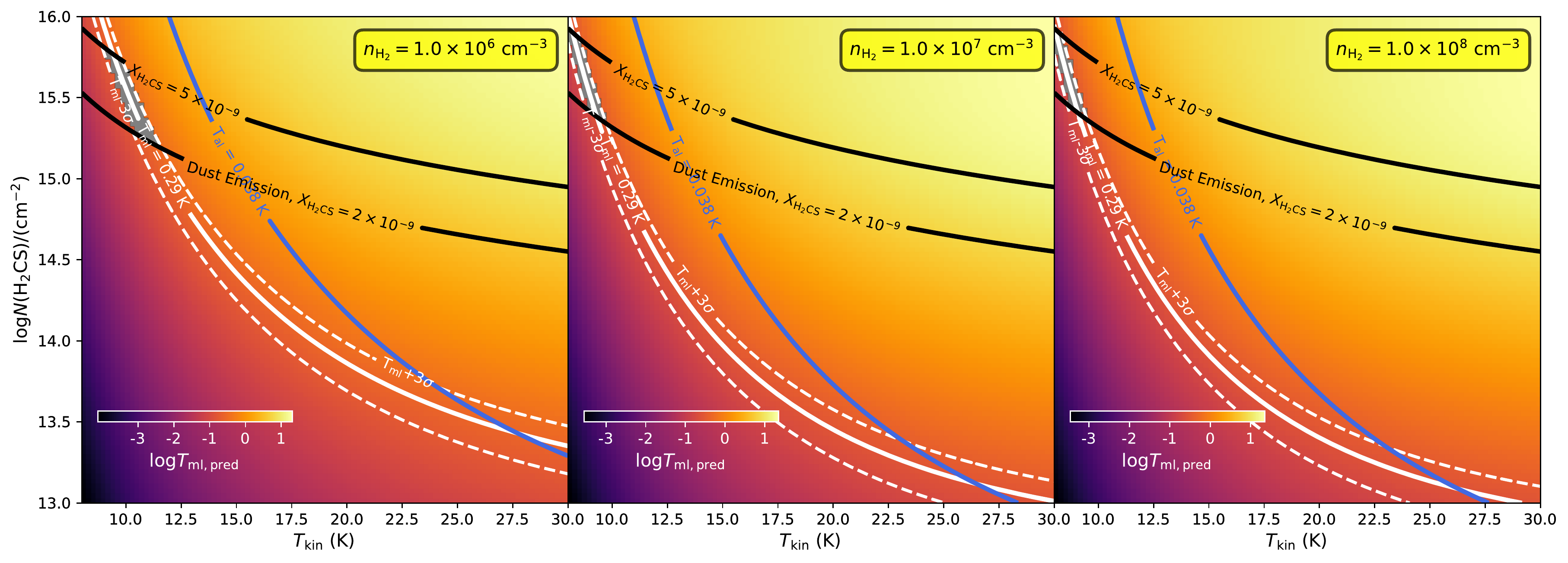}
\caption{The mock grid of \htcs~observations of S5. The three panels differ in input volume density of collisional partner, i.e., molecular hydrogen. The background color map shows the predicted brightness temperature of the main line, $T_\mathrm{ml}$. The white solid line indicates the same value as observed while the dashed lines indicate the $\pm3\sigma$ uncertainty range. The blue line indicates where the predicted brightness temperature of the associate line $T_\mathrm{al}$ is the same as RMS noise (0.038\,K). The double black lines indicates the prediction of dust continuum emission under the assumption of \htcs~abundance $2-5\times10^{-9}$. The grey shaded area shows the parameter range of $T_\mathrm{kin}$ and $N_\mathrm{H2CS}$. The predicted kinematic temperature is $9.5(\pm0.9$)\,K; the predicted column density is $3.8(\pm1.5)\times10^{15}$\,cm$^{-2}$. \label{fig:S5_MockGrid}}
\end{figure*}

Both two condensations S5 and S6 only have the $10_{0,10}-9_{0,9}$ transition of \htcs~(the main line) but no detection of higher transition from the $10_{2,9}-9_{2,8}$ (the associate line) above the rms level. Therefore, a upper limit of temperature should be given. Therefore, we perform a multi-parameter NLTE mock observations of the main and associate \htcs~line. The strategy here is to predict the brightness temperature of the main and associate line under the assumption of H$_2$ volume density, kinetic temperature, \htcs~column density, and the line width. For S5 and S6, the input linewidths of both main and associate lines are assumed to be the same as that of the observed main line. To constrain other parameters, we span the parameter space and mock a parameter grid with RADEX, a 1D NLTE radiation transfer code \citep{2007A&A...468..627V}. As shown in Figure\,\ref{fig:S5_MockGrid}, the background color map predicts brightness temperature of the main line while the white lines shows the observed value. The blue line gives the critical condition where the associate line is just hidden below the RMS noise. Therefore, only the left region of the blue line should satisfy the observation. We also extract the dust continuum flux and calculate the total gas column density based on the dust gray body emission as well as the assumption of $T_\mathrm{dust}=T_\mathrm{kin}$. We use the molecular abundance $2-5\times10^{-9}$ from a typical high mass star forming region Orion-KL (Minh et al. \hyperlink{cite}{1991}) and give another independent constraint on the $T_\mathrm{kin}-N_\mathrm{H2CS}$ plane by black lines. In each panel, the gray shaded region indicates the available values of parameters. The difference among three panels are the volume density of collisional partner, i.e., molecular hydrogen. We assume that S5 and S6 have the similar volume density with S1--4 of $10^{6-7}$\,cm$^{-3}$, so we take the grey shaded area in the left and middle panel as available parameter space. We note that the method is not sensitive to the volume density, so it's safe to assume the volume density in a wide range. One can also find the intersection between the blue and white lines should be the upper limit of temperature.

\addtocounter{figure}{-1}

\begin{figure*}
\includegraphics[width=1.0\linewidth]{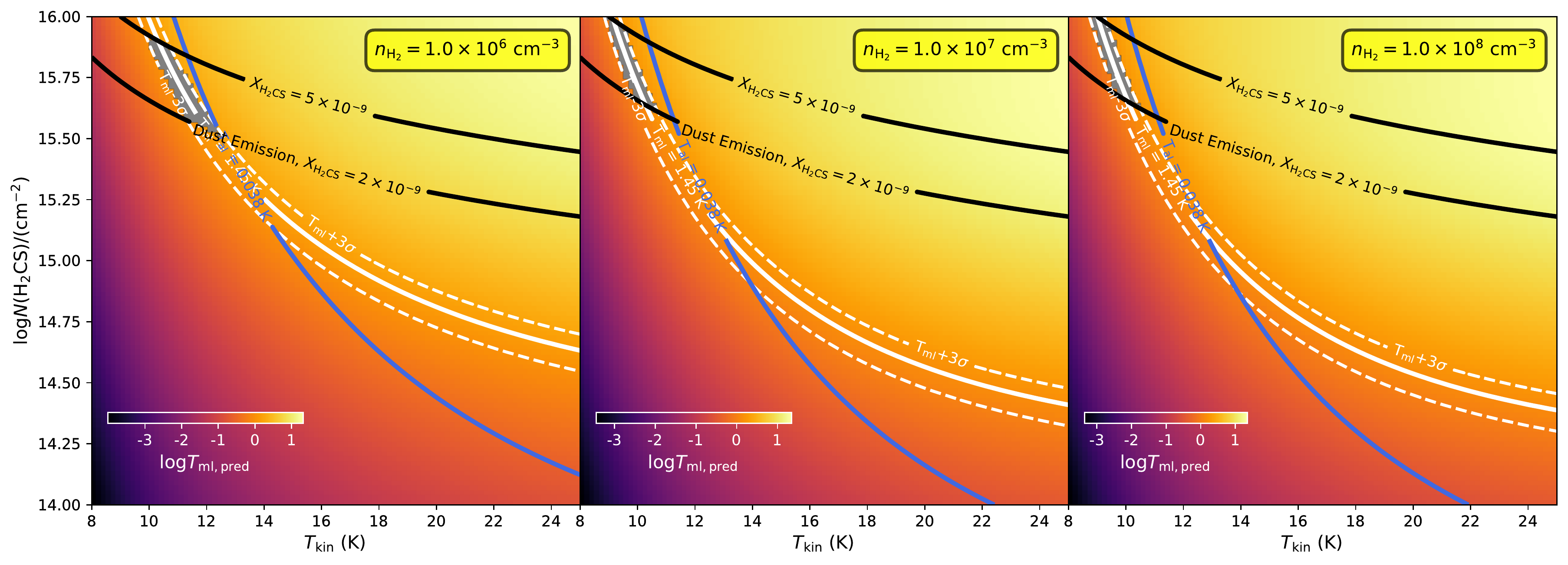}
\caption{The same but for S6. The predicted kinematic temperature is $11.2(\pm1.1$)\,K; the predicted column density is $4.6(\pm1.8)\times10^{15}$\,cm$^{-2}$. \label{fig:S6_MockGrid}}
\end{figure*}

\section{Separation from MST}
\label{app:MST}
\begin{figure}
\centering
\includegraphics[scale=0.36]{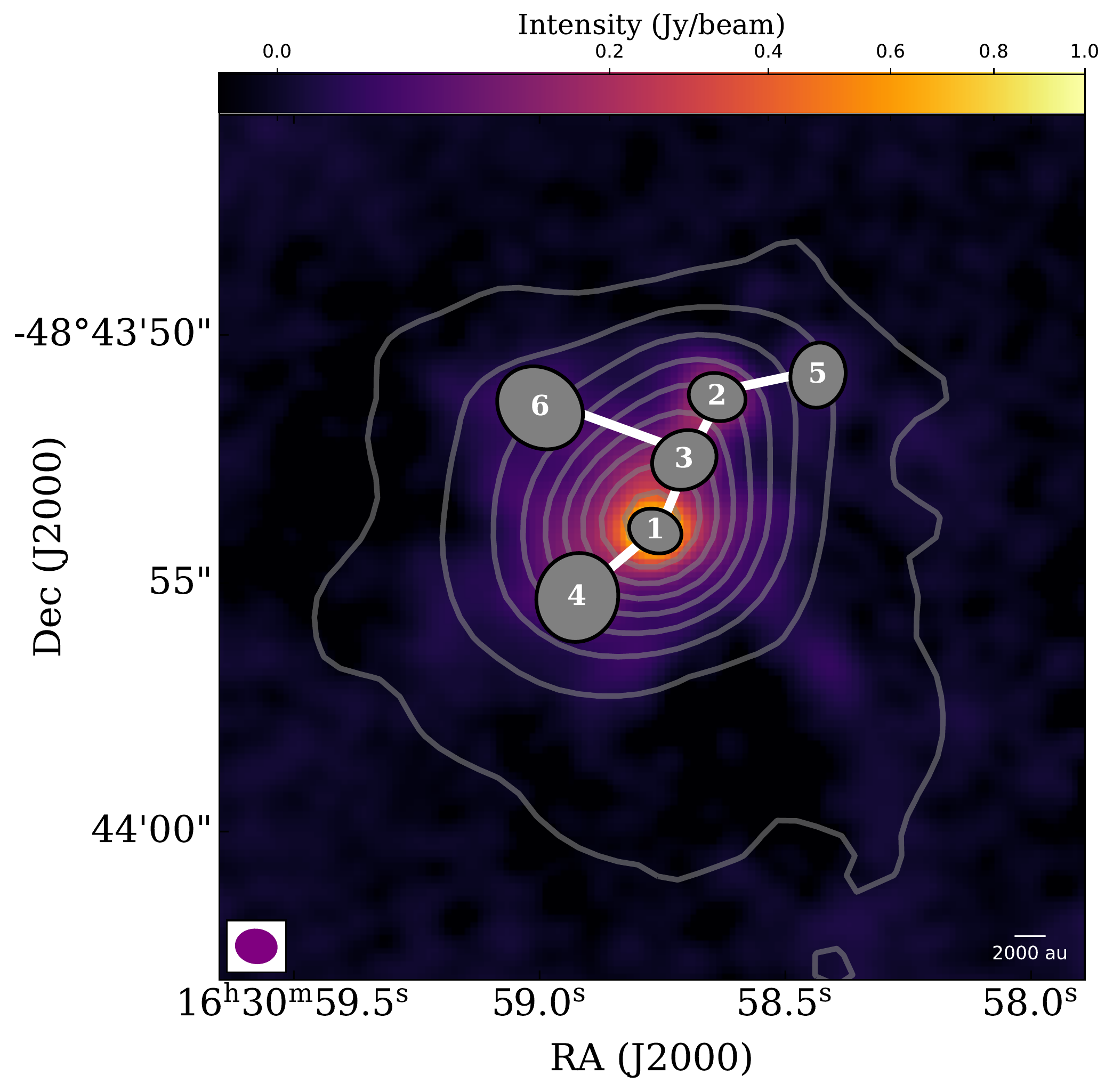}
\caption{The background color map shows the ALMA 0.87\,mm continuum emission SDC335-MM1 while the white contours of 3\,mm continuum emission follows the power-law levels of [1.0, 3.3, 7.6, 14.2, 23.4, 35.1, 49.6, 67.0]\,\mjybeam. Six condensations S1--S6 are marked with their fitted FWHM ellipses. The beam size of ALMA Band-7 observation is shown with the purple ellipse on the left bottom and the scale bar is shown on the right bottom. The connections from MST method are marked with white solid lines. \label{fig:connections}}
\end{figure}

Minimum spanning tree (MST), first developed for astrophysical applications by \citet{1985MNRAS.216...17B}, has been applied to simulations \citep[e.g.][]{2017ApJ...841...88W} and to observations \citep[e.g.][]{WangK2016,2019ApJ...886..102S,2022ApJS..259...36G}. 
In this paper, we use Prim's algorithm to find out the edges to form the tree including every node with the minimum sum of weights to form the MST. 
Prim's algorithm starts with the single source node and later explores all the nodes adjacent to the source node with all the connecting edges. 
During the exploration, we choose the edges with the minimum weight and those which cannot cause a cycle. 
The edge weight is set to be the length between two vertices \citep{1957BSTJ...36.1389P}.
Therefore, MST determines a set of straight lines connecting a set of nodes (condensations) that minimizes the sum of the lengths. Figure\,\ref{fig:connections} display the MST for the fragments inside SDC335-MM1 from ALMA Band-7 observation.

\section{Spectral line fitting}
\label{app:fit_spectral_lines}
\begin{figure}
\centering
\includegraphics[scale=0.48]{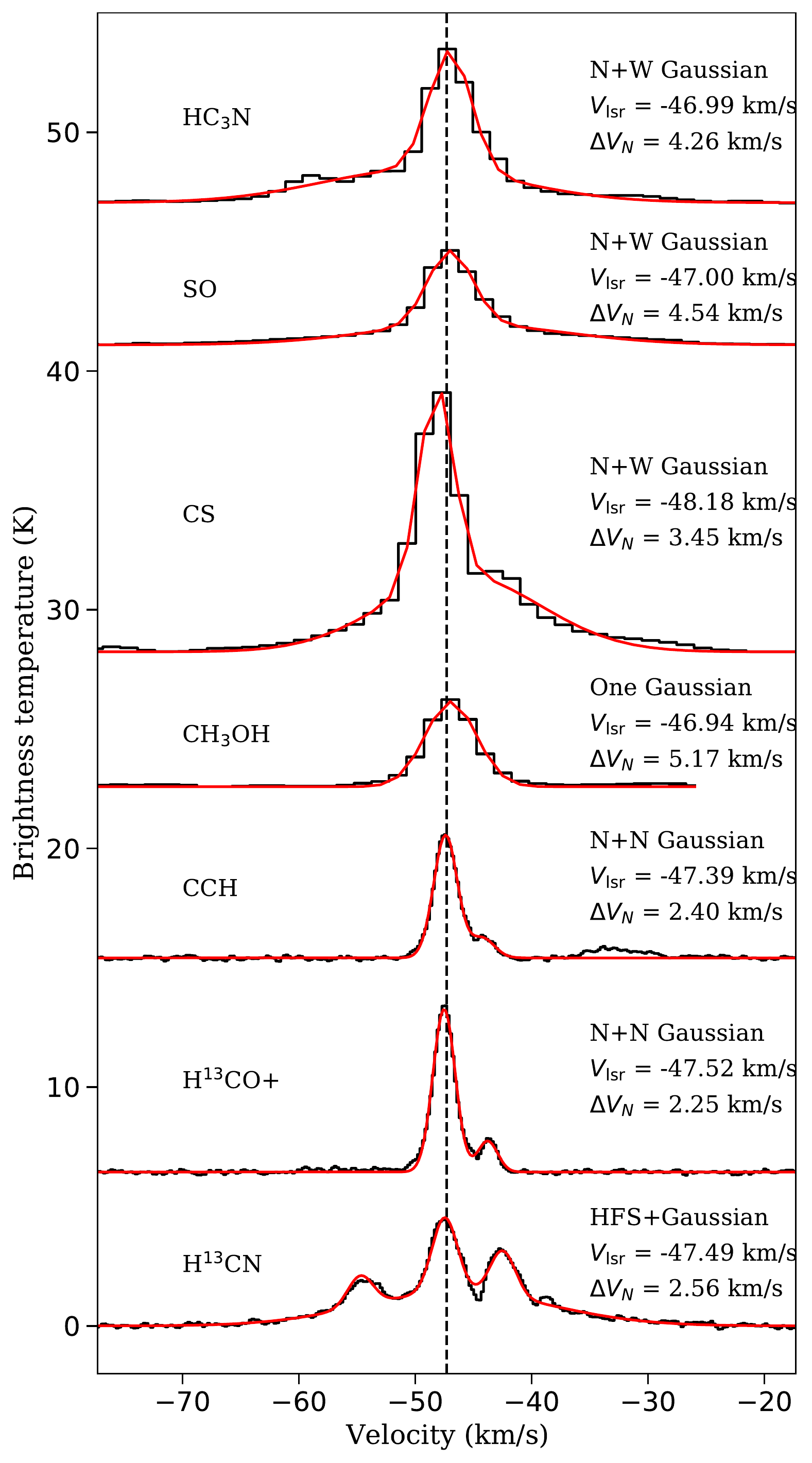}
\caption{The averaged spectra extracted from SDC335-MM1. 
The black solid lines are the original data and the red solid lines are fitting models. The models are marked on the top right. ``N+W Gaussian'' means one narrow and one wide Gaussian component where the wide one aims to fit the outflow wings. ``N+N Gaussian'' means two narrow Gaussian components. ``One Gaussian'' means a single Gaussian component. ``HFS+Gaussian'' means a narrow hyperfine structure component and a wide Gaussian component. The fitted line centroid velocity $V_\mathrm{lsr}$ and FWHM line width $\Delta V$ of the narrow componenet are shown on the right. If there are two narrow components, then the dominant one is shown. The black dashed line marks the velocity of -47.3\,\kms, the dominant componenet for all the lines. \label{fig:MM1_spectra}}
\end{figure}

\begin{figure}
\centering
\includegraphics[scale=0.48]{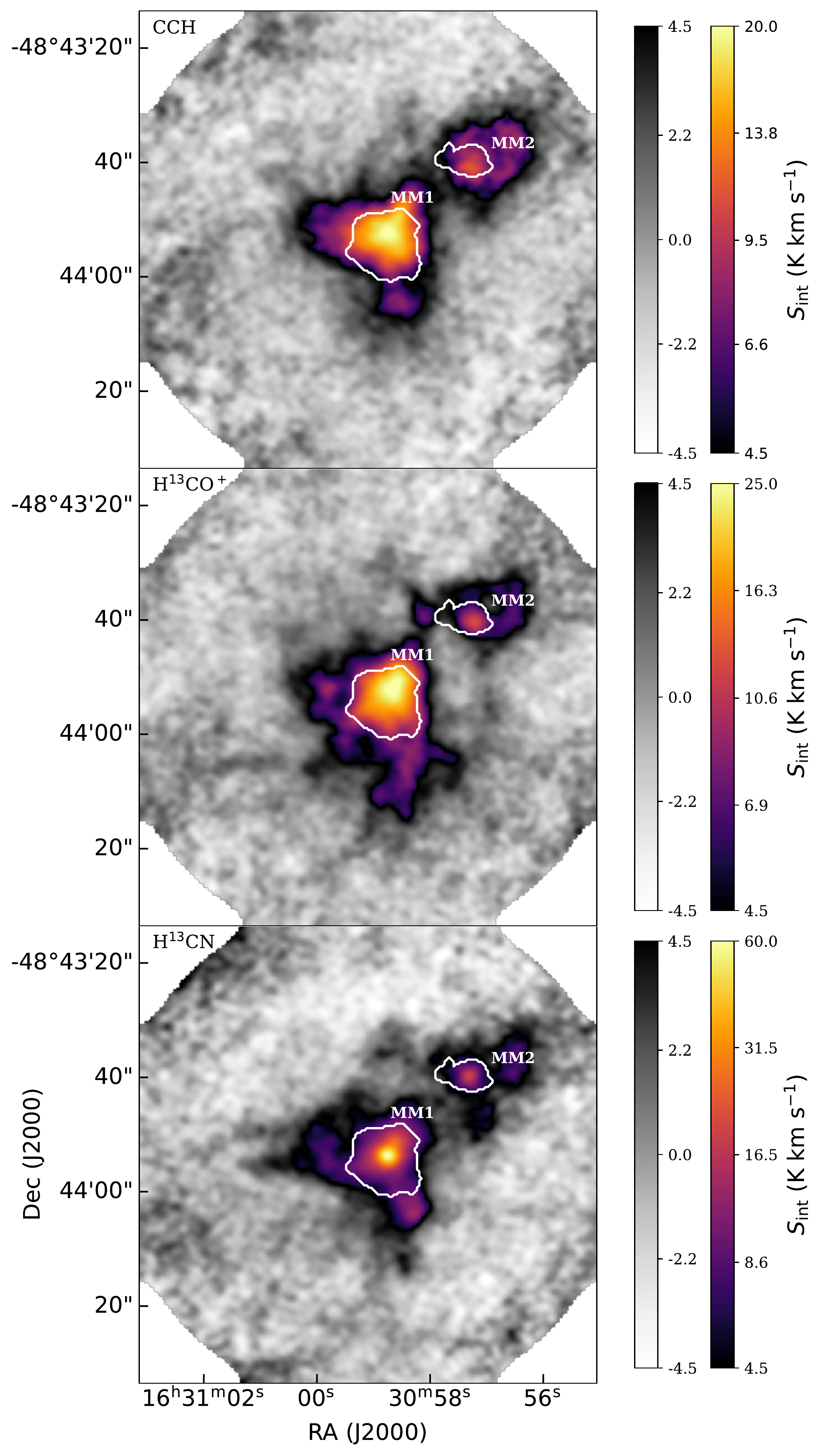}
\caption{Integrated maps of three dense gas tracers CCH (top), \htcop~(middle), and \htcn~(bottom). 
The integration ranges for three lines are specified by $V_\mathrm{lsr}-\Delta V \sim V_\mathrm{lsr}+\Delta V$ from the fitting results. 
The white solid lines mark the two dense cores MM1 and MM2 from ALMA Band-3 continuum ($F_\mathrm{cont,3mm}=1.0$\,\mjybeam). 
The plots are shown with two colorbars, the first one (grayscale) showing -15 to +15 times the noise on a linear scale, then a second one (color-scheme) showing the range +15 times the noise to the peak value of the image in an arcsinh stretch. 
The line species are labeled in the upper left corner. \label{fig:dense_gas}}
\end{figure}

Figure\,\ref{fig:MM1_spectra} shows the averaged spectra extracted from SDC335-MM1 as well as their fitting results. Figure\,\ref{fig:dense_gas} shows the integrated maps of three dense gas tracers CCH, \htcop, and \htcn, whose emission regions are in good spatial correlation with that of continuum emission.

To fit the spectral line of \htcn\,$J=1-0$, three HFS lines are included into the fitting models. 
The parameters for the three HFS lines are listed in Table\,\ref{tab:h13cn_parameters}. We note that further splitting (universally $\leq$\,0.1\,\kms) of the three lines can be ignored due to the limited velocity resolution and wide line width. So we only consider the splitting levels at quantum number $J,F$.

\begin{table}
\caption{The hyperfine structure of \htcn\,$J=1-0$}\label{tab:h13cn_parameters}
\centering
\scriptsize
\begin{tabular}{cccc} 
\hline
\hline
Resolved QNs\tablenotemarknew{a} & Frequency & Velocity\tablenotemarknew{b} & $\log I$\tablenotemarknew{c} \\
 & (GHz) & (\kms) & \\
\hline
$J=1-0$, $F=1-1$ & 86.3387352 & 5.00 & -3.02600 \\
$J=1-0$, $F=2-1$ & 86.3401666 & 0.00 & -2.54890	\\
$J=1-0$. $F=0-1$ & 86.3422543 & -7.22 & -3.50310 \\
\hline
\multicolumn{4}{l}{$^{a.}$ The split levels with the quantum numbers (QNs).}\\
\multicolumn{4}{l}{$^{b.}$ Relative to $J=1-0$, $F=2-1$.}\\
\multicolumn{4}{l}{$^{c.}$ The intensity in log scale (from CDMS).}\\
\end{tabular}
\end{table}

Towards MM1, the averaged \htcn\,$J=1-0$ line is suspected to contain two narrow velocity components, a strong and a weak one. Besides, \htcn~seems to be contaminated by strong and dense outflow from MM1. 
Therefore, we simply consider a fitting model with one component for narrow and strong HFS, and one Gaussian component for extended wings from outflow.

To fit the hyperfine structure of \htcn\,$J=1-0$, we make the following assumptions: 1) all the HFS lines share the same excitation temperature; 2) the opacity as a function of frequency (velocity) has a Gaussian profile; 3) all the lines share the same linewidth. 
Since we have three individual HFS lines (Table\,\ref{tab:h13cn_parameters}), the opacity of the $i$th component is written as,
\begin{equation}
\tau_i(v) = \tau_i \cdot \exp\left[-4\ln 2 \left(\frac{v-v_{0,i}}{\Delta v}\right)^2\right],
\end{equation}
where $\Delta v$ is the uniform FWHM of all components. The central velocity of component $i$ is $v_{0,i} = v_i + v_\mathrm{LSR}$, where $v_\mathrm{LSR}$ is the velocity of the reference component (i.e. the one with $v_i = 0$) at the Local Standard of Rest (LSR).

The opacity of the multiplet writes:
\begin{equation}
\tau(v) = \tau_\mathrm{tot} \sum\limits_{i=1}^{N} r_i \times \exp\left[-4\ln 2\left(\frac{v-v_\mathrm{LSR}-v_i}{\Delta v}\right)^2\right]
\end{equation}
where $\tau_\mathrm{tot}$ is the sum of the optical depths of all hyperfine lines and $r_i$ is the normalized relative intensity of the individual hyperfine line in the optically thin case under the condition of thermodynamical equilibrium (TE).

Given the opacity $\tau(v)$ and the amplitude $\mathcal{A}$, the brightness temperature is obtained from:
\begin{equation}
T_b(v) = \mathcal{A}(1-e^{-\tau(v)})
\end{equation}

In the optical thin regime, the brightness temperature $T_b(v)$ transforms into:
\begin{equation}
T_b(v) = \mathcal{A}\tau(v) = \mathcal{A}\tau_\mathrm{tot} \sum\limits_{i=1}^{N} r_i \times \exp\left[-4\ln 2\left(\frac{v-v_\mathrm{LSR}-v_i}{\Delta v}\right)^2\right]
\end{equation}

\begin{figure}
\centering
\includegraphics[scale=0.6]{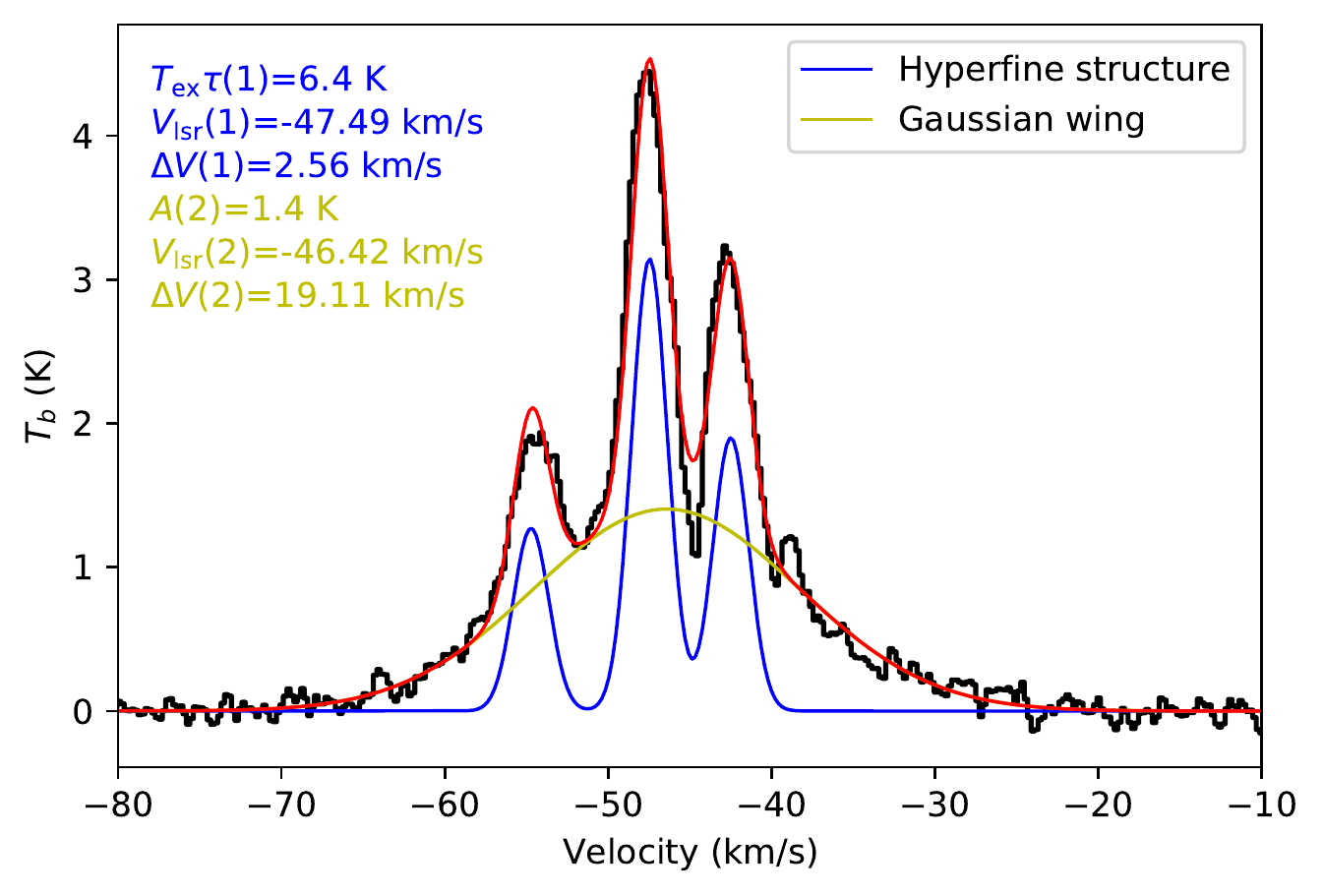}
\caption{Result of fitting \htcn\,$J=1-0$. 
The fitting model contains one hyperfine structure component and one Gaussian component. 
The former has narrow linewidth and strong emission. 
The latter has extended wings. 
The fitted parameters are listed on the left. 
Since the optical depth of \htcn\,$J=1-0$ is small, the amplitude (the function of excitation temperature $T_\mathrm{ex}$) and optical depth are coupled.
We use the parameter $T_\mathrm{ex}\tau$ to represent their product. \label{fig:h13cn_fit}}
\end{figure}

Each hyperfine component is considered as a single Gaussian profile at its centroid and neighborhood ($\sim 3 \sigma$). In other words, at the $\sim 3 \sigma$ range of the hyperfine component $i$, the brightness temperature can be written as:
\begin{equation}
T_b(v)|_{-3\sigma\,\leq\,v_{0i}\,\leq\,3\sigma} = \mathcal{A}\tau_\mathrm{tot} r_i \times \exp\left[-4\ln 2\left(\frac{v-v_\mathrm{LSR}-v_i}{\Delta v}\right)^2\right]
\end{equation}
and it is not possible to determine $\mathcal{A}$ and $\tau_\mathrm{tot}$ degenerately. To stabilize the numerical calculation of HFS fitting, we therefore define the product of the amplitude $\mathcal{A}$ and the total optical depth $\tau_\mathrm{tot}$ as a variable $T_\mathrm{ex}\tau = \mathcal{A}\tau_\mathrm{tot}$ \citep{2013A&A...553A..58L,2021ApJ...912..148L}. The other three variables are respectively $V_\mathrm{LSR}$, $\Delta V$, and $\tau_\mathrm{tot}$. The HFS fitting is performed by \textsc{\large python} package \textsc{\large lmfit} with ``leastsq'' method to minimize the $\chi^2$. The result is shown in Figure\,\ref{fig:h13cn_fit}.

\section{Optical depth of \texorpdfstring{\htcop\,$J=1-0$}. line}
\label{app:tau_h13cop}
To estimate the optical depth of \htcop~$J=1-0$, we assume local thermodynamic equilibrium (LTE), and therefore the brightness temperature $T_b$ can be derived from
\begin{equation}
\label{eq:radiation_transfer}
T_b = f\left[J_\nu(T_\mathrm{ex}) - J_\nu(T_\mathrm{bg})\right](1-e^{-\tau}),
\end{equation}
where $J_\nu(T) = \frac{h\nu}{k_\mathrm{B}}\frac{1}{\exp{(h\nu/kT)}\,-1}$.
The beam filling factor $f$ is taken to be 1 since ALMA can well resolve the structure. The cosmic background radiation $T_\mathrm{bg}$ is 2.73\,K. 

\begin{figure}
\centering
\includegraphics[scale=0.4]{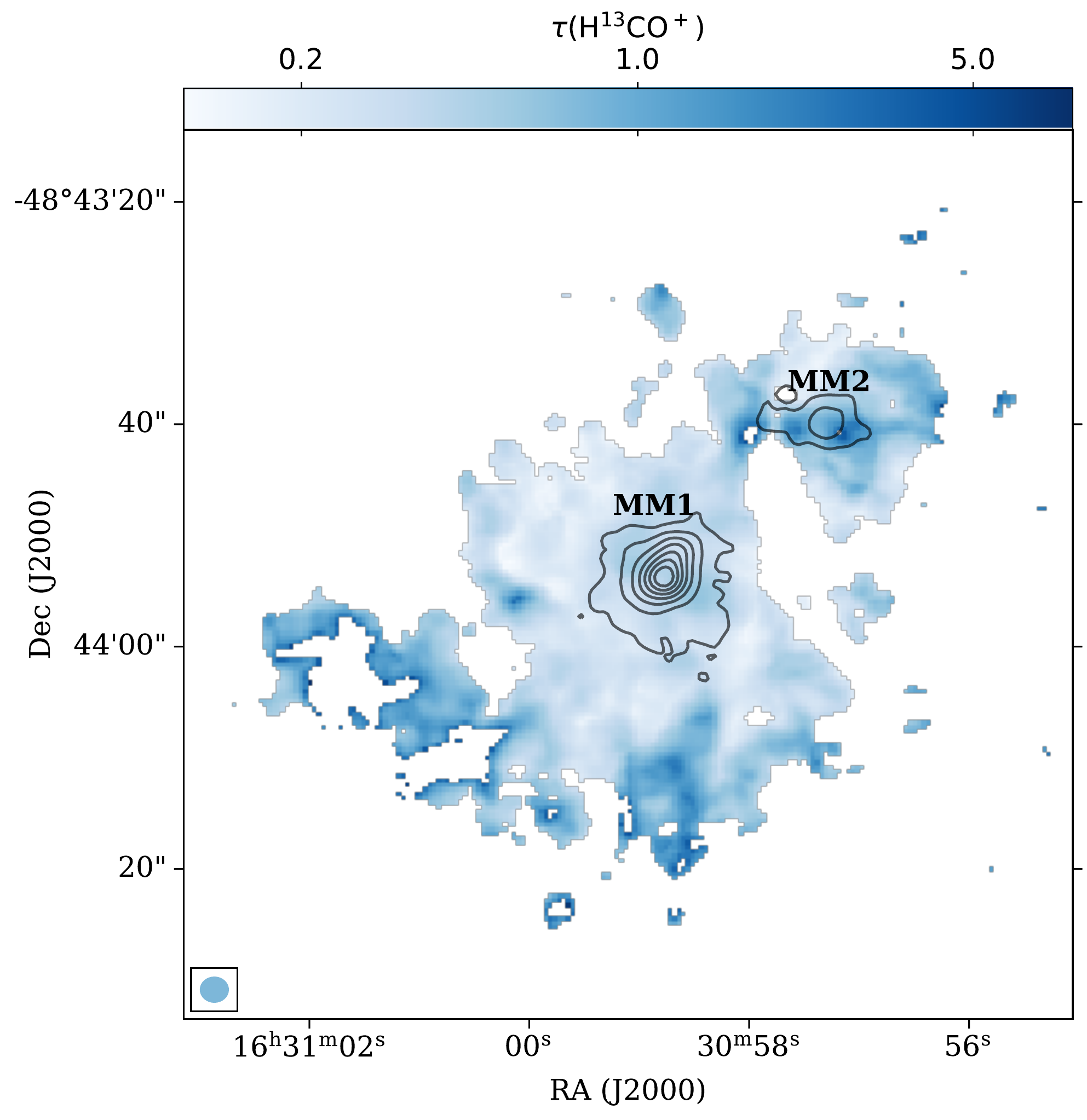}
\includegraphics[scale=0.65]{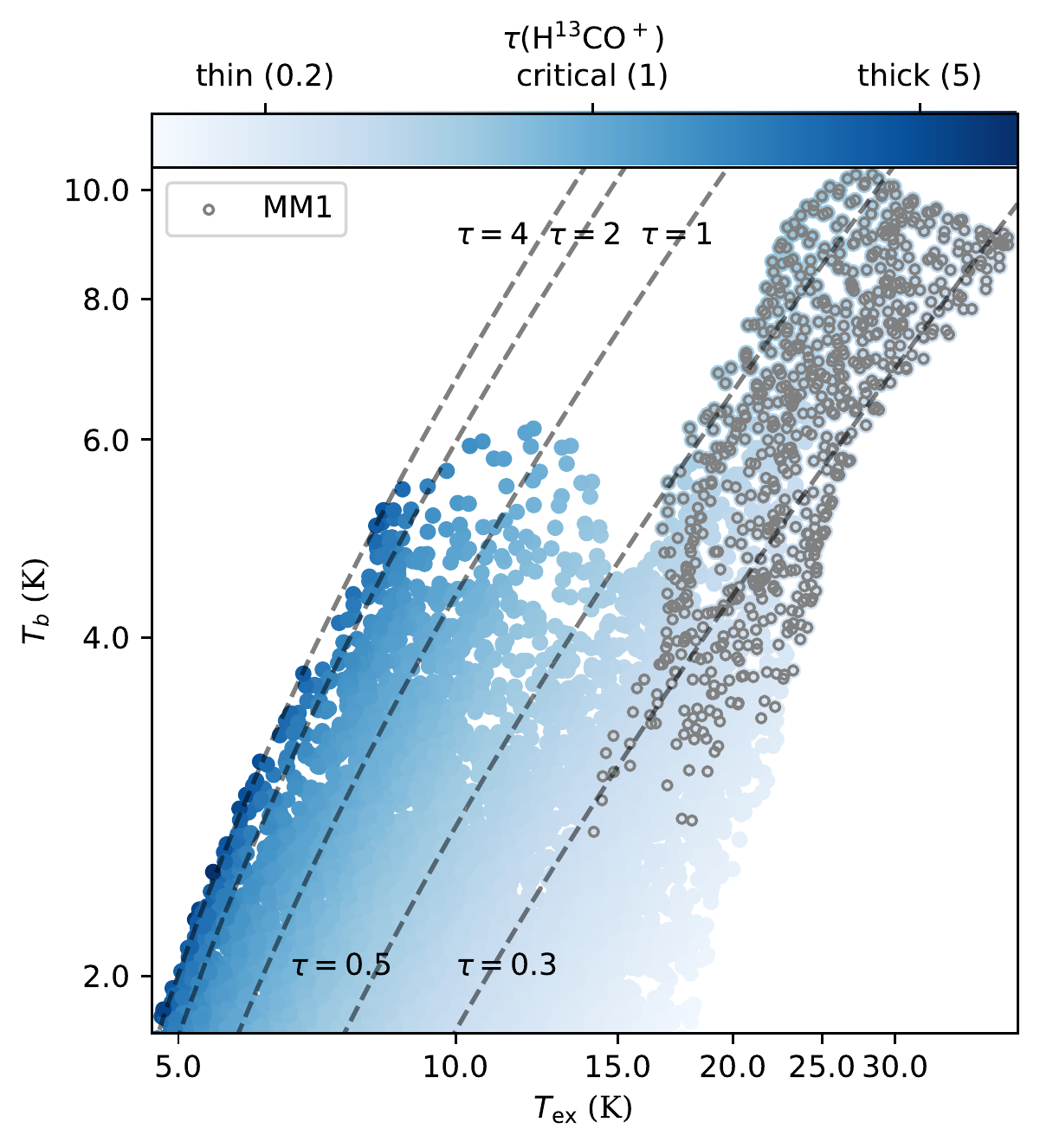}
\caption{Demonstration of the optical thinness of \htcop\,$J=1-0$. 
\textit{Top}: spatial distribution of optical depth $\tau$(\htcop) overlaid with ALMA Band-3 continuum contours. The white contours of 3\,mm continuum emission follows the power-law levels of [1.0, 3.3, 7.6, 14.2, 23.4, 35.1, 49.6, 67.0]\,\mjybeam. Two massive dense cores MM1 and MM2 are shown. The pixels with SNR < 5 are masked. The ALMA beam is shown at the lower left. 
\textit{Bottom}: distribution of $\tau$(\htcop) on the $T_b$-$T_\mathrm{ex}$ plane. 
$\tau=$0.3, 0.5, 1, 2, and 4 are shown with black dashed lines. 
The pixels within the MM1 mask ($F_\mathrm{cont,3mm} \geq 1.0$\,mJy) are shown with gray hollow circles. \label{fig:h13cop_tau}}
\end{figure}

\hcop\,$J=1-0$ is taken to be optically thick with $\tau \gg 1$. We simplify it as $T_b = J_\nu(T_\mathrm{ex}) - J_\nu(T_\mathrm{bg})$, with which we estimate the excitation temperature $T_\mathrm{ex}$ for \hcop\,$J=1-0$. We simply assume that the two molecules share the same $T_\mathrm{ex}$ and return to Equation\,\ref{eq:radiation_transfer}. 
For $\tau \ll 1$, the original equation becomes $T_b = \tau\left[J_\nu(T_\mathrm{ex}) - J_\nu(T_\mathrm{bg})\right]$. 
Substituting the parameters of \htcop\,$J=1-0$, $\tau$(\htcop) can be derived. 
As seen from Figure\,\ref{fig:h13cop_tau}, the optical depth $\tau$(\htcop) is universally thin, even for the densest part near the MM1.

We also test the optically thick assumption for \hcop\,$J=1-0$. Based on the radiation transfer equation\,\ref{eq:radiation_transfer}, we have,
\begin{equation}\label{eq:tau_12}
\begin{split}
\frac{T_{b,\mathrm{HCO}^+}(v)}{T_{b,\mathrm{H}^{13}\mathrm{CO}^+}(v)} & = \frac{1-\exp(-\tau_{v,12})}{1-\exp(-\tau_{v,13})} \\
& > \frac{1-\exp(-\tau_{v,12})}{\tau_{v,13}} = \frac{1-\exp(-\tau_{v,12})}{\tau_{v,12}}\frac{X_{12}}{X_{13}}
\end{split}
\end{equation}
where $\tau_{v,12}$ and $\tau_{v,13}$ are the optical depth at $v$ of the \hcop\,$J=1-0$ and \htcop\,$J=1-0$, and $X_{12}/X_{13}$ = [\hcop]/[\htcop] is the abundance ratio of \hcop~to \htcop, which is estimated to be 15--20 from Mopra observations and 1D non-LTE RATRAN radiative transfer modeling \citep{2013A&A...555A.112P}. 

Since the linewidth of \hcop\,$J=1-0$ is always larger than that of \htcop\,$J=1-0$, we only consider the peak brightness temperature at $v=v_\mathrm{peak}$. 
Among the whole map with solid data (both two lines should have SNR>5), $T_{b,\mathrm{HCO}^+}(v=v_\mathrm{peak})/T_{b,\mathrm{H}^{13}\mathrm{CO}^+}(v=v_\mathrm{peak})$ is 2.7 on average and 9 at maximum. 
By substituting these into equation\,\ref{eq:tau_12}, we estimate the lower limit of $\tau_{12}$ at $v=v_\mathrm{peak}$ to be 7.4 on average, and only is as low as $\sim$2 in extreme cases. 
In other words, the optically thick assumption, $\tau_{12}$ > 1, is always satisfied.

\section{SCOUSEPY decomposition of the ALMA \texorpdfstring{\htcop\,$J=1-0$}. map}
\label{app:decompose}
The first release version of \scousepy~is \textsc{\large scouse} \citep[short for Semi-automated multi-COmponent Universal Spectral-line fitting Engine][]{2016MNRAS.457.2675H}. \textsc{\large scouse} is an IDL package that fits complex spectral line data in a robust, systematic and efficient way, by manually fitting the spatially-averaged spectra into Gaussian components and then using them as input to the subsequent fitting on each individual spectrum. \citet{2019MNRAS.485.2457H} developed Python implementation \scousepy
\footnote{SCOUSEPY is publicly available: \href{https://github.com/jdhenshaw/scousepy}{https://github.com/jdhenshaw/scousepy}. Alternatively, the original IDL implementation can be downloaded here: \href{https://github.com/jdhenshaw/scouse}{https://github.com/jdhenshaw/scouse}.}
and its latest version \scousepy~v2.0.0 improves the workflow and operation interface. The procedure of \scousepy~is discussed in detail by \citet{2016MNRAS.457.2675H}, but we highlight the key points here.

\begin{figure}
\centering
\includegraphics[scale=0.48]{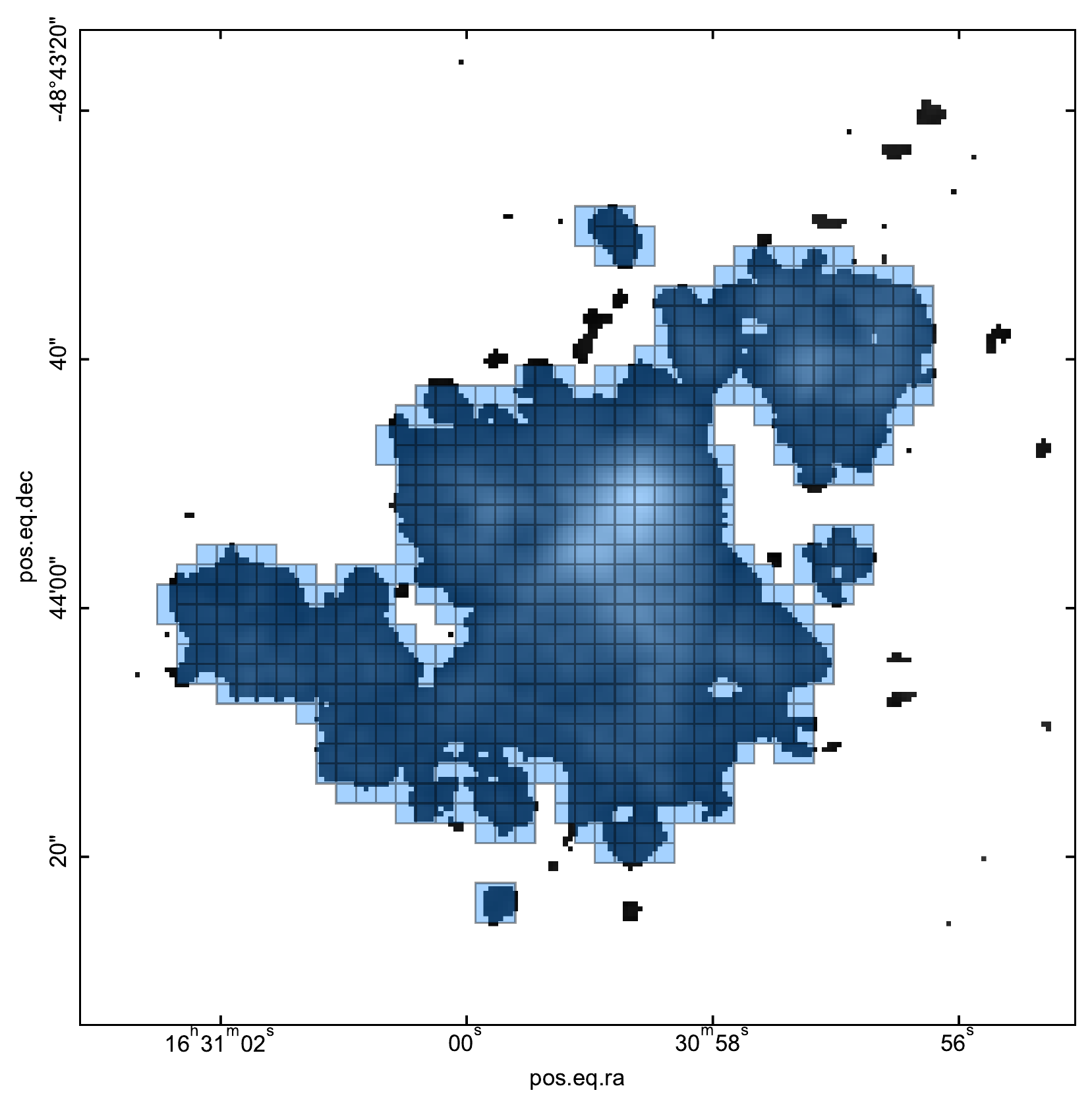}
\caption{The breaking-up stage of \scousepy~where the \hcop\,$J=1-0$ map is broken into 555 SAAs. \label{fig:h13cop_coverage}}
\end{figure}

First, we identify both the spatial and spectral coverage over which it will perform the fitting. We calculate the pixelwise noise from the baseline without \htcop~emission and mask the spectra whose peak flux is below $5\sigma$. 
The reason to do this is that the primary-beam corrected ALMA data has radial variation of noise. By doing so, the unmasked data has a uniform $\mathrm{SNR}>5$ for further spectral line fitting. 
We also cut the velocity range from -55 to -34\,\kms, both covering all the emission from \htcop\,$J=1-0$ and releasing the burden of loading data. 
Second, we set the Spectral Averaging Areas (SAAs) to be $8\times8$ pixel$^2$ and \scousepy~then breaks up the map (with 8873 pixels) into 555 small areas as shown in Figure\,\ref{fig:h13cop_coverage}. 
The spatially averaged spectra extracted from each SAA are then manually fitted. The spectra is assumed to be composed of a series of Gaussian profiles, as justified in Section\,\ref{ssubsec:justify}. 
Best-fitting solutions to the SAAs are then supplied to the fully automated fitting procedure that targets all of the individual spectra contained within each region. 
This process is controlled by a number of tolerance levels. Last, we check the fitting result and the quality due to the reduced $\chi^2$. Some minor manual adjustment can be done here.

\section{Clustering of decomposed Gaussian components by ACORNS}
\citet{2019MNRAS.485.2457H} developed a Python written algorithm, \acorns~(Agglomerative Clustering for ORganising Nested Structures)\footnote{ACORNS is publicly available at: \href{https://github.com/jdhenshaw/acorns}{https://github.com/jdhenshaw/acorns}.}. \acorns~is based on a technique known as hierarchical agglomerative clustering whose primary function is to generate a hierarchical system of clusters within discrete data. Refer to \citet{2019MNRAS.485.2457H} for the principle and details of the algorithm. In the following sections, we briefly describe how \acorns~characterizes the velocity structure of ``the Heart''.

We perform the \acorns~decomposition only on the most robust spectral velocity components extracted by \scousepy. 
We define ``robust'' as all velocity components whose peak flux density is greater than 5<$\sigma_\mathrm{rms}$> where <$\sigma_\mathrm{rms}$> is the averaged rms noise among the all pixels, which is $\sim$9.5\,\mjybeam. 
The selected data (15879 components out of 17143) constitute 92.6\% of the total data set extracted by \scousepy.

For the clustering, we set the minimum radius of a cluster to be 2.4\,arcsec, which is equal to the ALMA synthesized beam. 
This is to ensure that all identified clusters are spatially resolved. 
The consequence is that 545 data points are not assigned to clusters and isolated. Considering clustering in p--p--v space, for two data points to be classified as ``linked'', we specify that the Euclidean distance between the points and the absolute difference in both their measured centroid velocity and velocity dispersion can be no greater than 2.4\,arcsec and 0.21\,\kms, respectively\footnote{The velocity difference 0.21 km/s is equal to the velocity resolution of the \htcop~$J=1-0$ observations.}.

\begin{figure}
\centering
\includegraphics[scale=0.3]{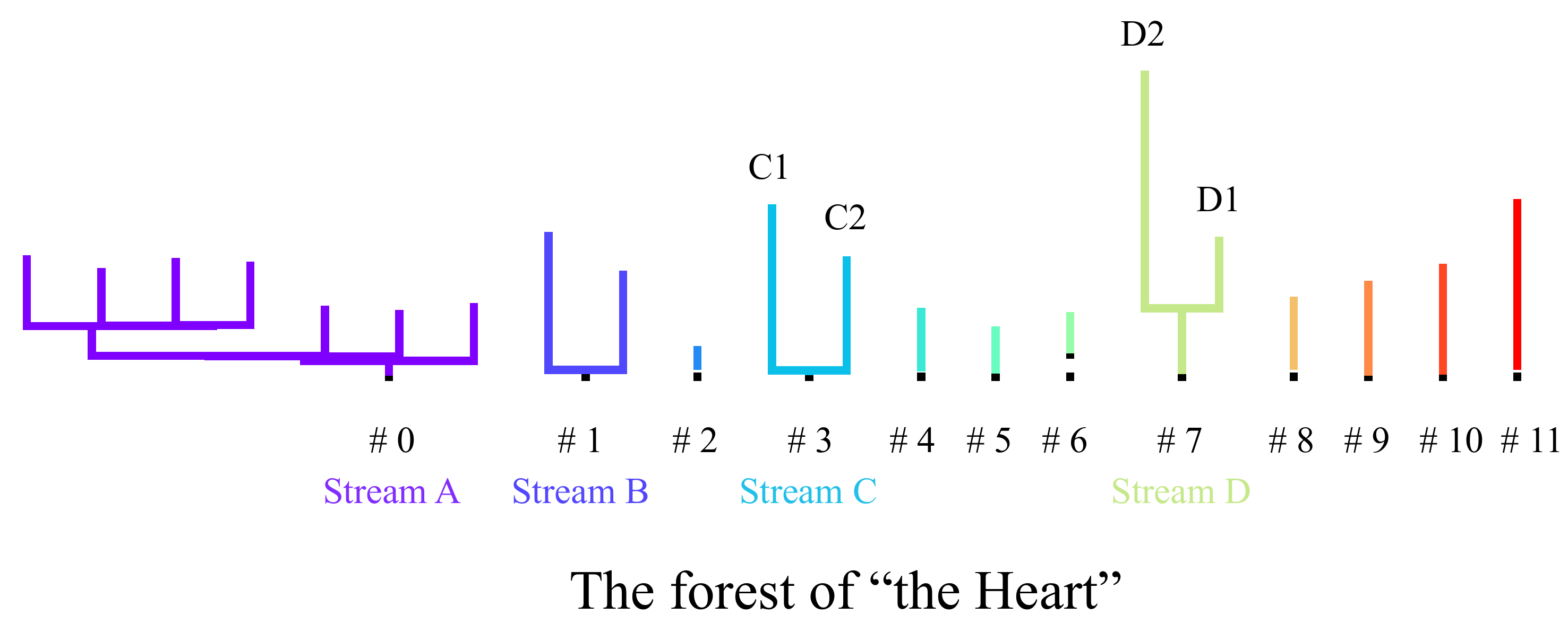}
\includegraphics[scale=0.4]{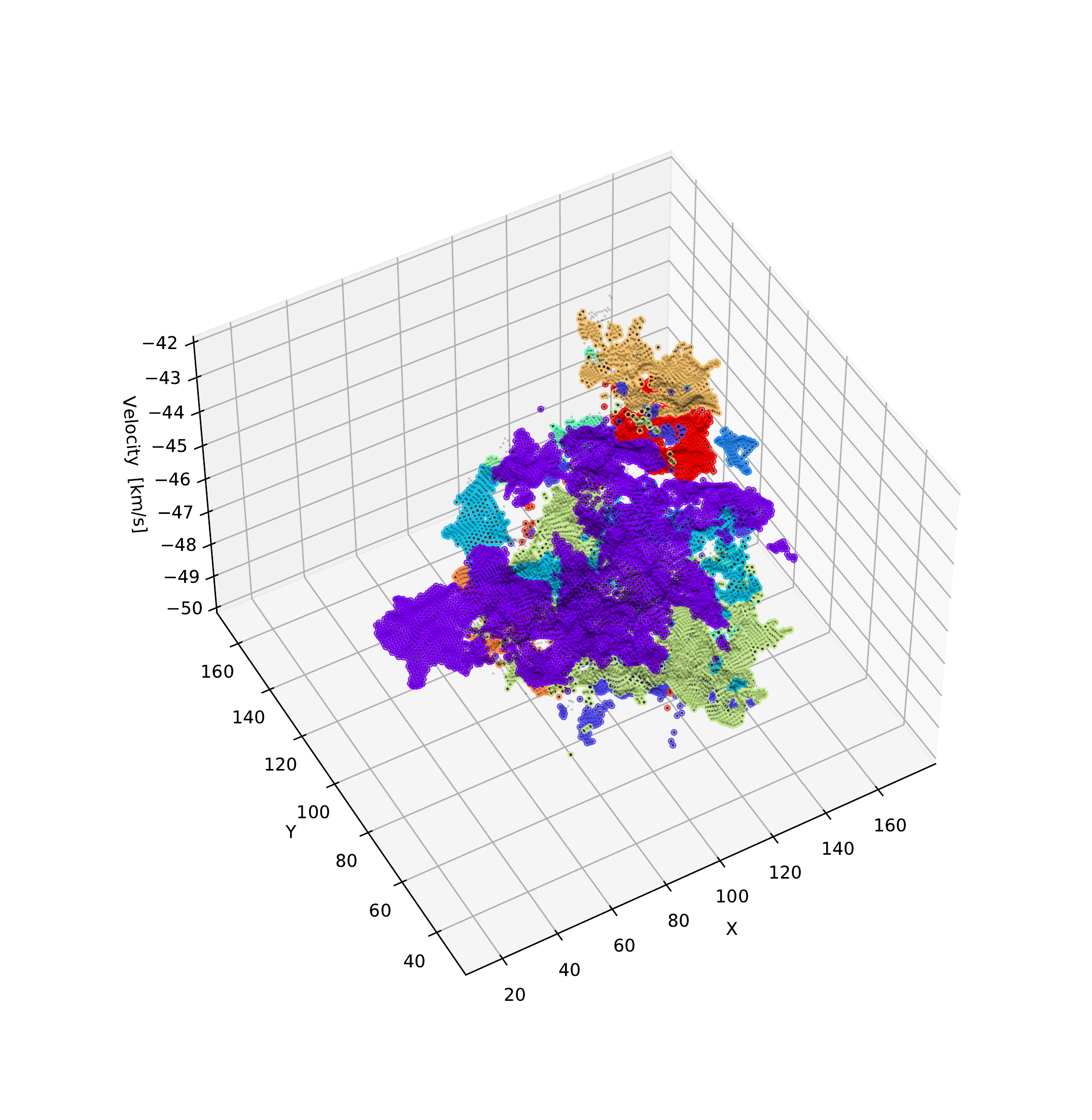}
\caption{The visualization of \acorns~outputs of ``the Heart''. \textit{Top}: \acorns~forest, the hierarchical structure of the data points from \acorns. In total, 12 trees are found and each is plotted in different color. 
Four are hierarchical (\#0, \#1, \#3, and \#7) and eight are non-hierarchical. 
The four hierarchical trees are named major streams A, B, C, and D, respectively. \textit{Bottom}: p--p--v 3D plot of the forest. The colors are the same as the top panel. \label{fig:acorns_output}}
\end{figure}

In the case of ``the Heart'', \acorns~builds a forest with a total of 12 trees, using 98 per cent of the selected ``robust'' data. 
As seen from Figure\,\ref{fig:acorns_output}. the forest consist of four hierarchical and eight non-hierarchical trees\footnote{\citet{2019MNRAS.485.2457H} expands on the nomenclature used in dendrograms. The whole tree system is called ``forest'', itself containing numerous trees. Each tree may or may not then be further subdivided into branches or leaves. 
Trees with no substructure are also classified as leaves.}. 
The forest is dominated by four hierarchical trees: \#0, \#1, \#3, and \#7, which in total contain 75.7\% of all robust data. 
Given the enormity of the data set, we call them as the major streams A, B, C, and D, respectively. ``Stream'' refers to the gas flowing (see Section\,\ref{subsec:stream} for its physical nature). 
The remaining data are classified as isolated points and are not included in the further analyses.

\begin{figure*}
\centering
\includegraphics[scale=0.4]{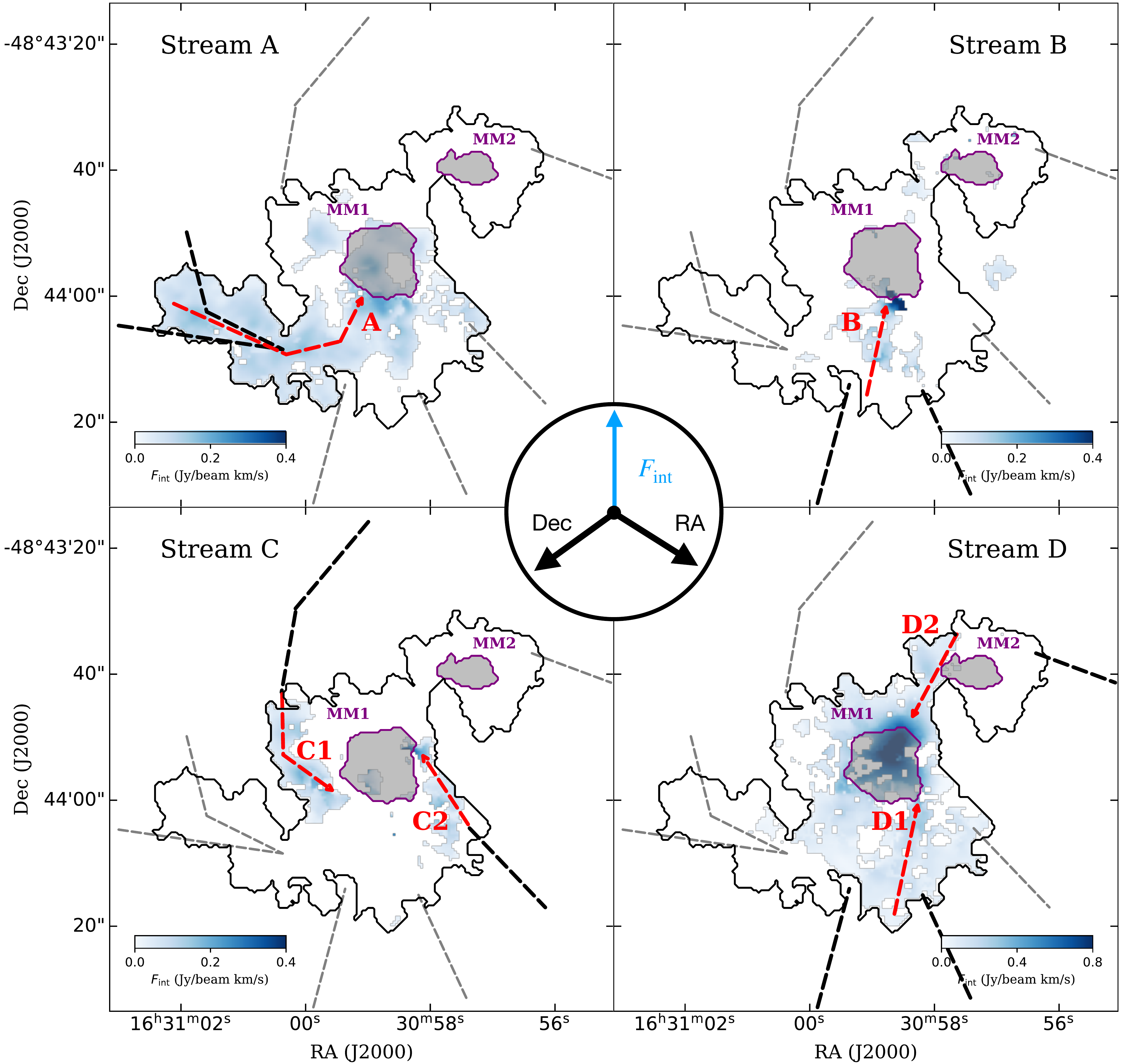}
\caption{The integrated flux maps of the four major streams are normalized to linear span from 0.0 to 0.4\,\jybeam\,\kms~(0.0 to 0.8\,\jybeam\,\kms~for the stream D) in colorscale. The integrated flux at each pixel is given by the integration of decomposed Gaussian component(s). The black solid contours depict the 5$\sigma$ level of \htcop\,$J=1-0$. The six groups of dashed gray lines are the filaments identified from the Spitzer 8\,$\mu$m extinction map (the same as in the middle panel of Figure\,\ref{fig:zoomin}). 
The black bold dashed lines mark the filaments which are assumed to be responsible for the streams in each panel. Streams including two major ones (A and B) and four minor ones (C1, C2, D1, and D2) are marked as red bold dashed vectors. The massive dense cores MM1 and MM2 together with their boundaries ($F_\mathrm{cont,3mm}=1.0$\,\mjybeam) are shown in purple. The central circle marks how we visualize the 3D data cube: blue axis is the collapsing axis, where integrated fluxes of pixels are projected on the plane spanned by two black axes. \label{fig:stream_integrate}}
\end{figure*}

Figure\,\ref{fig:stream_integrate} shows the integrated flux maps of the four major streams.

\clearpage
\onecolumn

\bsp	
\label{lastpage}
\end{document}